# Prediction of novel ordered phases in U-X (X= Zr, Sc, Ti, V, Cr, Y, Nb, Mo, Hf, Ta, W) binary alloys under high pressure

Xiao L. Pan[a,b], Hong X. Song[a,*], H. Wang[a], F. C. Wu[a], Y. C. Gan[a], Xiang R. Chen[b,*], Ying Chen[d], Hua Y. Geng[a,c,*]

[a] *National Key Laboratory of Shock Wave and Detonation Physics, Institute of Fluid Physics, China Academy of Engineering Physics, Mianyang, Sichuan 621900, P. R. China;*

[b] *College of Physics, Sichuan University, Chengdu 610065, P. R. China;*

[c] *HEDPS, Center for Applied Physics and Technology, and College of Engineering, Peking University, Beijing 100871, P. R. China;*

[d] *Fracture and Reliability Research Institute, School of Engineering, Tohoku University, Sendai 980-8579, Japan.*

**Abstract:** U-based binary alloys have been widely adopted in fast nuclear reactors, but their stability under extreme conditions of high-pressure is almost unknown, mounting up to latent risk in applications. Here, possible ordered phases in U-Zr system up to 200 GPa are comprehensively investigated by unbiased first-principles structure prediction. Stable $U_2Zr$, metastable $U_3Zr$ and $U_4Zr$ phases are discovered for the first time, which exhibit strong stability under compression. They all are metallic, with 5*f* electrons of uranium dominating the electronic density of states near the Fermi level. Prominent ionic interactions between U and Zr atoms, as well as covalent interactions between adjacent uranium atoms, are found. The same strategy is applied to explore the stability of ordered phases in other U-based binary transition metal alloys, U-X (X= Sc, Ti, V, Cr, Y, Nb, Mo, Hf, Ta, W). Stable and metastable ordered phases similar to U-Zr alloy are unveiled, all with similar electronic structures. For these alloys, we find that the structure of $U_2X$ (X=Zr, Ti, Hf) hosts a unique hybrid phase transition similar to $U_2Nb$, which is a superposition of a first-order transition and a second-order transition. The prediction of these novel phases not only refutes the stability of the long-believed ordered phase *I4/mmm*-$U_2Mo$, but also rewrites the phase diagrams of U-X (X= Zr, Sc, Ti, V, Cr, Nb, Mo, Hf, Ta) alloys under high pressure. All of these findings promote our understanding of the high-pressure behavior of the broad category of U-based binary alloys with transition metals.

* *Corresponding authors. E-mail: s102genghy@caep.cn; xrchen@scu.edu.cn; hxsong555@163.com*

# I. Introduction

Uranium-based binary alloys have attracted extensive research interests due to their excellent performance as a nuclear fuel in terms of thermal conductivity, burnup, and fuel-recycling compared to conventional ceramic type fuels such as $UO_2$ [1, 2]. Among them, U-Zr and U-Pu-Zr alloys are the most promising candidates for the next generation of nuclear reactor [3-5]. Addition of Zr into uranium improves the latter's corrosion resistance and increases the dimensional stability during thermal cycling [6]. Moreover, uranium alloyed with Zr stabilizes the body-centered cubic (bcc) $\gamma$ phase with excellent mechanical properties over a wide temperature range, and also makes up for the low melting point of $\gamma$-U [7]. Because of these advantages, a typical U-Zr alloy containing about 10 wt. % Zr (~23 at. %) has been irradiated in the Experimental Breeder Reactor (EBR) to validate its performance as a nuclear reactor fuel [4].

According to the phase diagram of U-Zr alloy [7-11], it has a solid solution region at high temperature (above 877 K at 0 GPa) with complete mutual solubility of $\gamma$-U and $\beta$-Zr (both in bcc structure). However, the solubility of Zr in $\alpha$-U becomes very low at room temperature, so that the α-U phase has a very small interval in the U-rich region of the phase diagram. In addition, a nonstoichiometric $\delta$ phase is also observed to exist at low temperature (below 893 K at 0 GPa) [12]. Homogeneity range and crystal structure of the $\delta$ phase were first determined by Akabori et al. [13, 14] using x-ray diffraction experiments. They found that the homogeneity range of the $\delta$ phase is from 44.4 wt. % (66.5 at. %) to 62 wt. % (80.2 at. %) Zr at 550°C. The crystal structure of $\delta$-$UZr_2$ phase was proposed to be the hexagonal $AlB_2$-type [15] lattice with space group

*P6/mmm*, where the Zr atoms completely occupy 1*a* (0, 0, 0) sites, while the 2*d* (1/3, 2/3, 1/2) sites are randomly occupied by U and Zr atoms. Shukla et al. reported that *δ*-$UZr_2$ can be stable up to 20 GPa in Diamond Anvil Cell (DAC) experiments [16]. The formation mechanism of *δ*-$UZr_2$ phase has also received extensive research attention [17-19]. These studies suggested that the transition from *γ*-(U, Zr) phase to *δ*-$UZr_2$ phase takes place by a collapse of the (111) plane of the *γ*-(U, Zr) phase.

For U-Zr alloy, only the *δ*-$UZr_2$ intermetallic phase has been reported. On the other hand, relevant U-based transition metal binary alloys, e.g., U-Ti and U-Mo alloys, were found to form ordered phases with a $U_2X$ (X=Ti, Mo) stoichiometry in the U-rich region [20-24]. $U_2Ti$ and $U_2Mo$ have been assumed to adopt the same $AlB_2$-type structure as *δ*-$UZr_2$, and have been reported being stable up to at least 100 GPa in theoretical studies [25, 26]. In addition, our recent study also showed that the same $AlB_2$-type structure exists in U-Nb alloy with a $U_2Nb$ stoichiometry under high pressure [27]. These preliminary results indicate that our current understanding of the phase diagram of U-based binary alloys is incomplete, which motivates us to explore all possible ordered phases of U-Zr and U-X (X= Sc, Ti, V, Cr, Y, Nb, Mo, Hf, Ta, W) alloys at high pressures. Importantly, the outcome of this exploration might be of great significance for improving the performance and safety of U-based alloy fuels.

To this end, we first carry out a systematic study of the possible ordered phases in U-Zr alloy using unbiased first-principles structure prediction, and then extend the analysis to other relevant U-X (X= Sc, Ti, V, Cr, Y, Nb, Mo, Hf, Ta, W) systems. Several previously unknown ordered intermetallic phases are predicted for U-Zr and U-X (X= Sc, Ti, V, Cr, Nb, Mo, Hf, Ta) alloys. All of them exhibit strong stability at high pressure. Our finding also refutes the well believed and widely studied *I4/mmm*-$U_2Mo$ as stable states at 0 GPa (It, however, might be metastable, or completely decomposes under certain conditions). The remainder of this paper is organized as follows. In next section, the computational method is briefly described, the results and discussion are presented in Section III, and Section IV summarizes the main findings and conclusions of the paper.

## II. Computational Method

The structure search for stable ordered phases of U-Zr system at 0, 25, 50, 75, 125, and 200 GPa were carried out with particle swarm optimization (PSO) method as implemented in CALYPSO [28]. Nine different stoichiometries (U:Zr = 6:1, 4:1, 3:1, 2:1, 1:1, 1:2, 1:3, 1:4, 1:5) with a size up to 4 formula units per cell for each composition were employed, while the structure prediction for U and Zr elements was performed with the maximum simulation cell of 8 formula units. There are 30 structures generated in each generation, with a total number of 30 generations for each search. The 20 candidate structures with the lowest enthalpy in each search were then carefully reoptimized with higher accuracy to determine the most stable structure. Structural relaxations and total energy calculations were performed using density functional theory (DFT) [29, 30] as implemented in the Vienna Ab initio Simulation Package (VASP) [31, 32]. The exchange-correlation functional is described by the Perdew-Burke-Ernzerhof (PBE) [33] parameterization of the generalized gradient approximation (GGA). The atomic coordinates and lattice vectors were fully optimized until the Hellmann–Feynman forces acting on each atom were less than $10^{-3}$ eV/Å and the convergence of the total energy was better than $10^{-7}$ eV. The cut-off energy of 550 eV for the plane-wave basis and the Monkhorst-Pack [34] k-meshes with a grid spacing of $2\pi \times 0.015$ Å$^{-1}$ were used to ensure that the energy of each structure converged to 1 meV per atom or better for all calculations. The cell shape is completely relaxed in the calculations of structural relaxation to achieve the hydrostatic conditions. In addition, spin polarization is taken into account in our calculations. Except for Cr metal, which exhibits antiferromagnetism at low pressure, as well as the $P6_3/mmc$-$U_2Zr$ phase in the U-Zr system and some ordered phases in the U-Y system, which display weak ferromagnetism at low pressure (less than $0.5\mu_B$ per uranium atom), all other compounds we studied are non-magnetic.

To determine the thermodynamic stability of U-X system at zero temperature, we further calculated the formation enthalpy of the most stable candidate structure for each fixed composition. The formation enthalpy per atom ΔH was calculated by following

definition:

$$\Delta H = (H_{total}^{U_xX_y} - yH^X - xH^U)/(x+y), \tag{1}$$

$$H = E + PV, \tag{2}$$

where E, $P$, and $V$ stand for the total internal energy, pressure and volume, respectively. $H_{total}^{U_xX_y}$ is the total enthalpy per formula unit of the candidate ordered phases in U-X system, with $x$ and $y$ represent the number of U and X atoms in each formula unit, respectively. In order to ensure the dynamic stability of the newly discovered ordered phases, the phonon dispersion curves were calculated using the small-displacement method as implemented in PHONOPY code [35]. A 2 × 2 × 2 supercell for *P6₃/mmc*-$U_2X$ phase (48 atoms), 2 × 2 × 4 supercell for *P6/mmm*-$U_2X$ (48 atoms), 3 × 1 × 2 supercell for *Cmcm*-$U_3X$ (96 atoms), 3 × 2 × 1 supercell for *Immm*-$U_4X$ (60 atoms), and 1 × 2 × 2 supercell for *Cmmm*-$U_6X$ (56 atoms) were employed in the calculations of phonon spectra.

The elastic constants of single crystal were calculated by using the energy-strain method as implemented in MyElas [36], where nine strains around the equilibrium lattice constant were employed. Hexagonal and orthorhombic lattices have six and nine independent components, respectively. The mechanical stability of the corresponding structure is determined by the following criteria [37]:

Hexagonal structure ($C_{11}$, $C_{33}$, $C_{44}$, $C_{12}$ and $C_{13}$):

$$C_{44} > 0,\ C_{11} > |C_{12}|,\ (C_{11} + 2C_{12})C_{33} > 2C_{13}^2;$$

Orthorhombic structure ($C_{11}$, $C_{22}$, $C_{33}$, $C_{44}$, $C_{55}$, $C_{66}$, $C_{12}$, $C_{13}$ and $C_{23}$):

$$C_{11} > 0, C_{22} > 0, C_{33} > 0, C_{44} > 0, C_{55} > 0, C_{66} > 0,$$
$$[C_{11} + C_{22} + C_{33} + 2(C_{12} + C_{13} + C_{23})] > 0,$$
$$(C_{11} + C_{22} - 2C_{12}) > 0, (C_{11} + C_{33} - 2C_{13}) > 0,$$
$$(C_{22} + C_{33} - 2C_{23}) > 0.$$

Considering the possible strong correlation interaction between the 5*f* electrons of U atoms, in addition to the standard DFT method, we also employ the DFT+*U* method with and without spin-orbit coupling (SOC) to verify the reliability of the predicted ordered phases by the DFT method. Specifically, we used the simplified DFT+*U* method proposed by Dudarev et al. [38] with only a single effective *U* parameter. The value of *U* was set to 1.24 eV for U-Zr and U-Mo alloys, which was proved being reliable by Xie et al. [39].

# III. Results and Discussion

## A. Structural stability of U-Zr alloy

It is well established that uranium at atmospheric pressure adopts an orthorhombic *α* phase with space group *Cmcm*, which is stable up to 270 GPa according to previous studies [40-42]. At higher pressures, *α*-U undergoes a structural phase transition into a body-centered tetragonal (bct) phase with space group *I4/mmm*. Here we compare the relative stability of these phases of U (*α*, *γ*, and bct) under pressure, along with high-temperature tetragonal *β*-U ($P4_2/mnm$) and *β*-Np (*P4/nmm*) phases, as shown in Fig. 1 (a). The *β*-Np phase of uranium spontaneously relaxes into the bct-U phase when beyond 35 GPa. Our calculations show that *α*-U is stable up to 271 GPa and then transforms to the bct phase, which is consistent with the results of the previous theoretical study [42, 43]. For Zr element, a high-pressure phase transition sequence has been widely accepted: α (hcp) → ω (*P6/mmm*) → *β* (bcc) [44-48]. However, a recent study suggests that Zr may have a bct phase at high pressure [49]. Therefore, we examined its phase stability under high pressure, including five phases *α*, *ω*, *β*, bct, and fcc (may exist at high pressure) of Zr, and the results are shown in Fig. 1(b). Our calculated results indicate that the *α*-*ω* transition occurs at 0.3 GPa, *ω*-bct transition at 26.6 GPa, bct-*β* transition at 33.7 GPa, and the *β* phase is always the most stable at higher pressure up to 200 GPa. At the same time, our structural search for U and Zr elements within 200 GPa also confirmed that these phases are the most stable under the corresponding pressures. The careful confirmation of the stability range of each individual phase of U and Zr elements within 200 GPa is necessary for the following determination of the ordered phases in U-Zr system.

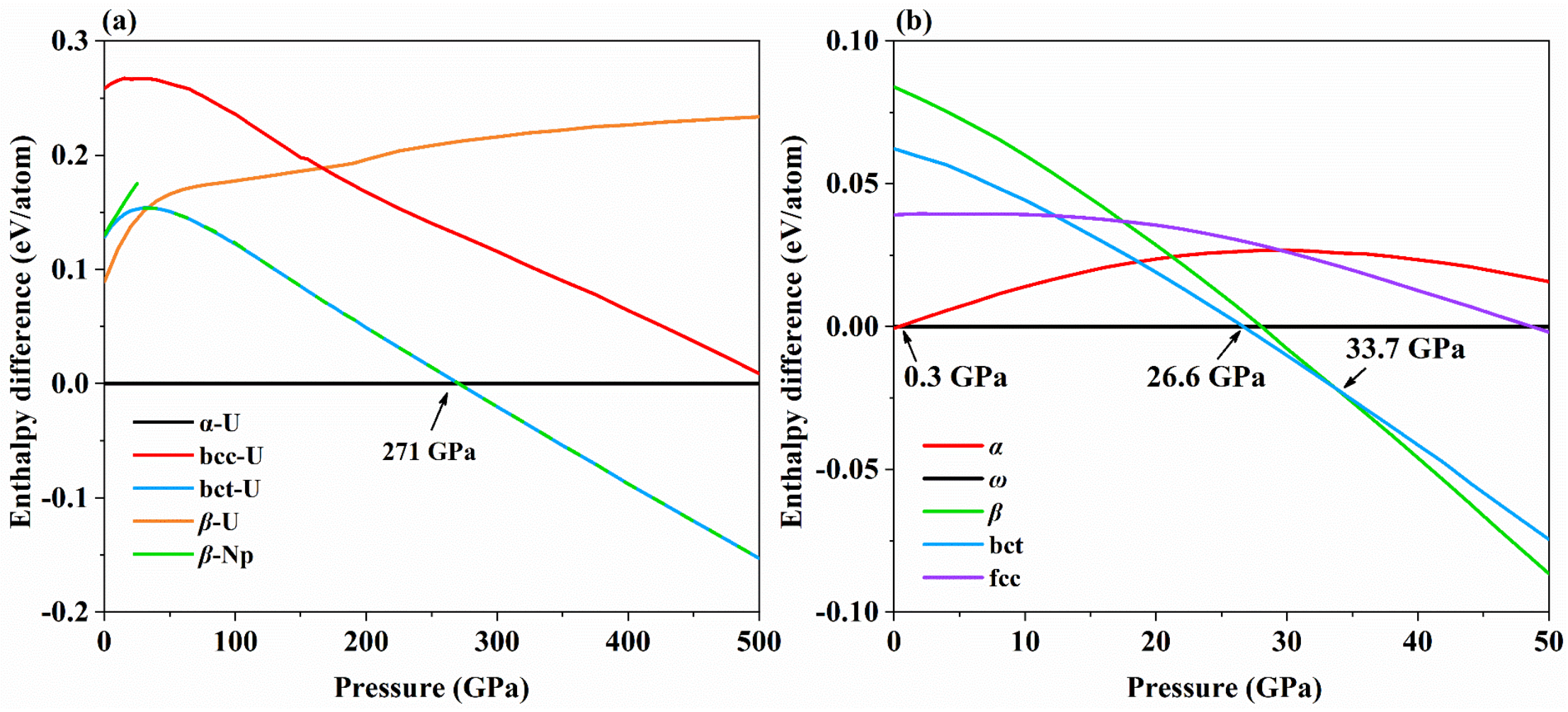

**Fig. 1** (color online) (a) Enthalpy difference of U in *γ*-U (bcc), bct-U, *β*-U, and *β*-Np phases with respect to *α*-U phase; (b) enthalpy difference of Zr in *α*, *β*, bct, and fcc phases with respect to *ω* phase, respectively.

The convex hull diagram of the predicted candidate structures in U-Zr system was constructed and shown in Fig. 2 (a). From the convex hull at ambient pressure, we can conclude that the U-Zr alloy has positive formation enthalpies in the whole composition space. In addition, we also calculated the formation enthalpy of both the partially ordered *δ*-$UZr_2$ phase, which was observed in experiments and long believed to be a stable phase at zero pressure, and the fully ordered *δ*-$UZr_2$ phase for comparison. Specifically, the structure of the partially ordered *δ*-$UZr_2$ phase was modelled by special quasi-random structure (SQS) method. This phase has been shown to be dynamically stable by previous study [19]. In the crystal structure of the fully ordered *δ*-$UZr_2$ phase, one Zr atom occupy the 1*a* (0, 0, 0) site, while U atom and another Zr atom occupy the 2*d* sites at coordinates (1/3, 2/3, 1/2) and (2/3, 1/3, 1/2), respectively. From our structure prediction, the partially ordered *δ*-$UZr_2$ is more stable than all other ordered candidate structure at this 1:2 stoichiometry, which is consistent with previous experimental results that $UZr_2$ stoichiometry tends to form a partially ordered *δ*-$UZr_2$ structure [13, 14, 50]. The formation enthalpy of the candidate structure obtained through our structure prediction in the $UZr_2$ stoichiometry is much lower than that of the fully ordered *δ*-$UZr_2$ phase.

Furthermore, in Zr-rich region, the formation enthalpy of the ordered candidate structures in all stoichiometries increases with increasing pressure, which means that no ordered phase can be formed at high pressure in Zr-rich region. The widely studied $\delta$-$UZr_2$ should not be the ground state at low temperature. It is metastable and will decompose at appropriate conditions (note that all non-stoichiometric compounds are thermodynamically unstable at low temperature).

In the U-rich region, $U_2Zr$ compound has the lowest formation enthalpy. The positive ΔH shows that U-Zr alloys do not have any thermodynamically stable ordered phase at ambient pressure. Upon further compression, the stability of the U-rich region increases gradually, as evidenced by the decreasing of the formation enthalpy with increasing pressure: at 75 GPa, the formation enthalpy of the most stable structure of $U_2Zr$ becomes negative, suggesting that it can be thermodynamically stable at this pressure. At 125 GPa, the formation enthalpy of the most stable structure of $U_3Zr$ also becomes negative. In addition, the formation enthalpy of the most stable structure of $U_4Zr$ is only 0.002 eV/atom at 125 GPa. Further compression to 200 GPa, the formation enthalpy of all stoichiometries in the U-rich region becomes positive but still smaller than 0.02 eV/atom.

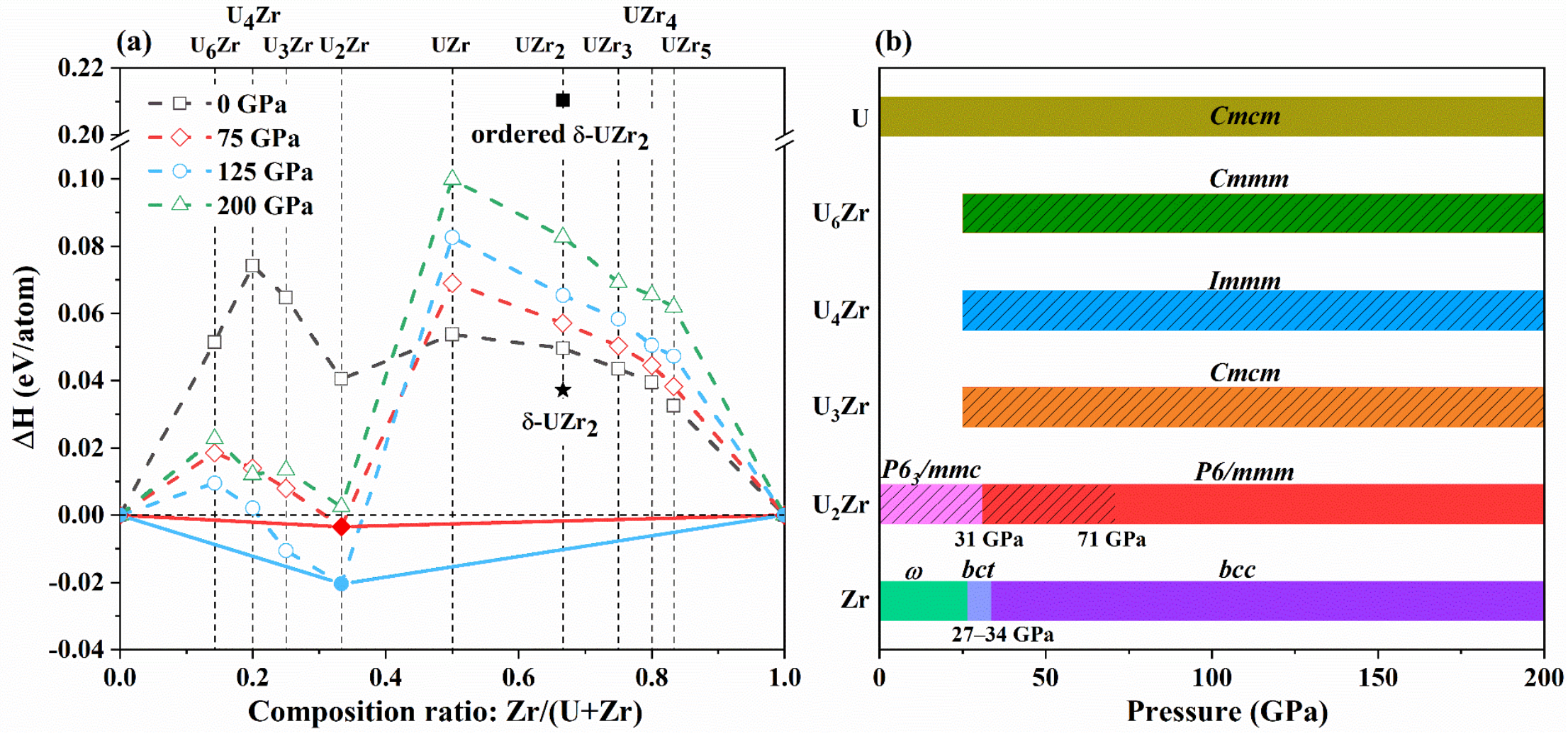

**Fig. 2** (color online) (a) Thermodynamic convex hull diagram of candidate structures in U-Zr system. Dashed line connections indicate unstable or metastable phases, while solid line connections indicate stable phases. The solid star and square indicate the formation enthalpy of the partially ordered $\delta$-$UZr_2$ and fully ordered $\delta$-$UZr_2$ phases at zero pressure, respectively. (b) Pressure–composition phase diagram of U-Zr alloy. Different color regions indicate different structures, and black slashed regions indicate that the corresponding phases is metastable within this pressure range.

In Fig. 2(b), we show the pressure-composition phase diagram of U-Zr alloy. At ambient pressure, $U_2Zr$ can be metastable in a hexagonal $CaIn_2$-type structure with space group $P6_3/mmc$, as shown in Fig. 3(a). Here, we define a metastable phase based on the fact that the structure is not on the convex hull but close to it and exhibits dynamic stability, which is determined by the phonon spectra without any imaginary frequency, as shown in Fig. 4. When the pressure exceeds 31 GPa, it transforms into the $AlB_2$-like structure with space group *P6/mmm* (Fig. 3(b)). The variation of enthalpy difference is presented in Fig. 5. It is noteworthy that the smooth merge of the enthalpy curves of the two phases beyond 31 GPa is due to the spontaneous relaxation of the $P6_3/mmc$ structure into *P6/mmm* structure (the same phenomenon was observed in $U_2Nb$ [27]). For other stoichiometries in the U-rich region, the corresponding phases with the lowest formation enthalpy of $U_3Zr$, $U_4Zr$, and $U_6Zr$ are also dynamically stable and can be metastable from 25 to 200 GPa. They all have orthorhombic lattice with space groups of *Cmcm, Immm,* and *Cmmm* (Figs. 3 (c-e)), respectively. The dynamical stability of these structures is confirmed by the calculated phonon dispersion curves that have no any imaginary frequency, as shown in Fig. 4. Furthermore, the mechanical stability of these structures is also supported by the calculated elastic constants according to the mechanical stability criterion [37] (Table S3). Considering the possible strong correlation of the localized 5*f* electrons of uranium [51-54] , GGA+*U* method was also used to check the results of GGA in terms of the predicted structures, as shown in the supplementary material (SM) (Fig. S2) [55]. The results show that the shape of the convex hull calculated by the GGA+*U* method is consistent with that of GGA, which means that the stable and metastable structures predicted by the two methods are completely consistent.

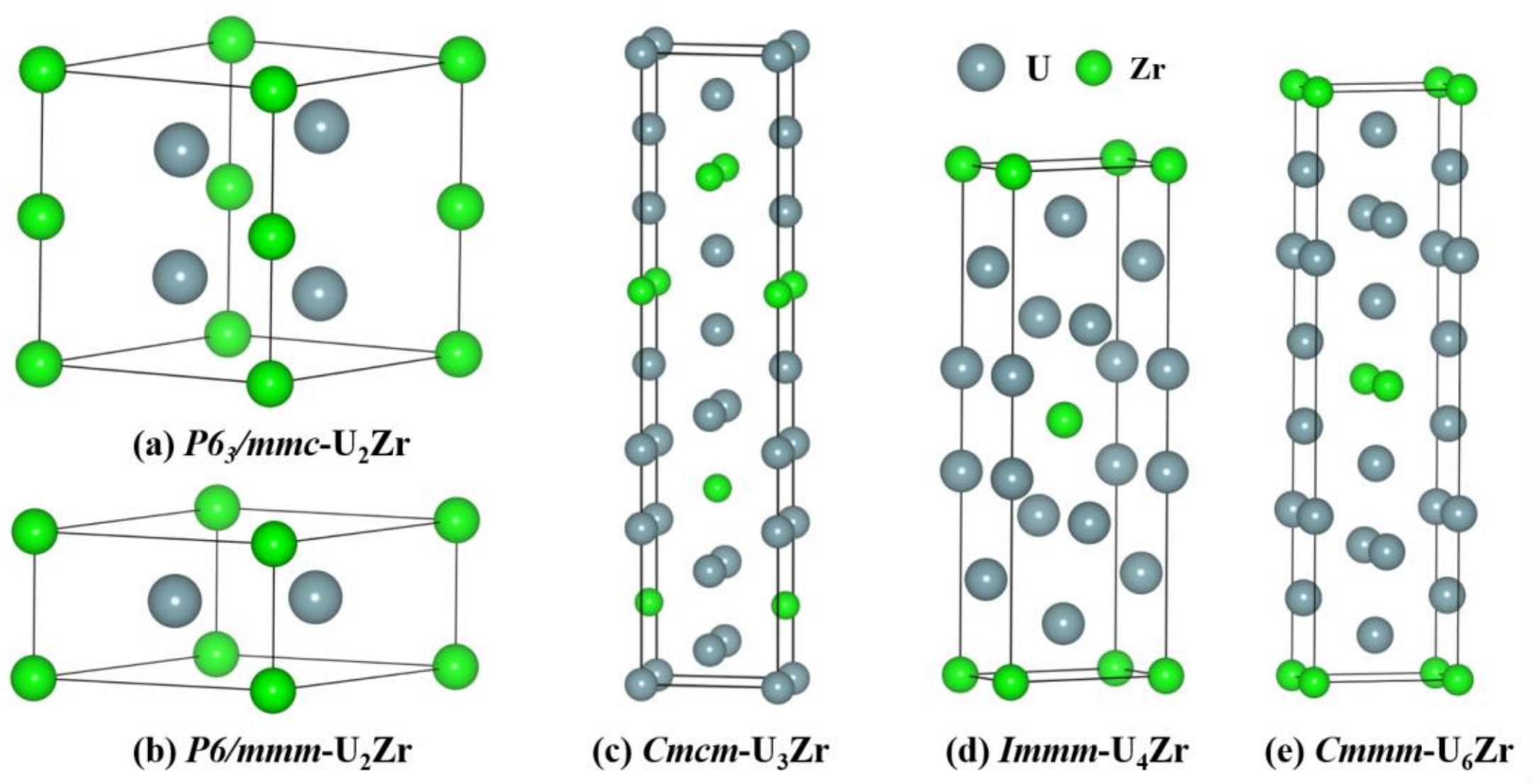

**Fig. 3** Predicted stable and metastable phases for U-Zr alloy: (a) *P6$_3$/mmc*-$U_2Zr$ at 0 GPa, (b) *P6/mmm*-$U_2Zr$ at 75 GPa, (c) *Cmcm*-$U_3Zr$ at 125 GPa, (d) *Immm*-$U_4Zr$ at 125 GPa, and (e) *Cmmm*-$U_6Zr$ at 125 GPa.

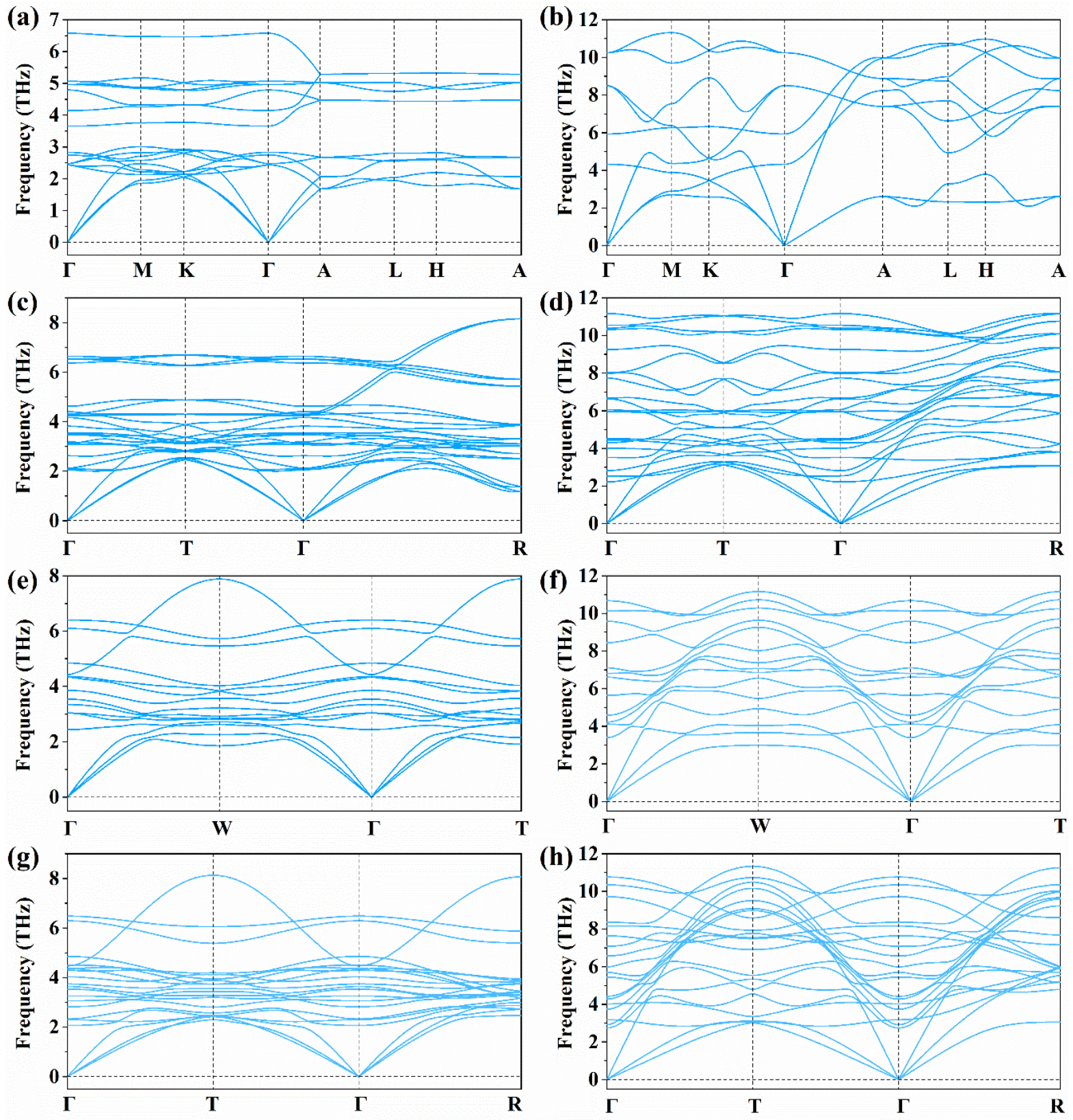

**Fig. 4** Calculated phonon spectra for (a) *P6$_3$/mmc*-$U_2Zr$ at 0 GPa, (b) *P6/mmm*-$U_2Zr$ at 200 GPa, *Cmcm*-$U_3Zr$ at (c) 25 GPa and (d) 200 GPa, *Immm*-$U_4Zr$ at (e) 25 GPa and (f) 200 GPa, and *Cmmm*-$U_6Zr$ at (g) 25 GPa and (h) 200 GPa, respectively.

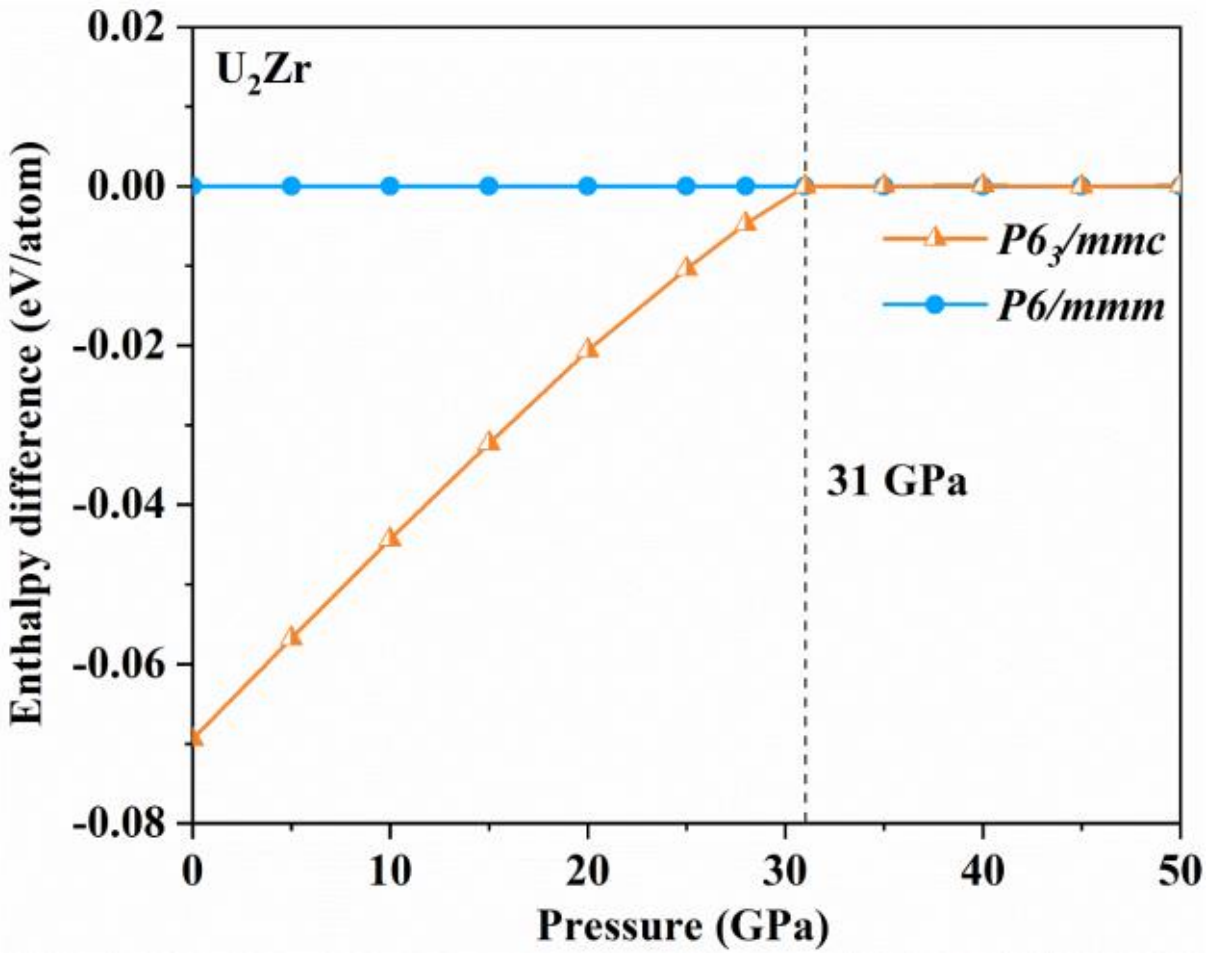

**Fig. 5** (color online) Enthalpy difference of $U_2Zr$ in *P6₃/mmc* phase with respect to the *P6/mmm* phase. The dashed-line indicates the pressure of the phase transition.

## B. Electronic structure and bonding properties of U-Zr alloy

The electronic projected density of states (PDOS) of these ordered phases are plotted in Fig. 6. All of them are metallic with U-5*f* electrons dominating the DOS near the Fermi level. The contribution of Zr-4*d* to the total DOS is very small, suggesting a prominent charge transfer from Zr-4*d* to U-5*f* orbitals. The residual Zr-4*d* electrons hybridize with U-6*d* or U-5*f* states slightly. The charge transfer between U and Zr atoms was estimated by Bader charge analysis [56], as shown in Fig. 7. From the figure, it can be seen that the charge transfer between U and Zr atoms is more noticeable in *P6/mmm*-$U_2Zr$ than in *Cmcm*-$U_3Zr$, *Immm*-$U_4Zr$, and *Cmmm*-$U_6Zr$. The mean average transfer (averaged by atoms of the same type in the unit cell) is -0.76*e* for Zr atoms and +0.38*e* for U atoms in the *P6/mmm*-$U_2Zr$ at 25 GPa. As pressure increases, the charge transfer from Zr atoms to U atoms in these three structures increases slightly at first, then decreases slightly when the pressure exceeds 25 GPa. However, the charge transfer maintains an overall stable level within 200 GPa. This result indicates that there are prominent ionic interactions in U-Zr alloys with these novel structures.

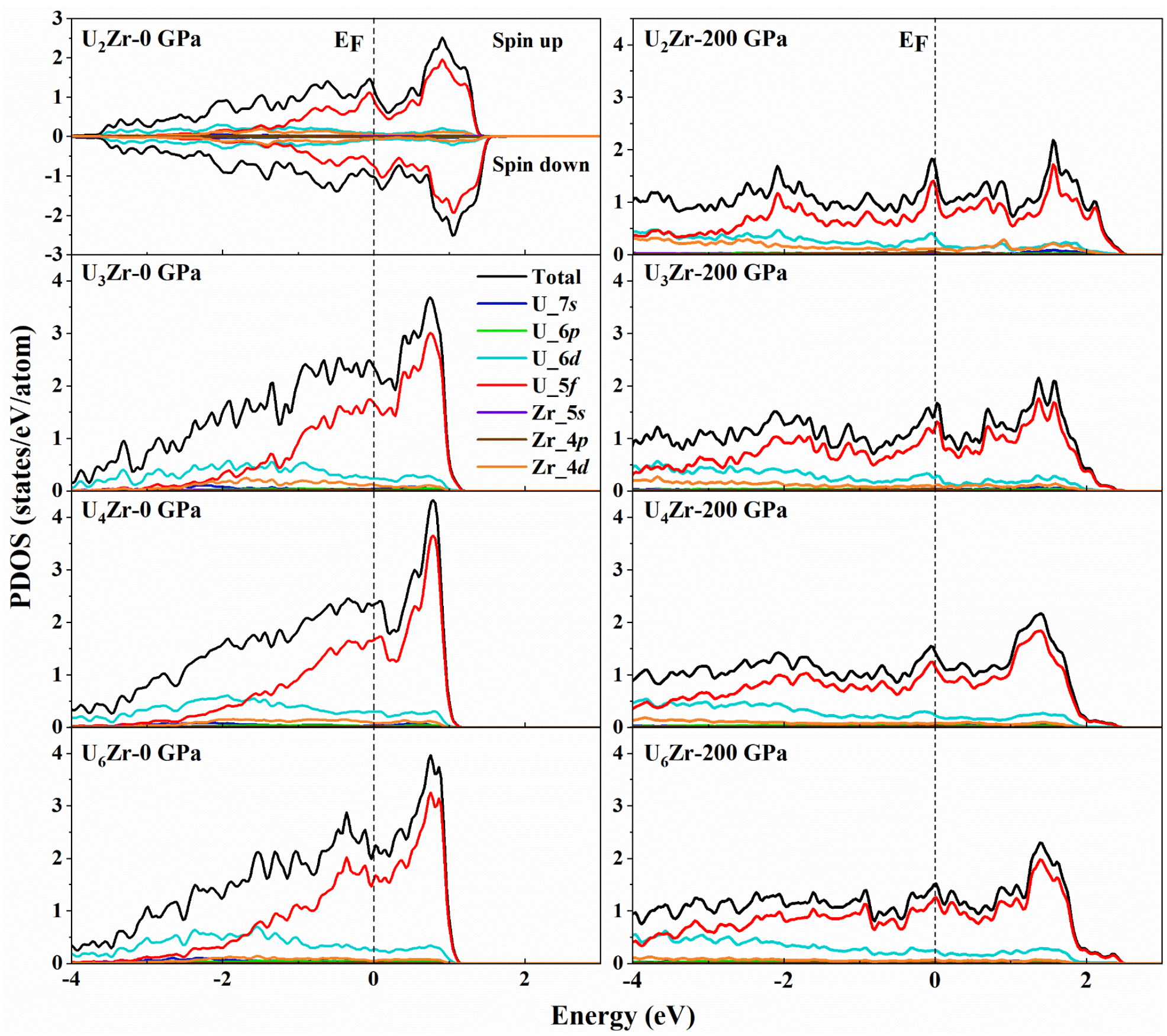

**Fig. 6** (color online) Calculated electronic projected density of states (PDOS) for *P63/mmc*-$U_2Zr$ at 0 GPa, *P6/mmm*-$U_2Zr$ at 200 GPa, and *Cmcm*-$U_3Zr$, *Immm*-$U_4Zr$, and *Cmmm*-$U_6Zr$ at 0 and 200 GPa, respectively.

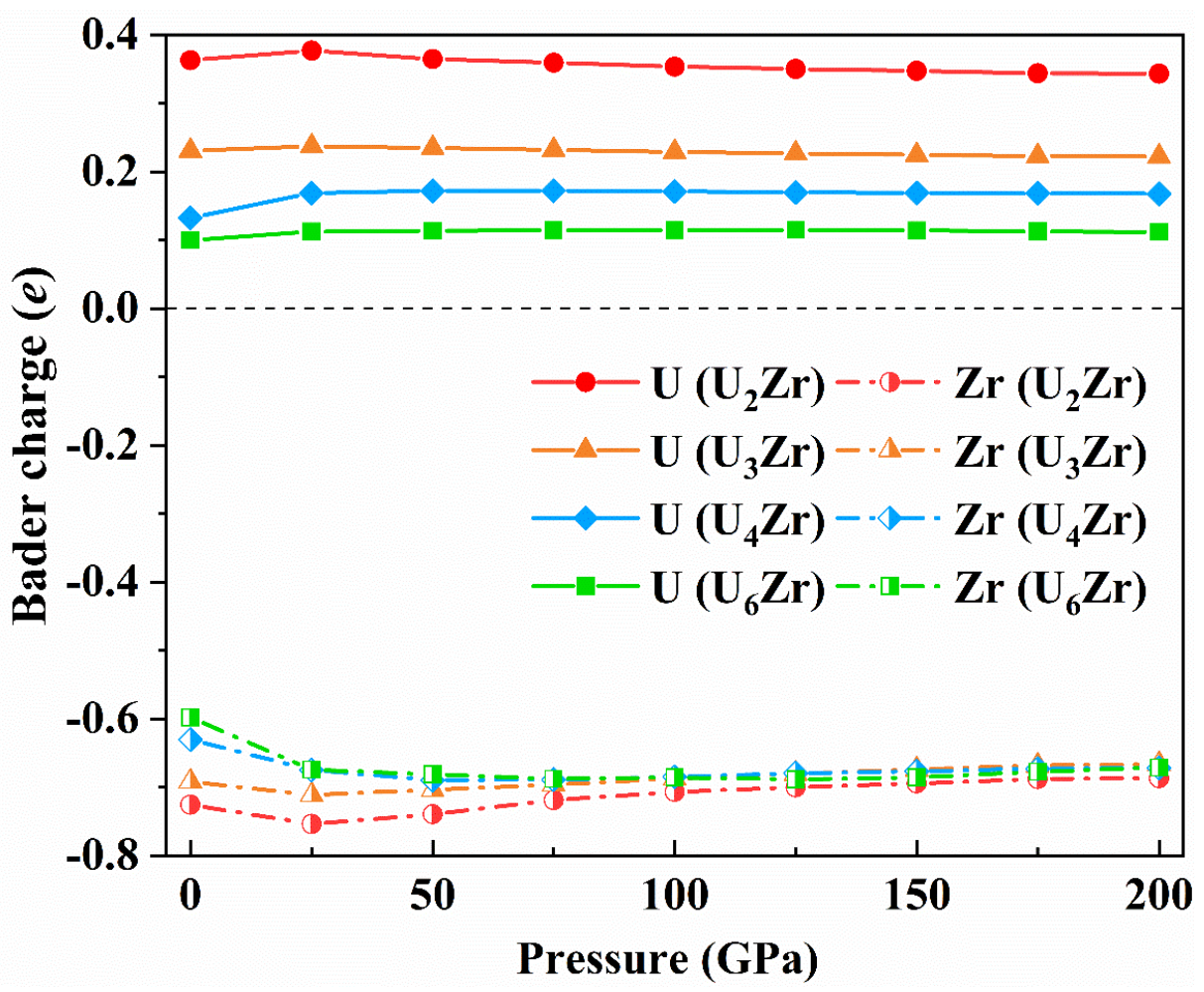

**Fig. 7** (color online) The pressure dependence of the averaged Bader charge of U and Zr atoms for *P6/mmm*-$U_2Zr$, *Cmcm*-$U_3Zr$, *Immm*-$U_4Zr$, and *Cmmm*-$U_6Zr$, respectively.

The electron redistribution between U and Zr atoms is further characterized by differential charge density as shown in Fig. 8: Zr atom loses electrons and the charges gather together between two neighboring U atoms to form a covalent U-U bond. For *P63/mmc*-$U_2Zr$, the neighboring U atoms between different layers along [001] direction manifest evident covalent feature even at ambient pressure. In the *P6/mmm*-$U_2Zr$ phase, besides for the inter-layer U-U bonds, the neighboring U atoms within the same layer also form covalent bonds that resembling the hexagonal ring in graphene. Notably, similar inter-layer and intra-layer U-U bonds are also observed in *Cmcm*-$U_3Zr$, *Immm*-$U_4Zr$, and *Cmmm*-$U_6Zr$ phases at high pressures, as shown in Figs. 8 (c-e).

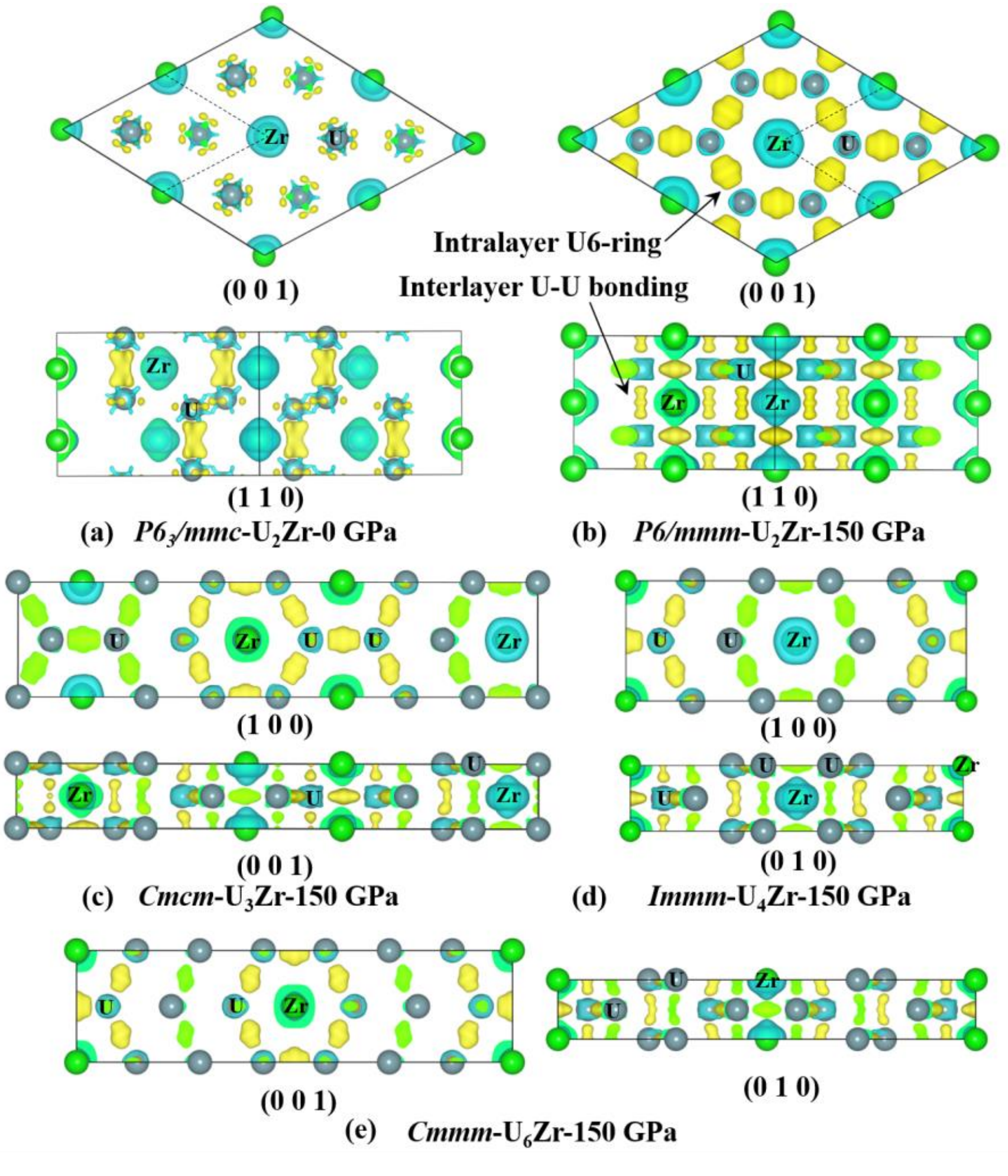

**Fig. 8** (color online) Differential charge density (isosurface = 0.02 $e$/Bohr$^3$) for (a) *P63/mmc*-$U_2Zr$ at 0 GPa, (b) *P6/mmm*-$U_2Zr$ at 150 GPa, (c) *Cmcm*-$U_3Zr$ at 150 GPa, (d) *Immm*-$U_4Zr$ at 150 GPa, and (e) *Cmmm*-$U_6Zr$ at 150 GPa, respectively. Blue color indicates loss of electrons, whereas yellow color represents the gain of electrons between neighboring U atoms.

To further quantify the covalent bonding feature of $U_2Zr$ in detail, we calculated the crystal orbital Hamiltonian populations (COHP) [57, 58] of *P6₃/mmc*-$U_2Zr$ at 0 GPa and *P6/mmm*-$U_2Zr$ at 200 GPa, as shown in Fig. 9 (a). The -COHP of both $P6_3/mmc$-$U_2Zr$ and *P6/mmm*-$U_2Zr$ in the energy region below the Fermi level is mainly contributed by the bonding states, suggesting that these two phases have strong local stability. The strength of the inter-layer and intra-layer U-U bonds is significantly greater than that of the U-Zr and Zr-Zr bonds. We also calculated the integrated COHP (ICOHP) of $U_2Zr$ to examine quantitatively how these bond interactions change with pressure, and the results are shown in Fig. 9 (b). In the $P6_3/mmc$-$U_2Zr$ phase, the inter-layer U-U bonds are always stronger than all other bonds. However, in the *P6/mmm*-$U_2Zr$ phase, the strength of the intra-layer U-U bond is greater than that of inter-layer when below 61 GPa. Above 61 GPa, the situation becomes opposite. We note that there is a big jump of the inter-layer U-U bonding strength when $P6_3/mmc$-$U_2Zr$ transforms to *P6/mmm*-$U_2Zr$, suggesting these bonds are crucial for the stability of the former at low pressure. In addition, in the *P6/mmm*-$U_2Zr$ phase region, all covalent bonding interactions get enhanced as the pressure increases, with U-U bonds being the constant dominance. At 150 GPa, the -ICOHP of interlayer U-U bonds reaches a record value of 16.98 eV/pair. By comparison, it is 9.6 eV/pair for the typical covalent C-C bond in diamond at 0 GPa.

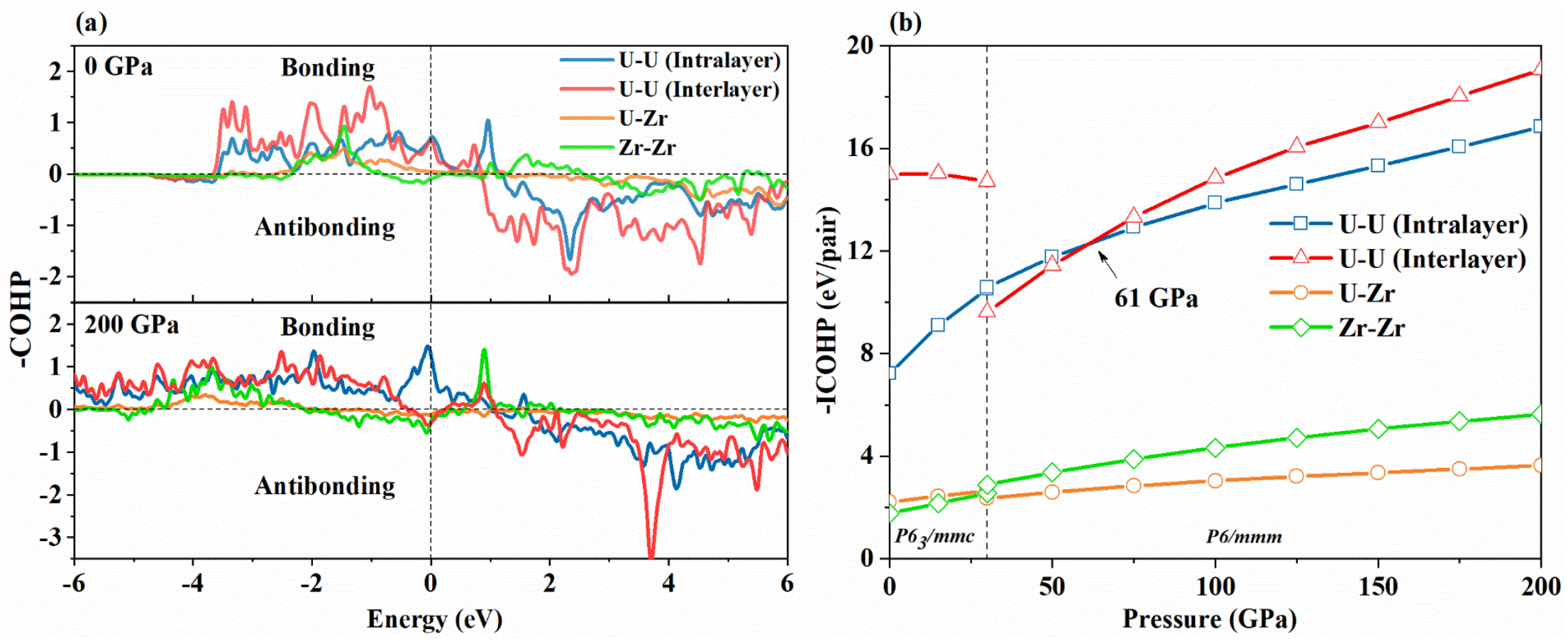

**Fig. 9** (color online) (a) Calculated crystal orbital Hamiltonian populations (COHP) of $U_2Zr$ at 0 and 200 GPa, respectively. (b) The integrated COHP (ICOHP) of $U_2Zr$ as a function of pressure.

## C. Ordered phases in other uranium-based transition metal binary alloys

Considering the structural similarity of transition metals Zr, Ti, V, Cr, Nb, Mo, Hf, Ta, and W (all adopting the bcc structure within a specific high-pressure range), as well as the predicted structures of the U-Zr system being similar to that of the U-Nb system [27], We can infer that other uranium-based binary alloys containing transition metals with their outermost *d* orbitals filled with electrons up to half or fewer might exhibit ordered phases similar to those predicted in the U-Zr and U-Nb alloys. Especially, current research also has indicated that the high-pressure structure of $U_2Zr$ is consistent with that of $U_2Nb$, $U_2Ti$, and $U_2Mo$ (all adopting the $AlB_2$-type structure), so we conjecture that U-Ti, U-Nb, U-Mo and U-Zr alloys should exhibit the same ordered phases under high pressure. To verify this hypothesis, we systematically investigate the stability of $P6_3/mmc$-$U_2X$, $P6/mmm$-$U_2X$, $Cmcm$-$U_3X$, $Immm$-$U_4X$, and $Cmmm$-$U_6X$ phases in U-X (X= Sc, Ti, V, Cr, Y, Nb, Mo, Hf, Ta, W) transition metal binary alloys (we note that, from the results of U-Zr and U-Nb systems, the occurrence of other type of ordered phase is highly unlikely).

First, we calculated the formation enthalpy of these ordered phases in the different alloy systems with respect to $\alpha$-U phase and the most stable phase of the respective transition metals as reported at the corresponding pressures [59-66]. The thermodynamic convex hulls and pressure–composition phase diagrams of U-X (X= Sc, Ti, V, Cr, Nb, Mo, Hf, Ta) alloys are plotted in Figs. 10 and 11, respectively. Notably, we display only the convex hulls for the U-rich region of concern. From the convex hull diagrams, it can be seen that the formation enthalpies of the most stable structure of $U_2X$, $U_3X$, $U_4X$, and $U_6X$ stoichiometries in U-X (X= Sc, Ti, V, Cr, Nb, Mo, Hf, Ta) decrease noticeably with increasing pressure, and almost all become negative, indicating that these structures all have better stability under high pressures in these systems. For U-X (X= Ti, V, Nb, Sc, Cr) alloys, the formation enthalpies of these stoichiometries first decrease as the pressure increases, but then increase again slightly at 200 GPa. At high pressures, the *P6/mmm*-$U_2X$ (X= Sc, Ti, V, Cr, Nb, Mo Hf, Ta) phase always locates on the convex hulls of their respective systems. Furthermore, the *Cmcm*-$U_3X$ phase in U-X (X= Ti, Mo, Sc) alloys, and the *Immm*-$U_4Sc$ and *Cmmm*-$U_6Sc$ phases in U-Sc alloy also sit on their respective convex hulls over a certain pressure range, showing that they also could be ground states. In addition, the formation enthalpy of the most stable structure of $U_2X$ stoichiometry is negative at zero pressure in U-X (X= Ti, V, Nb, Ta) systems, while that of $U_2X$ stoichiometry in U-X (X= Mo, Hf, Sc, Cr) system becomes negative when at and beyond 50 GPa. These results show that U-X (X= Sc, Ti, V, Cr, Nb, Mo, Hf, Ta) alloys have the same U-rich ordered structures as U-Zr alloy at high pressures.

On the other hand, the formation enthalpies of these structures in both U-Y and U-W alloys are all positive in the whole pressure range we studied (Figs. S36 and S37), but they all are dynamically stable over a certain pressure range [55], indicating that these structures might be only metastable in either system.

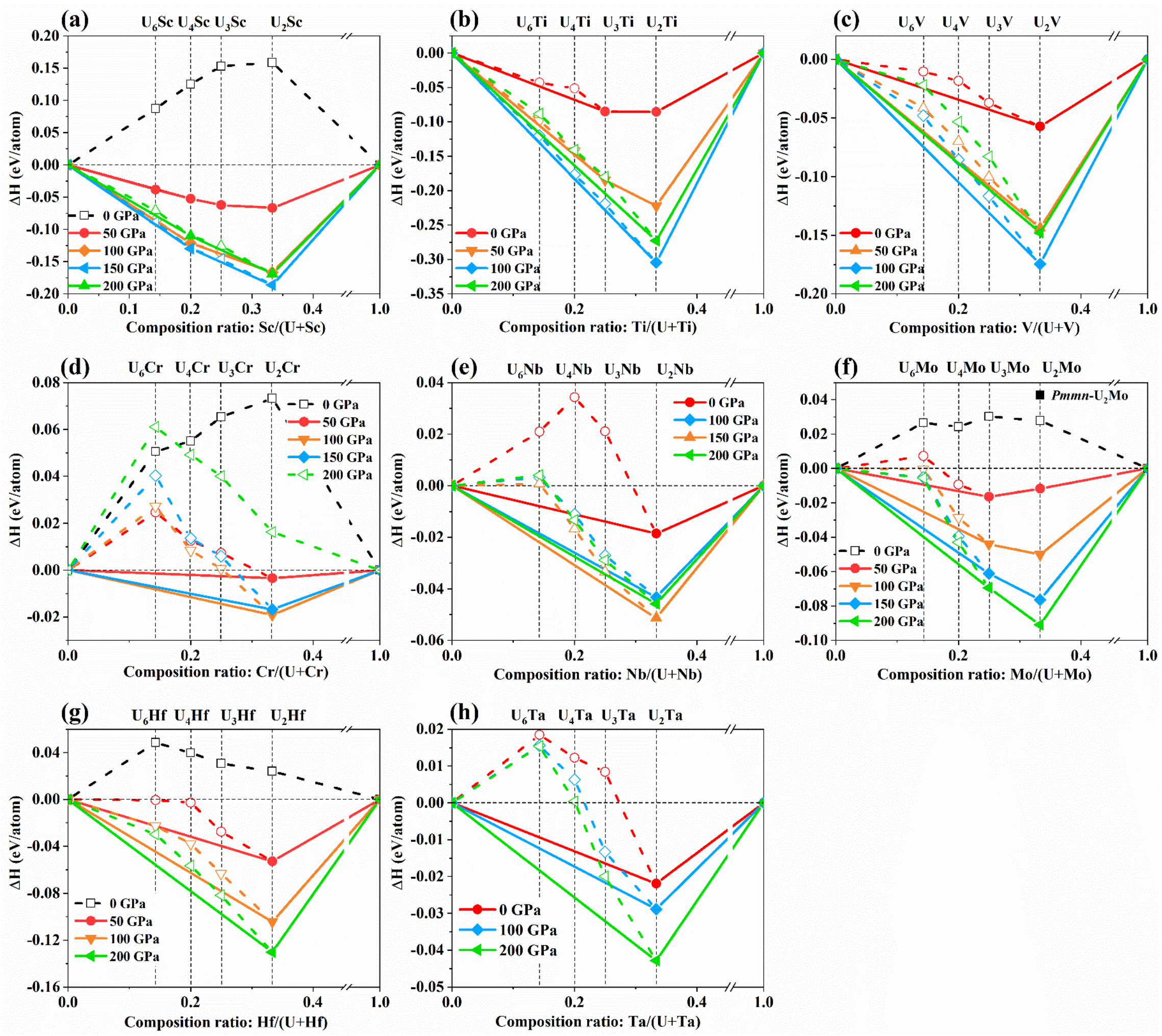

**Fig. 10** (color online) Thermodynamic convex hulls of U-X (X= Sc, Ti, V, Cr, Nb, Mo, Hf, Ta) alloys: (a) U-Sc alloy; (b) U-Ti alloy; (c) U-V alloy; (d) U-Cr alloy; (e) U-Nb alloy; (f) U-Mo alloy, in which the solid square indicates the formation enthalpy of the previously proposed ordered *Pmmn*-$U_2Mo$ phase at 0 GPa [23]; (g) U-Hf alloy; (h) U-Ta alloy. Dashed line connections indicate unstable or metastable phases, while solid line connections indicate stable groundstate phases.

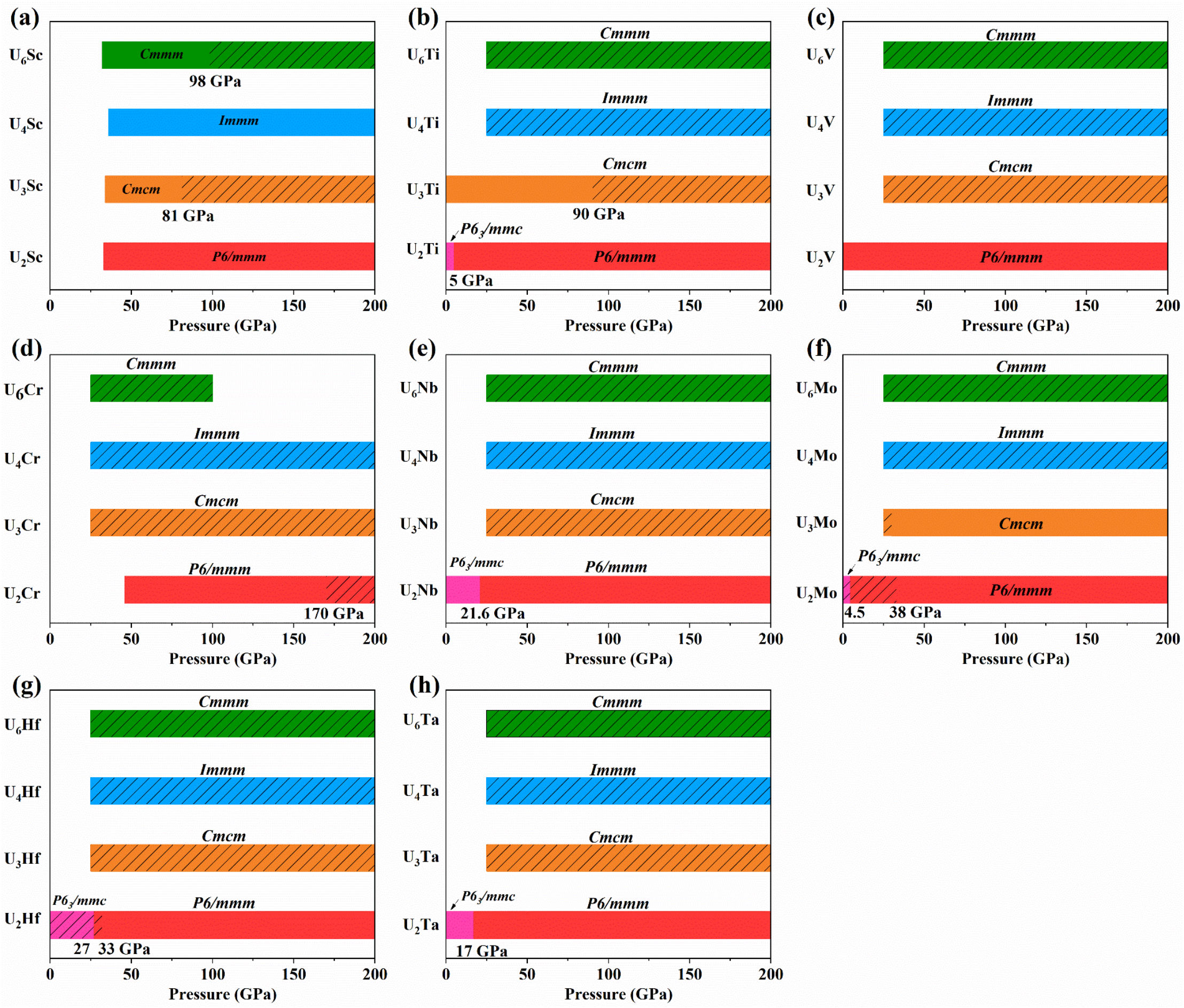

**Fig. 11** (color online) Pressure–composition phase diagrams of U-X (X= Sc, Ti, V, Cr, Nb, Mo, Hf, Ta) alloys: (a) U-Sc alloy; (b) U-Ti alloy; (c) U-V alloy; (d) U-Cr alloy; (e) U-Nb alloy; (f) U-Mo alloy; (g) U-Hf alloy; (h) U-Ta alloy. Different color regions indicate different structures, and black slashed regions indicate that the corresponding phases can be metastable within this pressure range.

The stable pressure-range of these ordered phases in various alloy systems is shown in detail in the pressure–composition phase diagrams (Fig. 11), where the slash-filled region indicates that the phase is only dynamically stable and can be thermodynamically metastable within the given pressure range. The dynamical stability of these ordered phases in U-X (X= Sc, Ti, V, Cr, Y, Nb, Mo, Hf, Ta, W) alloys is carefully examined by phonon spectra analysis [55] where all phonon modes are positive. Except that the *Cmcm* phase in U-X (X= Ti, Mo, Sc) alloy and the *Immm* and *Cmmm* phases in U-Sc alloy are thermodynamically stable under certain pressure ranges (0–90 GPa for *Cmcm*-$U_3Ti$, 30–200 GPa for *Cmcm*-$U_3Mo$, 34–81 GPa for *Cmcm*-$U_3Sc$, 36–200 GPa for *Immm*-$U_4Sc$, and 32–98 GPa for *Cmmm*-$U_6Sc$), in other systems the *Cmcm, Immm,* and *Cmmm* phases are only metastable from 25 to 200 GPa (except for *Cmmm*-$U_6Cr$, which is only metastable between 25 and 100 GPa). In addition, $U_2X$ (X=Ti, V, Nb, Ta) is stable between 0 and 200 GPa, while $U_2X$ (X=Mo, Hf, Sc, Cr) is thermodynamically stable only at pressures above 33 GPa or higher.

It is worth mentioning that ordered phases at the 2:1 stoichiometry in U-Ti and U-Mo alloy systems have been studied before, but no ordered phase has been reported experimentally and theoretically in the U-X (X= Sc, V, Cr, Y, Hf, Ta, W) alloys. Previous experimental studies [20, 24] proposed that $U_2Ti$ and $U_2Mo$ adopt the hexagonal *P6/mmm* and body-centered tetragonal *I4/mmm*-$U_2Mo$ structures at zero pressure, respectively. However, our results indicate that $U_2Ti$ should be in the *P6₃/mmc* phase, which transitions to the *P6/mmm* phase at pressures higher than 5 GPa, in agreement with a recent theoretical study [67]. For $U_2Mo$, our calculations indicate that the *P6/mmm*-$U_2Mo$ phase has a lower energy than the proposed *I4/mmm*-$U_2Mo$ phase between 0 and 200 GPa [55], which is consistent with previous theoretical studies [22, 25, 68-70]. Nonetheless, it is just metastable rather than being the ground state at 0 GPa, in sharp contrast with previous assessment [22, 23, 70]. This discrepancy is due to the wrong ground state of the uranium employed in Refs [23, 70]. They used bcc-U rather than the more stable α-U, which led to a negative formation enthalpy and an incorrect conclusion. We recalculated the formation enthalpy of the same structure as Ref [23] but with the α-U as the reference state. The obtained result is plotted in Fig. 10(f) for comparison. Furthermore, our results also demonstrate that the newly predicted *P6₃/mmc*-$U_2Mo$ phase is more stable than the previously proposed *P6/mmm*-$U_2Mo$ phase when the pressure ranges from 0 to 4.5 GPa.

Of the alloys we studied, only $U_2X$ (X=Ti, Zr, Nb, Mo, Hf, Ta) adopts the *P6₃/mmc* structure at zero pressure, which implies that a pressure-driven phase transition from *P6₃/mmc*-$U_2X$ to *P6/mmm*-$U_2X$ similar to that of $U_2Nb$ [27] also can present in these alloys. Figure 12 shows the detailed enthalpy difference between *P6₃/mmc* and *P6/mmm* phases for these alloys (see Fig. 5 for $U_2Zr$). The *P6₃/mmc* structure in $U_2X$ (X= Ti, Zr,

Hf, Ta) is spontaneously relaxed into the *P6/mmm* symmetry at high pressures, indicating that they also host a unique hybrid phase transition as $U_2Nb$, which was termed as $1\frac{1}{2}$-order transition in Ref [27], and demonstrated that it must have concomitant elastic anomaly [27]. In contrast, unambiguous and sharp first-order phase transition from $P6_3/mmc$ to *P6/mmm* phase is predicted for $U_2X$ (X= V, Mo, Ta), in which $P6_3/mmc$-$U_2Mo$ phase shows a very strong metastability so that the $P6_3/mmc$-$U_2Mo$ structure can be maintained up to 20 GPa far beyond the phase transition pressure (4.5 GPa). It is also worth mentioning that this phase transition exists in the negative pressure region (at -6 GPa) for $U_2V$.

It is evident from above analysis that our results not only chart new high pressure phase diagrams for these U-based binary alloys, but also rewrite their existing phase diagrams at 0 GPa significantly. In particular, for all of these alloys, previous experimental information at low temperature and 0 GPa is qualitatively correct only for U-X (X= Sc, Cr, Hf, Y, W), but U-X (X= Sc, Cr, Hf) all form intermetallic compounds at high pressures by our investigations. On the other hand, previous theoretical investigations only gave a correct structure and phase transition in $U_2Ti$, whereas wrongly assessed the stability of $U_2Mo$ and missed $U_3Ti$. In U-X (X= V, Nb, Mo, Ta), either wrong composition ($UNb_3$) was reported, or the correct ground state compounds were totally missed in all currently available literature at 0 GPa.

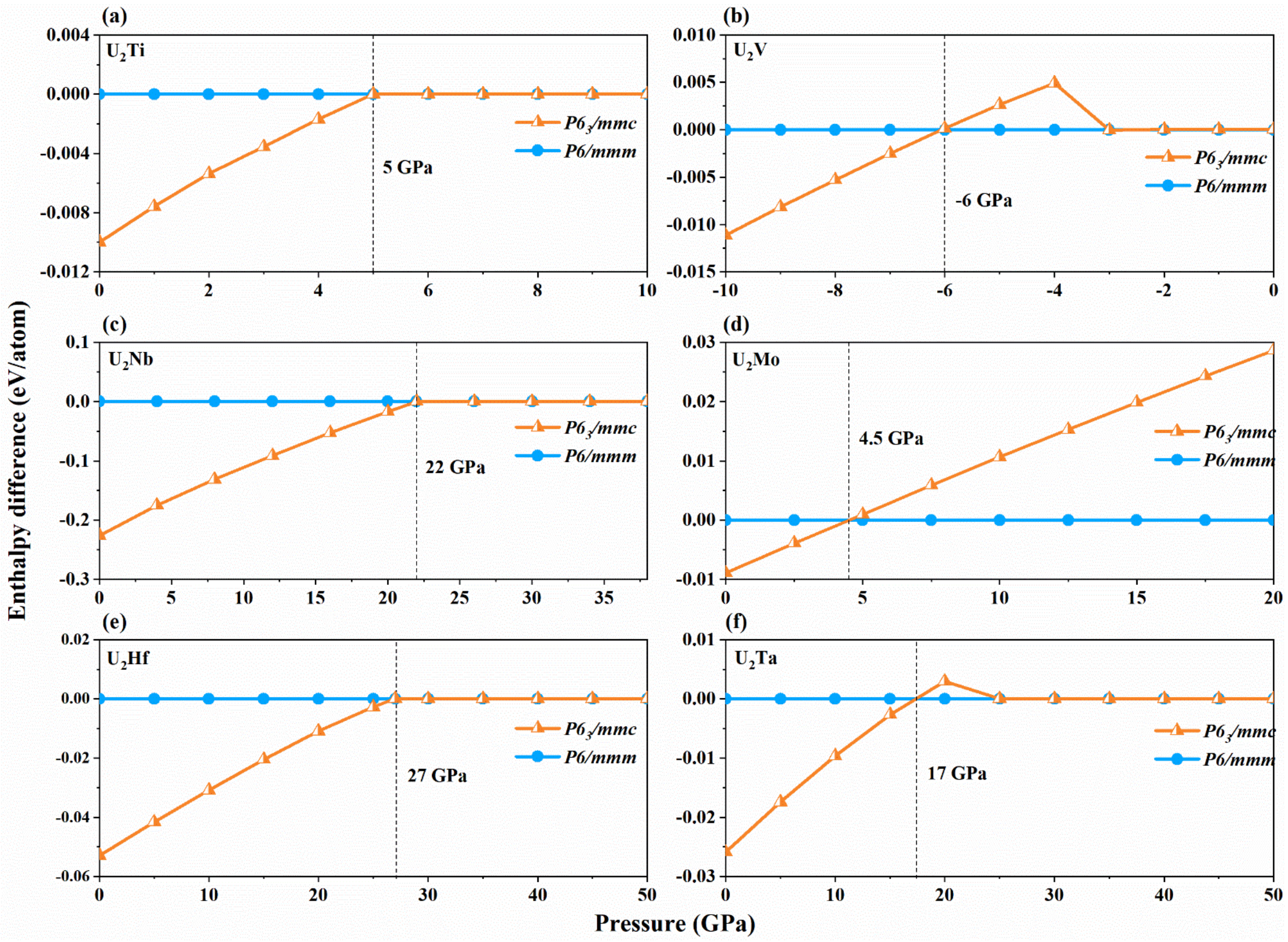

**Fig. 12** (color online) Enthalpy difference of $U_2X$ (X=Ti, V, Nb, Mo, Hf, Ta) in *$P6_3/mmc$* phase with respect to the *P6/mmm* phase: (a) $U_2Ti$, (b) $U_2V$, (C) $U_2Nb$, (d) $U_2Mo$, (e) $U_2Hf$, and (f) $U_2Ta$, respectively.

In addition to the structural stability and phase transition of all of these ordered phases, their electronic structure and bonding properties in different alloy systems were further investigated [55]. They all are metallic, with the 5*f* electrons of U atoms dominating the electronic DOS near the Fermi level. As pressure increases, the localization of the *f* electrons decreases, which is similar to that of U-Zr alloy. Further, The Bader charges of these phases in different alloys reveal that there is an evident charge transfer from the transition metal atoms to U atoms under high pressures, suggesting strong ionic interactions in these alloys (Fig. S35). The variation of charge state from $U_2X$ to $U_3X$, $U_4X$, and $U_6X$ mainly comes from U atoms, rather than the transition metal atoms. On the other hand, the differential charge density also shows that the *P6/mmm*-$U_2X$, *Cmcm*-$U_3X$, *Immm*-$U_4X$ and *Cmmm*-$U_6X$ phases (X= Sc, Ti, V, Cr, Nb, Mo, Hf, Ta, Y, W) all exhibit strong inter-layer and intra-layer U-U covalent bonding at high pressures, which is similar to U-Zr alloy.

## IV. Conclusion

All possible ordered phases in U-Zr system were systematically explored using unbiased first-principles structure prediction. $U_2Zr$, $U_3Zr$, $U_4Zr$, and $U_6Zr$ were predicted to be stable and metastable within 200 GPa. Calculations of the PDOS show that all of these newly discovered phases are metallic, with the 5*f* electrons of U atoms dominating the electronic DOS near the Fermi level. The results of the Bader charge analysis and differential charge density reveal significant charge transfer between U and Zr atoms in these ordered phases, and prominent inter-layer and intra-layer covalent U-U bonding were also observed under high pressure, which is stronger than C-C bond in diamond at 0 GPa.

Furthermore, we extended the predicted structures of U-Zr alloy to other uranium-based U-X (X= Sc, Ti, V, Cr, Y, Nb, Mo, Hf, Ta, W) transition metal alloys. We found all *P6/mmm*-$U_2X$ (X= Sc, Ti, V, Cr, Nb, Mo, Hf, Ta) exhibits excellent stability at high pressures, while *P6₃/mmc*-$U_2X$ phase is stable only in U-X (X=Ti, Nb, Mo, Hf, Ta) alloys at low pressures. A structural transition from $P6_3/mmc$ to *P6/mmm* phase, similar to that observed in the U-Zr alloy, is also present in $U_2X$ (X=Ti, Nb, Mo, Hf, Ta) alloys, in which it is a unique hybrid phase transition in $U_2X$ (X=Ti, Zr, Nb, Hf), that possesses a special superposition of a first-order transition and a second-order transition and termed as $1\frac{1}{2}$-order transition in Ref [27], and is pure first-order in $U_2Mo$ and $U_2Ta$. Except for the *Cmcm*-$U_3X$ phases in U-X (X= Ti, Mo, Sc), and the *Immm*-$U_4Sc$ and the *Cmmm*-$U_6Sc$ phases in U-Sc alloy that are thermodynamically stable at certain pressure intervals, all *Cmcm*, *Immm,* and *Cmmm* phases in other systems studied here are metastable between 25 and 200 GPa (except for *Cmmm*-$U_6Cr$, which is only metastable between 25 and 100 GPa). The electronic structure and bonding properties of *P6/mmm*-$U_2X$, *Cmcm*-$U_3X$, *Immm*-$U_4X$, and *Cmmm*-$U_6X$ (X= Sc, Ti, V, Cr, Nb, Mo, Hf, Ta, Y, W) exhibit similar features as U-Zr system, including the charge transfer and covalent

U-U bonding.

In addition, our studies clarify the controversy over $U_2Mo$ as the stable state at ambient conditions. In line with most theoretical studies, our studies indicate that the *P6/mmm*-$U_2Mo$ phase is more stable compared to the *I4/mmm* and *Pmmn* phases. However, some previous studies had erroneously taken the *P6/mmm* or *Pmmn* phases as ground states due to the incorrect choice of bcc-U as the reference state rather than the more stable α-U, leading to false negative formation enthalpies. We corrected this error by choosing the correct reference state to calculate the formation enthalpies for the *P6/mmm* and *Pmmn* phases, which are in fact positive at zero pressure. Furthermore, we predicted a new $P6_3/mmc$-$U_2Mo$ phase, which is more stable than the previously proposed *P6/mmm*-$U_2Mo$ phase within the pressure range of 0 to 4.5 GPa.

These findings and insights not only charted the new high-pressure phase diagrams for U-X (X= Zr, Sc, Ti, V, Cr, Nb, Mo, Hf, Ta) alloys, but also rewrote their low temperature phase diagrams at 0 GPa, which further deepened the understanding about the electronic structure in ordered phases and their common features in broad U-based transition metal alloys.

## Acknowledgments

This work was supported by National Key R&D Program of China under Grant No. 2021YFB3802300, the NSAF under Grant Nos. U1730248 and U1830101, the National Natural Science Foundation of China under Grant Nos.12202418, 11872056, 11904282, 12074274. Part of the simulation was performed on resources provided by the Center for Comput. Mater. Sci. (CCMS) at Tohoku University, Japan.

## Author Contribution

Xiao L. Pan: Calculation, Analysis, Writing, Original draft preparation. Hong X. Song: Analysis, Validation, Editing. Hao Wang: Analysis, Validation. F. C. Wu: Analysis, Validation. Y. C. Gan: Analysis, Validation. Xiang R. Chen: Writing, Reviewing and Editing, Fund. Ying Chen: Analysis, Validation. Hua Y. Geng: Idea conceiving, Project design, Writing, Reviewing and Editing, Fund.

# Supplementary material for " Prediction of novel ordered phases in U-X (X= Zr, Sc, Ti, V, Cr, Y, Nb, Mo, Hf, Ta, W) binary alloys under high pressure"

Xiao L. Pan[1, 2], Hong X. Song[1]*, H. Wang[1], F. C. Wu[1], Y. C. Gan[1], Xiang R. Chen[2]*, Ying Chen[4], Hua Y. Geng[1,3]*

[1] *National Key Laboratory of Shock Wave and Detonation Physics, Institute of Fluid Physics, China Academy of Engineering Physics, Mianyang, Sichuan 621900, P. R. China;*

[2] *College of Physics, Sichuan University, Chengdu 610065, P. R. China;*

[3] *HEDPS, Center for Applied Physics and Technology, and College of Engineering, Peking University, Beijing 100871, P. R. China.*

[4] *Fracture and Reliability Research Institute, School of Engineering, Tohoku University, Sendai 980-8579, Japan.*

## SI. Structure stability of U-Zr system

To verify the completeness of ordered phases predicted by CALYPSO in the U-Zr system, we also performed another independent structural search by using variable-chemical-composition genetic method as implemented in USPEX to check possible ground states that have unusual compositions. This structural search allows each cell to contain up to 20 atoms. The first generation includes 100 structures and for the remaining generations, each contains 60 structures. A total of 50 generations are generated for every search. We performed a full-composition structural search for U-Zr system at three different pressures of 0, 100 and 200 GPa by using USPEX, and the results are shown in Fig. S1. No structure with negative formation enthalpy was obtained at zero pressure. At 100 GPa, the most stable structure is predicted to be *P6/mmm*-$U_2Zr$; *Cmcm*-$U_3Zr$ is close to the convex hull. At 200 GPa, the lowest formation enthalpy is close to zero, and the corresponding structure is *P6/mmm*-$U_2Zr$. These results are consistent with those of fixed composition prediction by using

* *Corresponding authors. E-mail: s102genghy@caep.cn; xrchen@scu.edu.cn; hxsong555@163.com*

CALYPSO, which further confirms the reliability of our structural search.

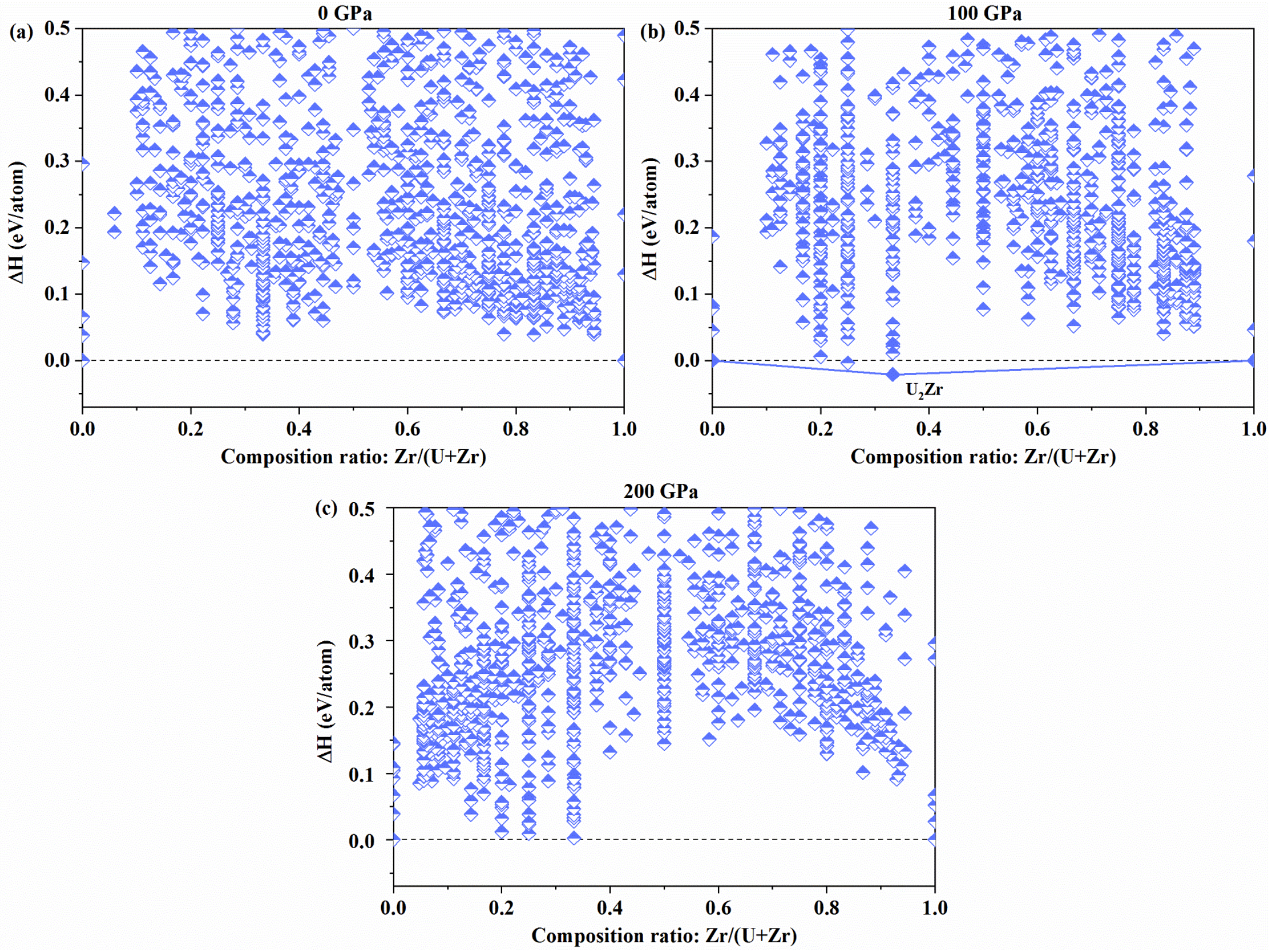

**Fig. S1** Variable-composition structure search results by using USPEX at (a) 0 GPa, (b) 100 GPa, and (c) 200 GPa, respectively.

Furthermore, in order to determine whether different calculation methods affect the stability of different stoichiometric structures on the convex hull diagram, we also calculated the convex hull diagram using DFT+*U* and DFT+*U*+SOC at 125 GPa, as shown in Fig. S2. Obviously, in addition to the reduced formation enthalpy for all stoichiometric structures, *P6/mmm*-$U_2Zr$ is still located on the convex hull, while *Cmcm*-$U_3Zr$ is still close to the convex hull, which means that it is reasonable to consider only the results of DFT method to make a reliable structural prediction for U-Zr system.

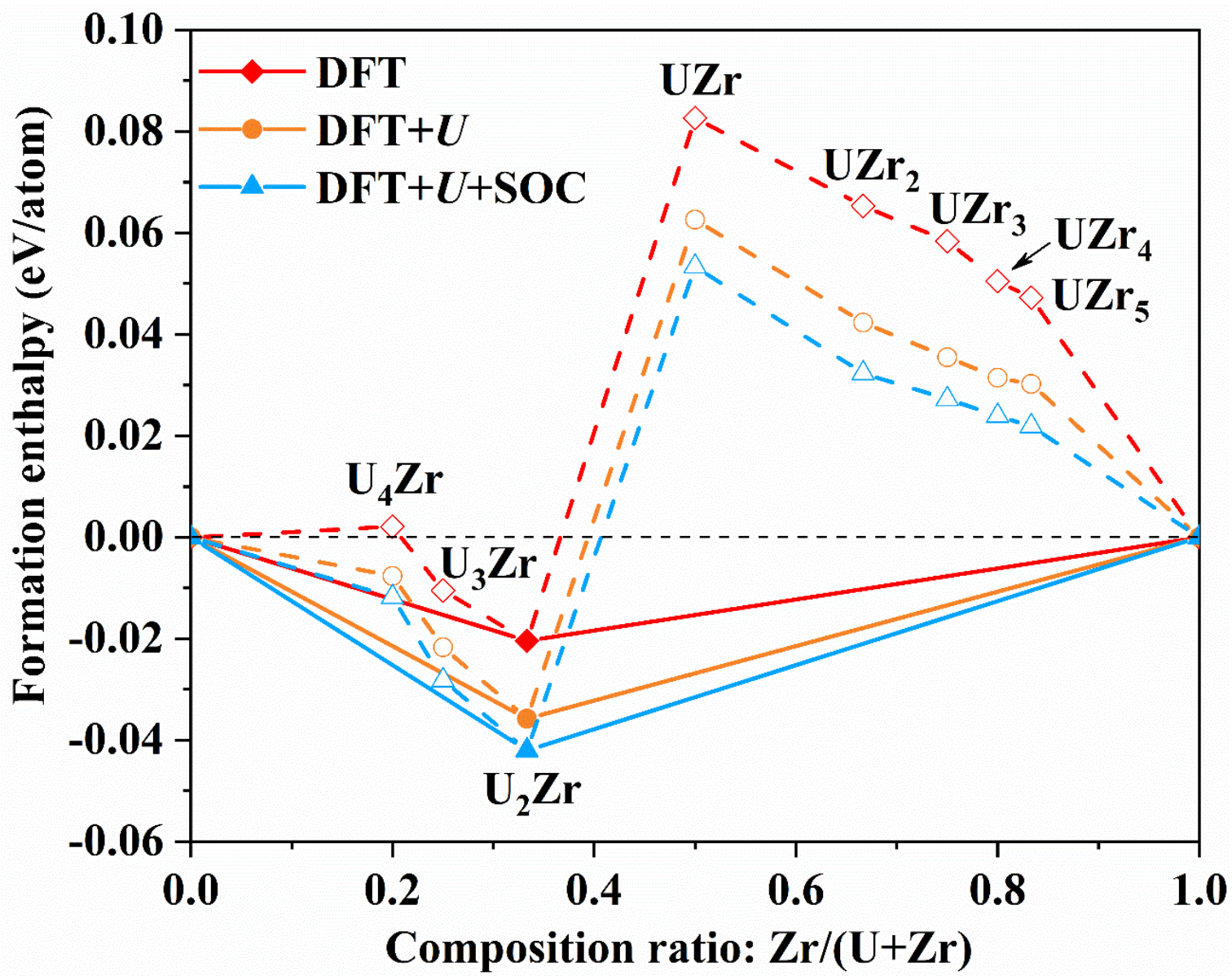

**Fig. S2** Convex hull diagram of candidate structures of U-Zr alloy at 125 GPa calculated by different methods: DFT, DFT+*U*, and DFT+*U*+SOC.

We also studied the stability of the partially ordered and fully ordered $\delta$-$UZr_2$ under high pressure, and the change of their formation enthalpies with pressure is shown in Fig. S3. Compared with the results calculated by DFT, the formation enthalpies of $\delta$-$UZr_2$ calculated by DFT+*U* decrease significantly in the whole pressure range we studied, and that of calculated by DFT+*U*+SOC decrease further. All three methods show that the formation enthalpy of the partial ordered $\delta$-$UZr_2$ increases significantly with the increasing pressure, suggesting that its stability decreases with increasing pressure. The formation enthalpy of the partially ordered $\delta$-$UZr_2$ is significantly lower than that of the fully ordered $UZr_2$ in the pressure range of 0 to 45 GPa, indicating that $UZr_2$ stoichiometry tends to adopt the partially ordered $\delta$-$UZr_2$ phase at low pressure. Furthermore, the atomic volume and formation enthalpy of $\delta$-$UZr_2$ at zero pressure calculated by these methods are compared with the results of previous experiment [4], as shown in Table S1. The formation enthalpy of the partially ordered $\delta$-$UZr_2$ given by DFT+*U*+SOC method are close to the experimental value, but the atomic volume is too larger. By the way, a recent experimental study [1] has casted doubt on this experimental value of formation enthalpy, and the authors called for further accurate measurement. The volumes of the partially ordered $\delta$-$UZr_2$ calculated by DFT and DFT+*U* methods are closer to the experimental data, while that of the fully ordered $\delta$-$UZr_2$ is deviated from the experimental value. This verifies the reliability of the DFT results again.

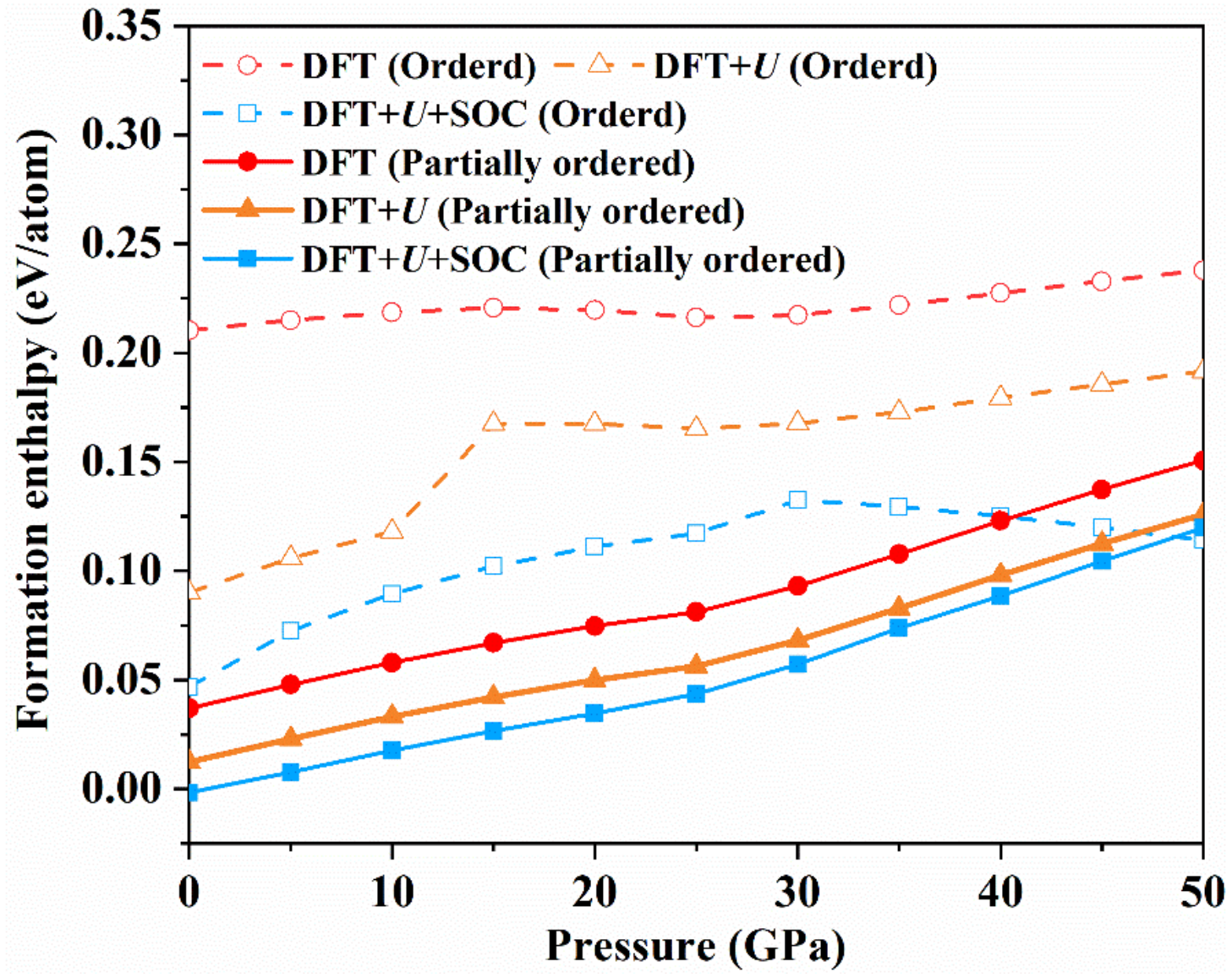

**Fig. S3** Variation of the formation enthalpy with pressure for the partially and fully ordered $\delta$-$UZr_2$ as calculated by DFT, DFT+*U* and DFT+*U*+SOC. The reference states are the most stable phase of U and Zr metal at the corresponding pressures, respectively.

**Table S1.** Calculated atomic volume *V* and formation enthalpy ΔH of $\delta$-$UZr_2$ at 0 GPa.

| | | DFT | DFT+*U* | DFT+*U*+SOC | Expt. [2] |
|---|---|---|---|---|---|
| Partially ordered | ΔH(eV/atom) | 0.037 | 0.012 | -0.002 | −0.04± 0.11 |
| $\delta$-$UZr_2$ | *V* (Å$^3$) | 22.482 | 22.708 | 22.868 | 22.580 |
| Fully ordered | ΔH(eV/atom) | 0.2104 | 0.090 | 0.047 | |
| $\delta$-$UZr_2$ | *V* (Å$^3$) | 22.31 | 23.01 | 23.47 | |

## SII. Phonon spectra for other U-based transition metal alloys

Phonon spectra of *P6₃/mmc*, *P6/mmm*, *Cmcm*, *Immm* and *Cmmm* structures in other U-based transition metal alloys, U-X (X= Sc, Ti, V, Nb, Cr, Mo, Hf, Ta, Y, W), are presented in Figs. S4-S13. Except for phonon spectra of $U_2Sc$ at 200 GPa and $U_2Cr$ at 50 GPa that have slight imaginary frequencies, all of these phases in U-based transition metal alloys we studied are dynamically stable within the corresponding pressure range.

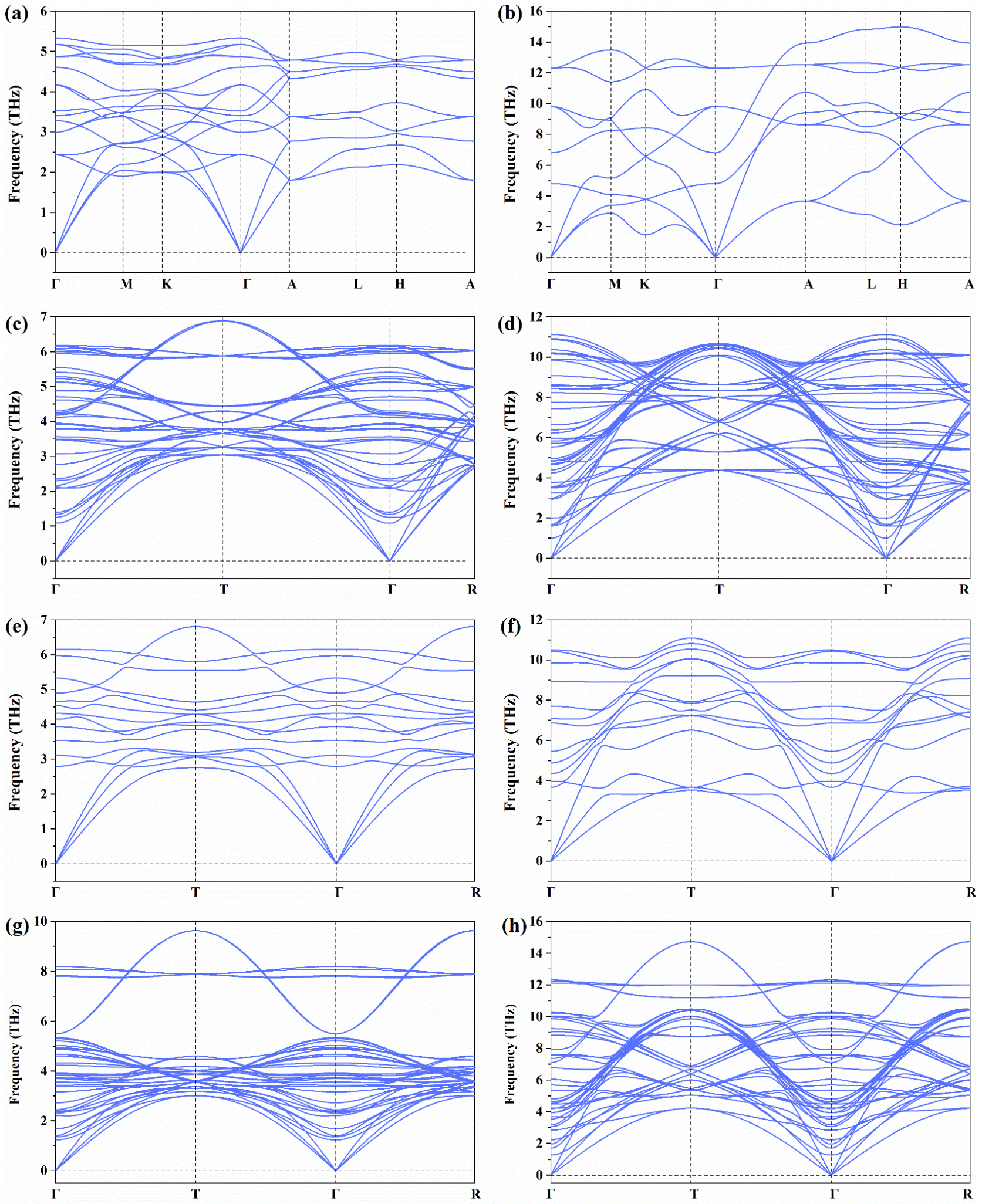

**Fig. S4** Calculated phonon spectra for (a) *P6₃/mmc*-$U_2Ti$ at 0 GPa, (b) *P6/mmm*-$U_2Ti$ at 200 GPa, *Cmcm*-$U_3Ti$ at (c) 25 GPa and (d) 200 GPa, *Immm*-$U_4Ti$ at (e) 25 GPa and (f) 200 GPa, and *Cmmm*-$U_6Ti$ at (g) 25 GPa and (h) 200 GPa, respectively.

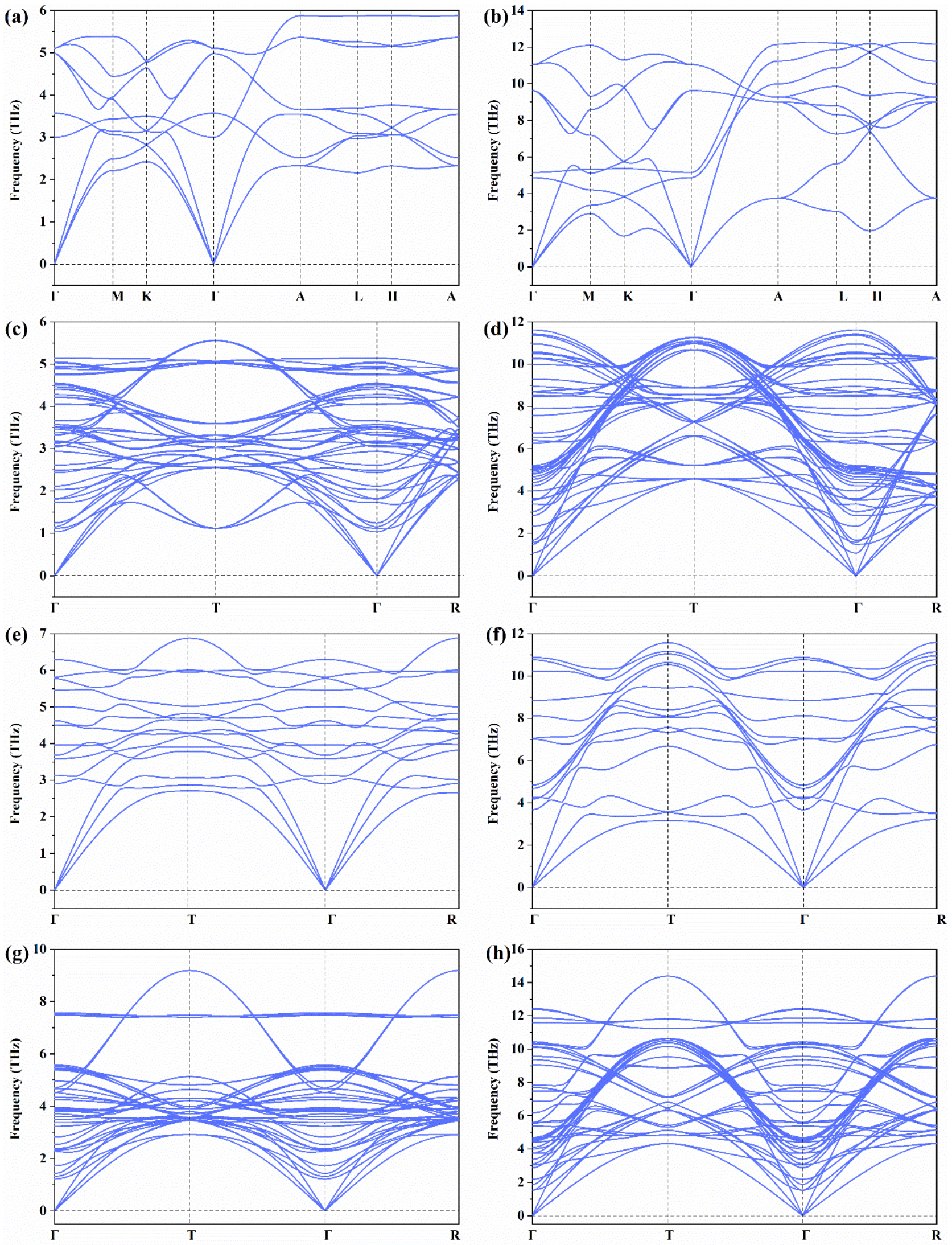

**Fig. S5** Calculated phonon spectra for *P6/mmm*-$U_2V$ at (a) 0 GPa and (b) 200 GPa, *Cmcm*-$U_3V$ at (c) 25 GPa and (d) 200 GPa, *Immm*-$U_4V$ at (e) 25 GPa and (f) 200 GPa, and *Cmmm*-$U_6V$ at (g) 25 GPa and (h) 200 GPa, respectively.

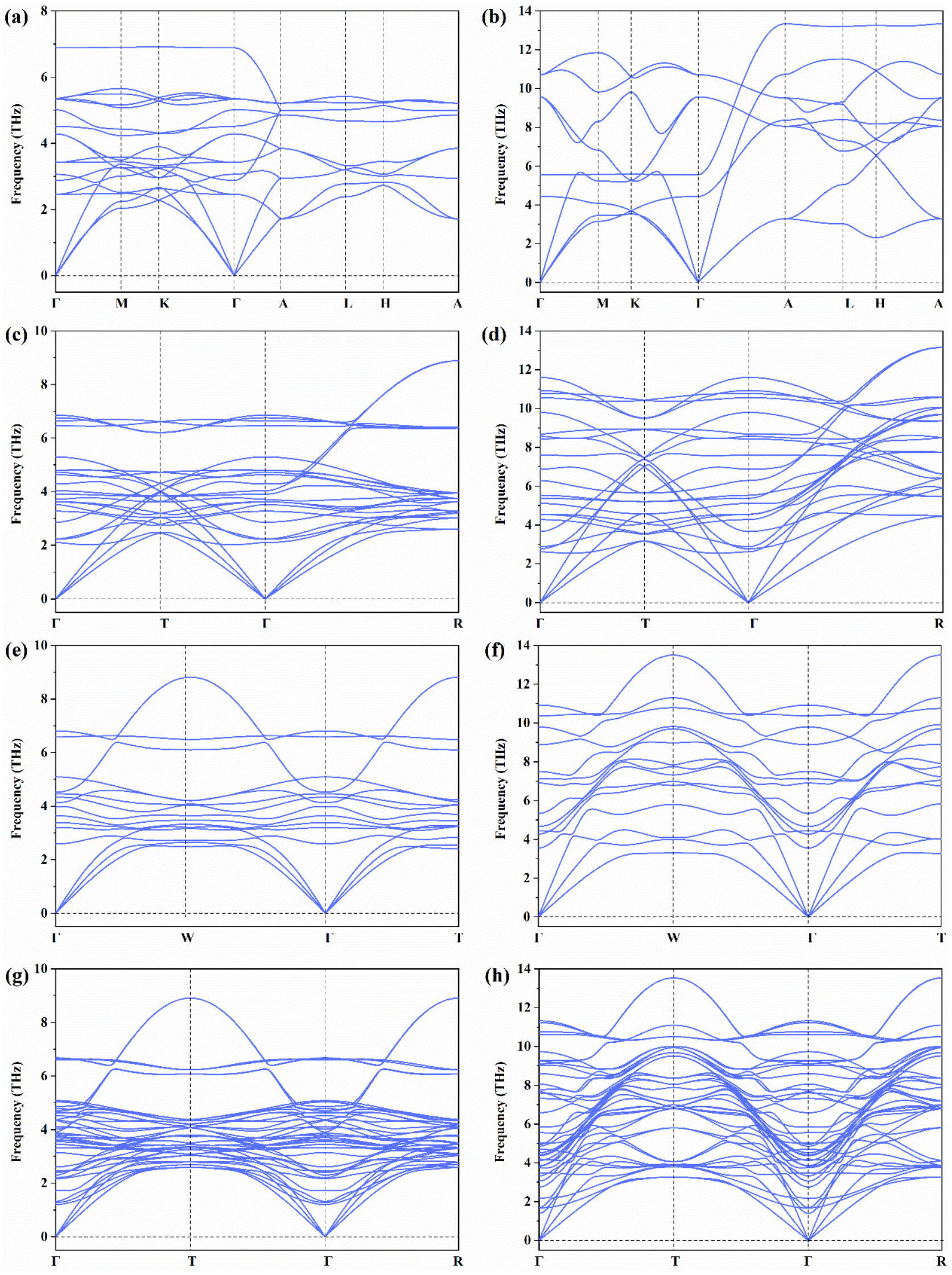

**Fig. S6** Calculated phonon spectra for (a) *P6₃/mmc*-$U_2Nb$ at 0 GPa, (b) *P6/mmm*-$U_2Nb$ at 200 GPa, *Cmcm*-$U_3Nb$ at (c) 25 GPa and (d) 200 GPa, *Immm*-$U_4Nb$ at (e) 25 GPa and (f) 200 GPa, and *Cmmm*-$U_6Nb$ at (g) 25 GPa and (h) 200 GPa, respectively.

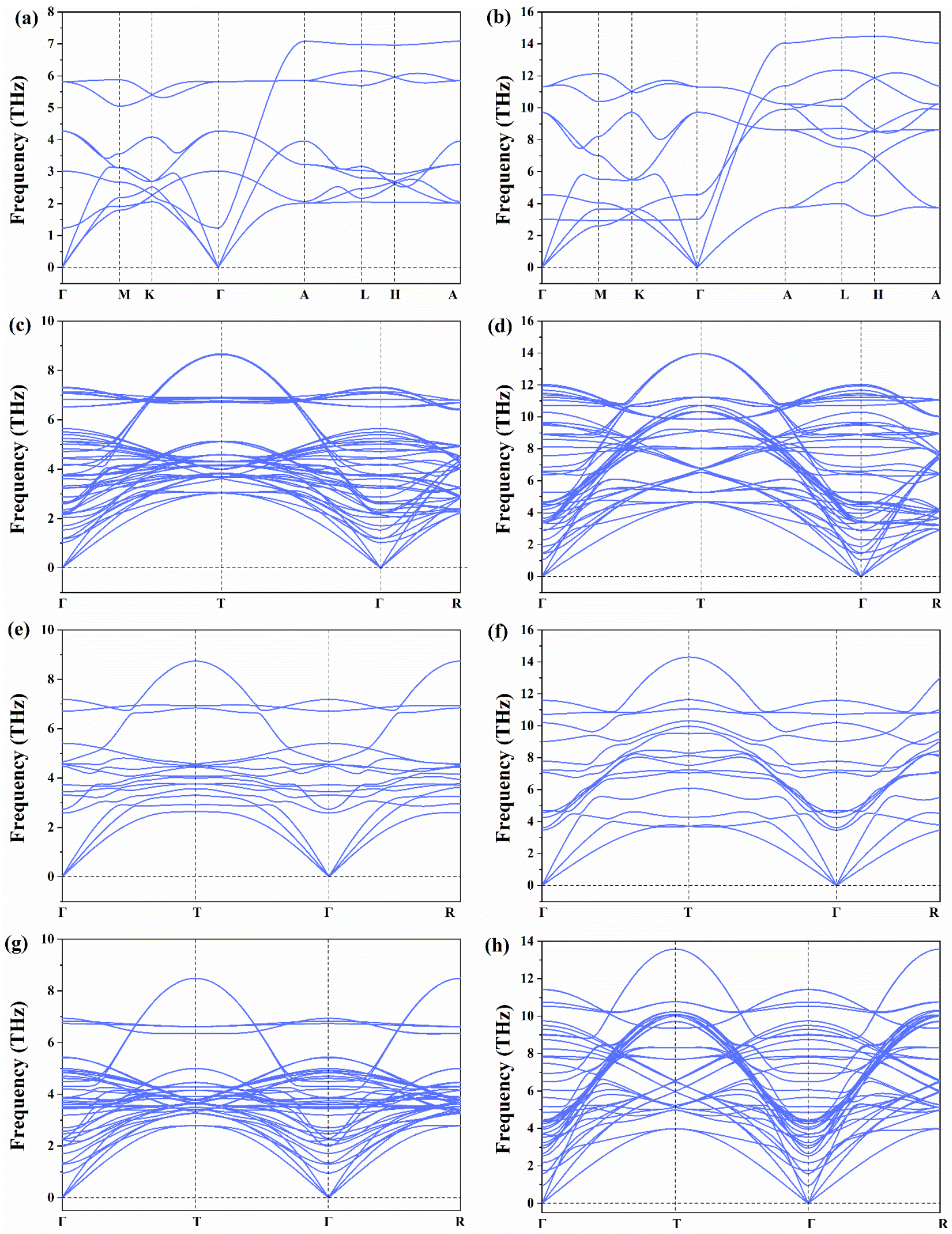

**Fig. S7** Calculated phonon spectra for (a) *P6₃/mmc*-$U_2Mo$ at 0 GPa, (b) *P6/mmm*-$U_2Mo$ at 200 GPa, *Cmcm*-$U_3Mo$ at (c) 25 GPa and (d) 200 GPa, *Immm*-$U_4Mo$ at (e) 25 GPa and (f) 200 GPa, and *Cmmm*-$U_6Mo$ at (g) 25 GPa and (h) 200 GPa, respectively.

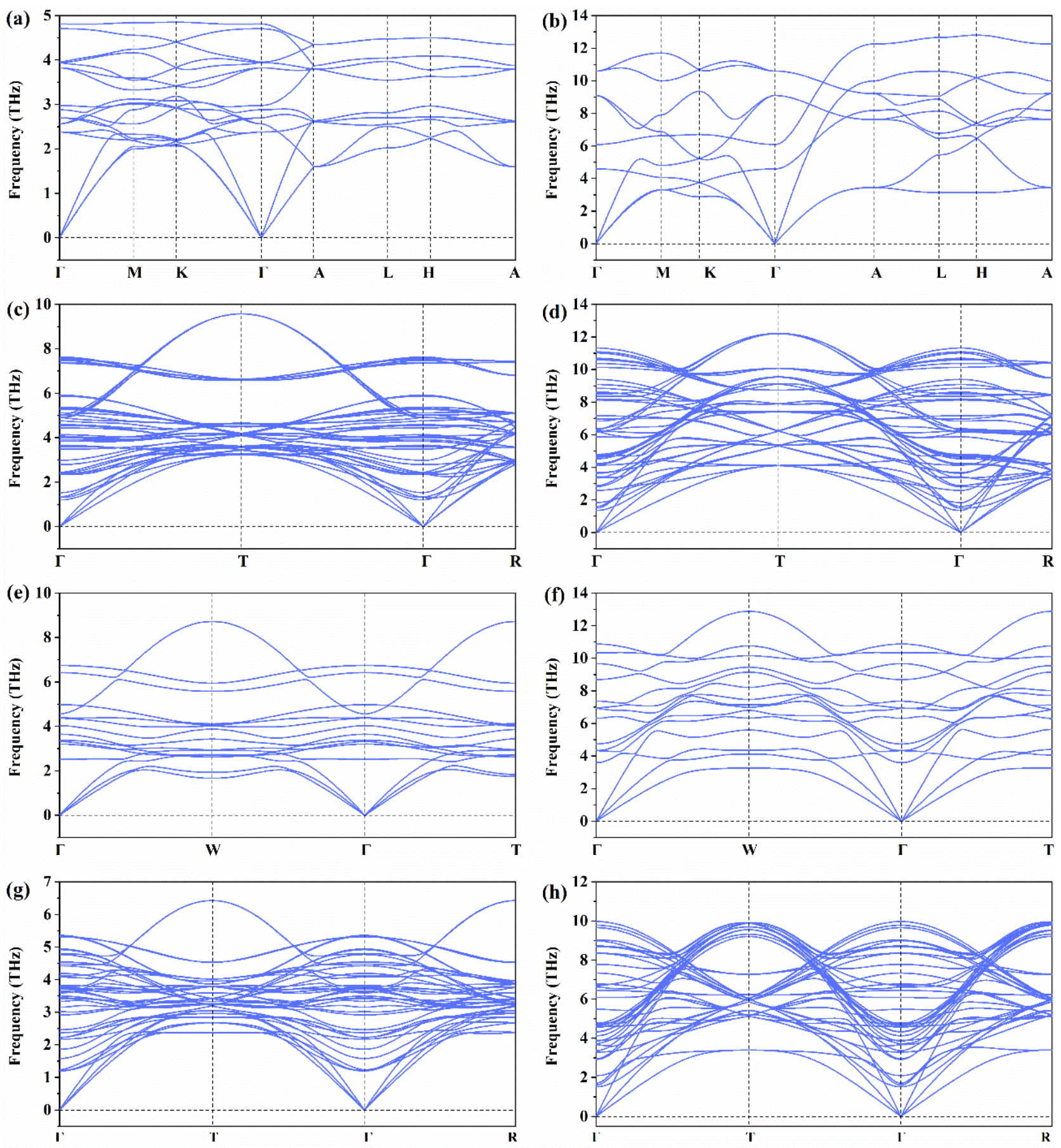

**Fig. S8** Calculated phonon spectra for (a) *P6₃/mmc*-$U_2Hf$ at 0 GPa, (b) *P6/mmm*-$U_2Hf$ at 200 GPa, *Cmcm*-$U_3Hf$ at (c) 25 GPa and (d) 200 GPa, *Immm*-$U_4Hf$ at (e) 25 GPa and (f) 200 GPa, and *Cmmm*-$U_6Hf$ at (g) 25 GPa and (h) 200 GPa, respectively.

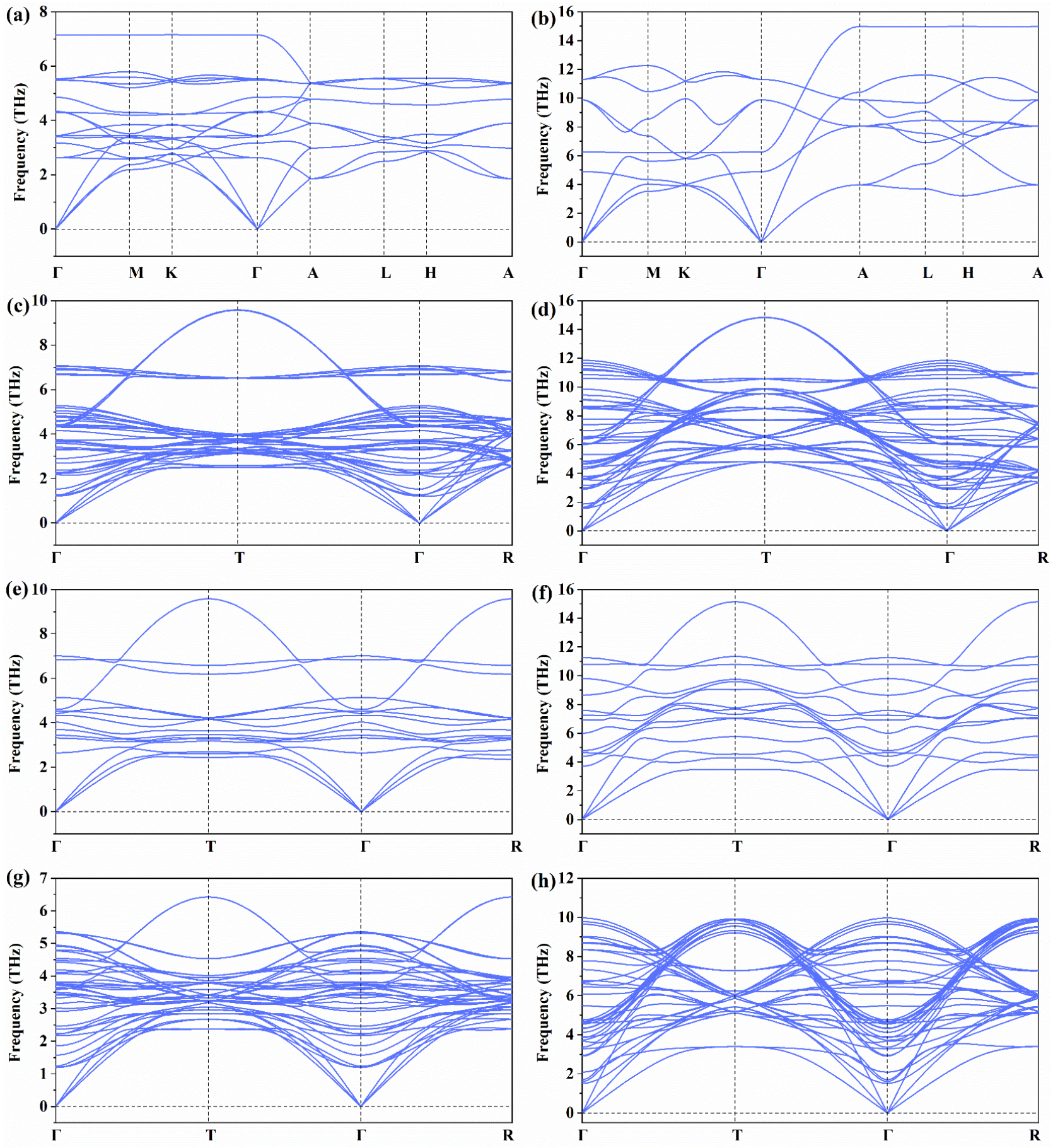

**Fig. S9** Calculated phonon spectra for (a) *P6₃/mmc*-$U_2Ta$ at 0 GPa, (b) *P6/mmm*-$U_2Ta$ at 200 GPa, *Cmcm*-$U_3Ta$ at (c) 25 GPa and (d) 200 GPa, *Immm*-$U_4Ta$ at (e) 25 GPa and (f) 200 GPa, and *Cmmm*-$U_6Ta$ at (g) 25 GPa and (h) 200 GPa, respectively.

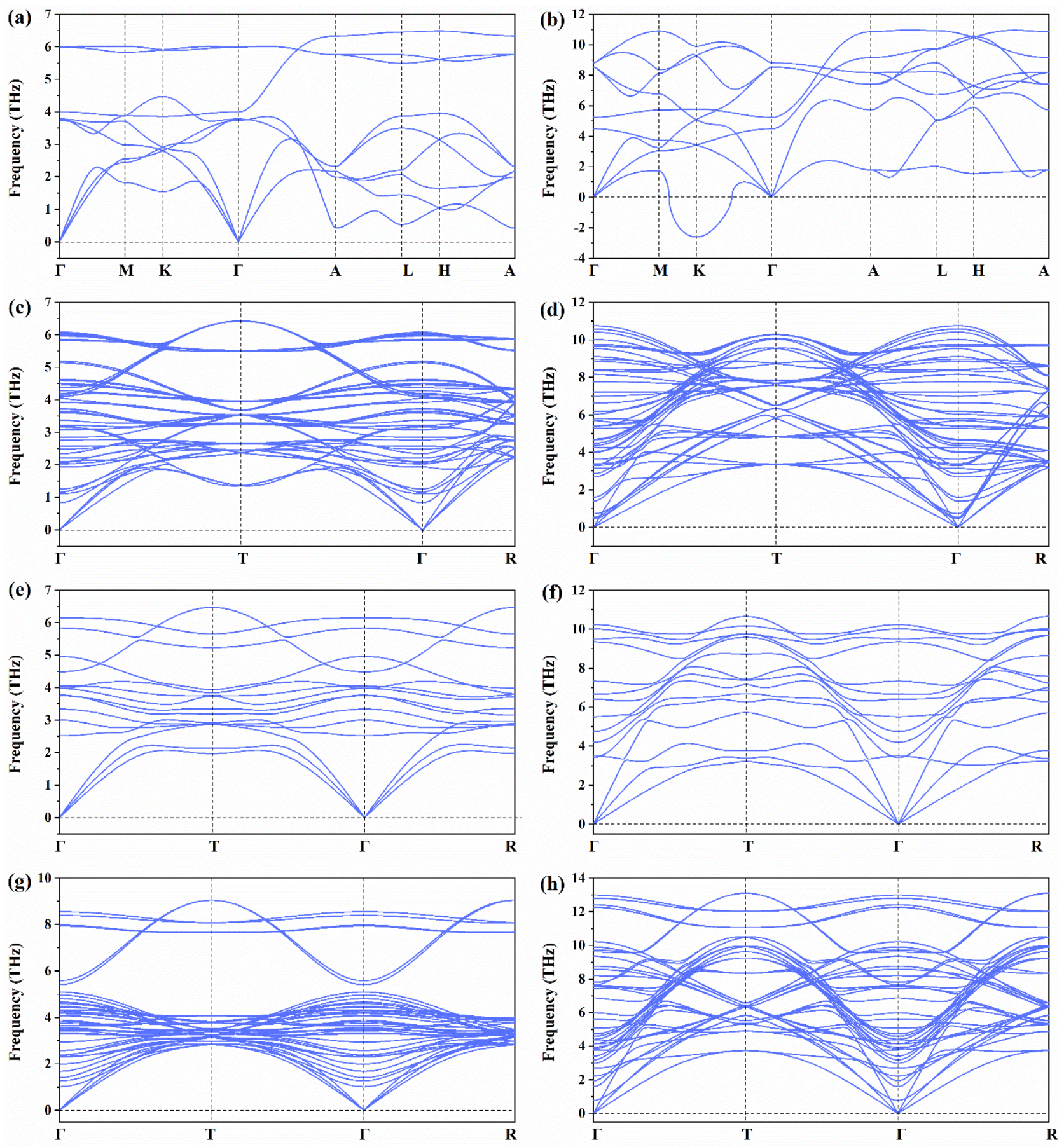

**Fig. S10** Calculated phonon spectra for *P6/mmm*-$U_2Sc$ at (a) 25 GPa and (b) 200 GPa, *Cmcm*-$U_3Sc$ at (c) 25 GPa and (d) 200 GPa, *Immm*-$U_4Sc$ at (e) 25 GPa and (f) 200 GPa, and *Cmmm*-$U_6Sc$ at (g) 25 GPa and (h) 200 GPa, respectively.

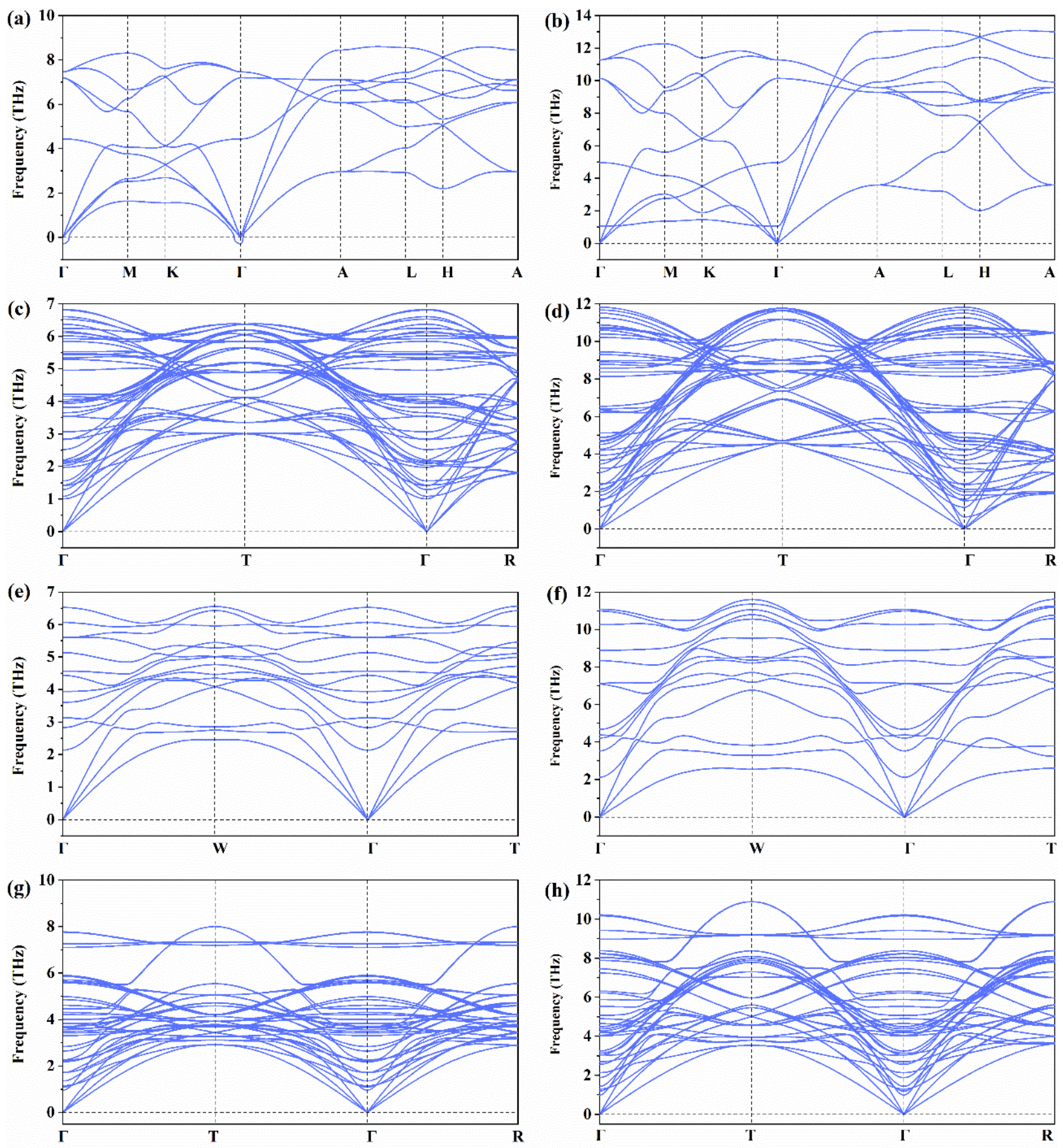

**Fig. S11** Calculated phonon spectra of *P6/mmm*-$U_2Cr$ at (a) 50 GPa and (b) 200 GPa, *Cmcm*-$U_3Cr$ at (c) 25 GPa and (d) 200 GPa, *Immm*-$U_4Cr$ at (e) 25 GPa and (f) 200 GPa, and *Cmmm*-$U_6Cr$ at (g) 25 GPa and (h) 100 GPa, respectively.

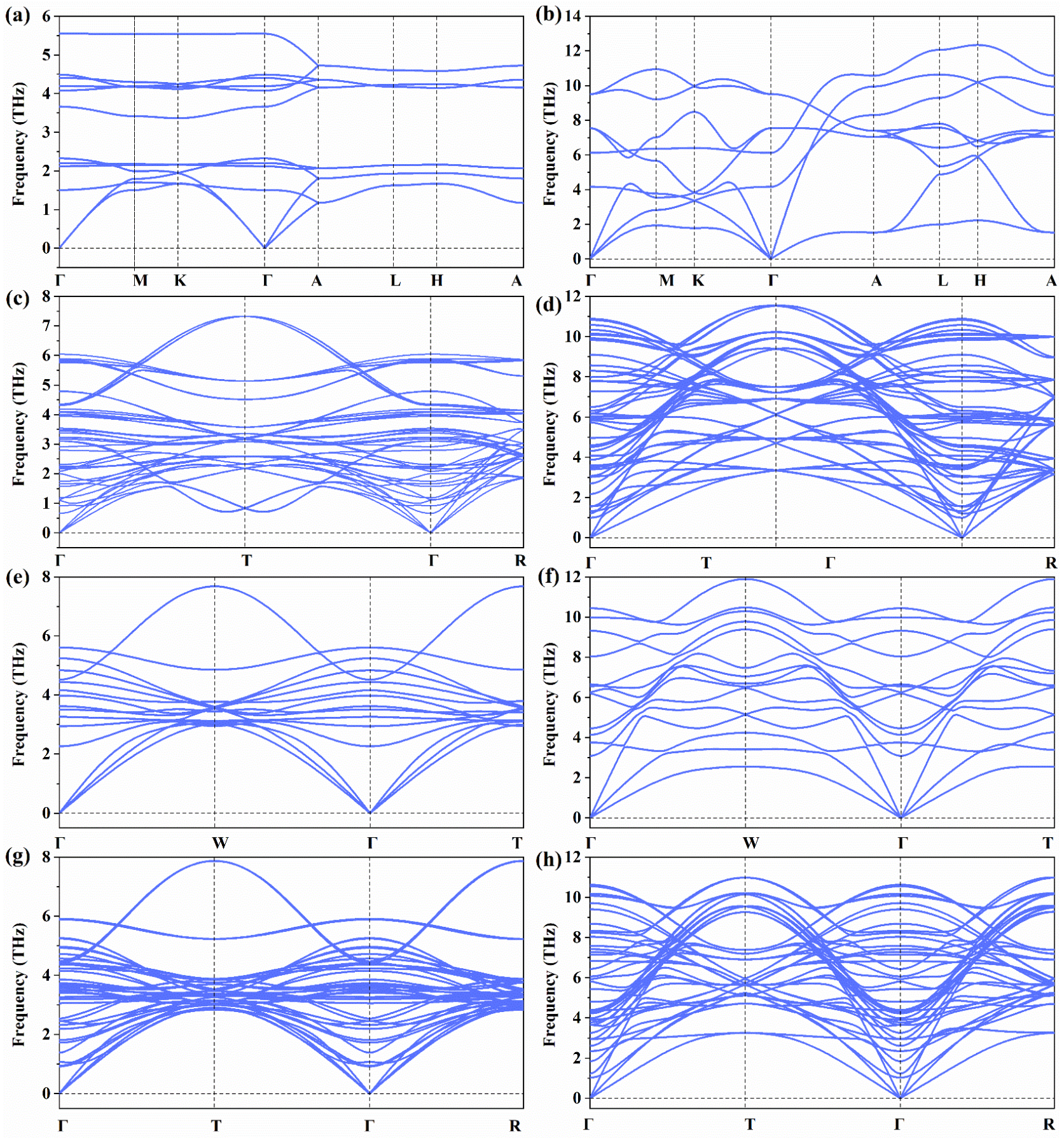

**Fig. S12** Calculated phonon spectra for (a) *P6₃/mmc*-$U_2Y$ at 0 GPa, (b) *P6/mmm*-$U_2Y$ at 200 GPa, *Cmcm*-$U_3Y$ at (c) 25 GPa and (d) 200 GPa, *Immm*-$U_4Y$ at (e) 25 GPa and (f) 200 GPa, and *Cmmm*-$U_6Y$ at (g) 25 GPa and (h) 200 GPa, respectively.

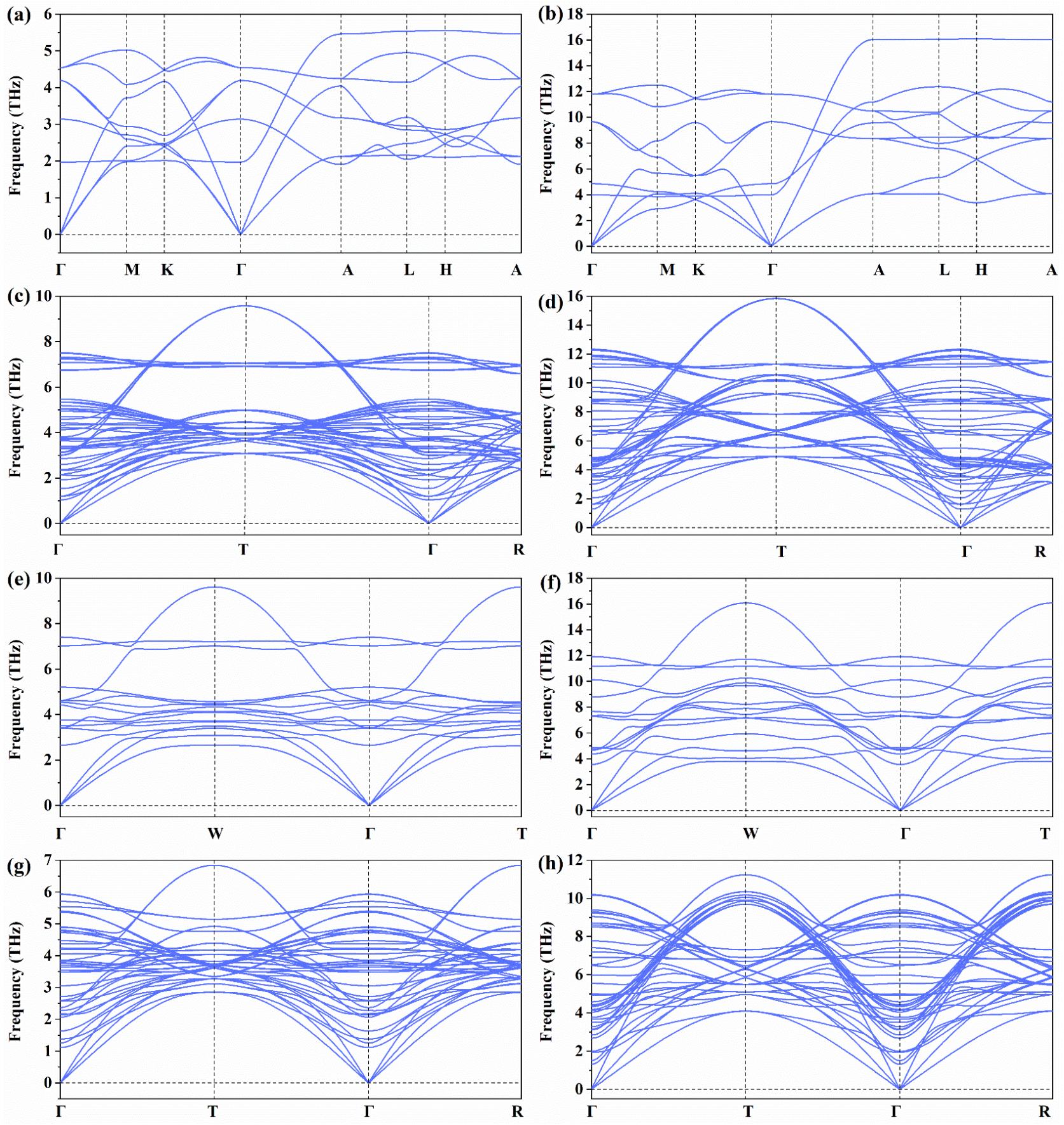

**Fig. S13** Calculated phonon spectra of *P6/mmm*-$U_2W$ at (a) 50 GPa and (b) 200 GPa, *Cmcm*-$U_3W$ at (c) 25 GPa and (d) 200 GPa, *Immm*-$U_4W$ at (e) 25 GPa and (f) 200 GPa, and *Cmmm*-$U_6W$ at (g) 25 GPa and (h) 200 GPa, respectively.

## SIII. Electronic structure and bonding properties for other U-based transition metal alloys

The calculated electronic projected density of states (PDOS) for other U-based transition metal alloys, U-X (X= Sc, Ti, V, Nb, Cr, Mo, Hf, Ta, Y, W), are presented in Figs. S14-S23. At zero pressure, *P6₃/mmc*-$U_2X$ or *P6/mmm*-$U_2X$, *Cmcm*-$U_3X$, *Immm*-$U_4X$, and *Cmmm*-$U_6X$ are all metallic with the 5*f* electrons of U atoms dominating the DOS near the Fermi level. Under high pressure, the dominance of the 5*f* electrons is weakened and the total DOS tends to be delocalized. The differential charge density of the ordered phases in different alloying systems is also calculated and shown in Figs. S24-S33. Similar inter-layer and intra-layer U-U covalent bonds are also observed at high pressures for these phases in different alloying systems.

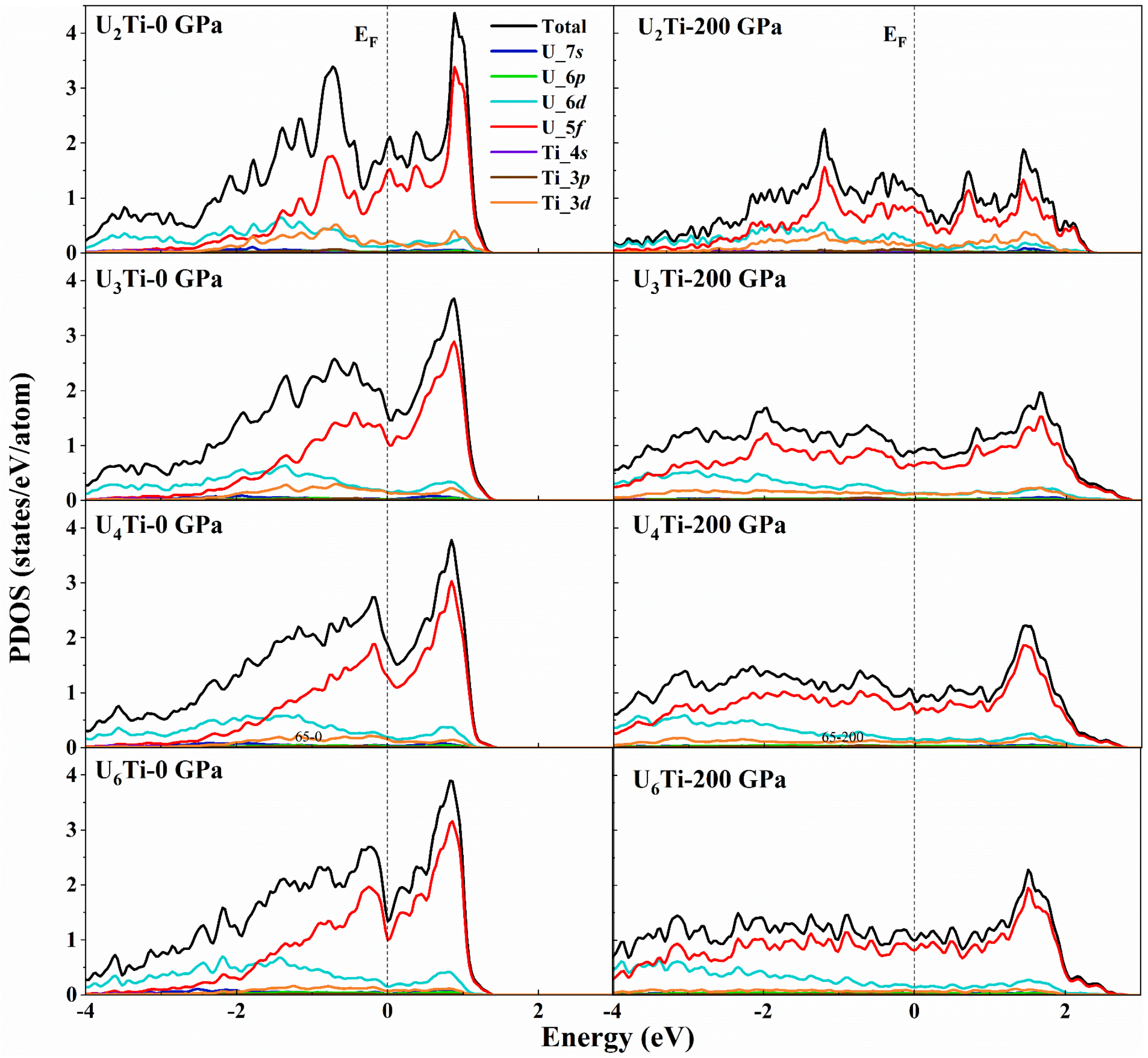

**Fig. S14** Calculated projected density of states (PDOS) for *P6₃/mmc*-$U_2Ti$ at 0 GPa and *P6/mmm*-$U_2Ti$ at 200 GPa, and *Cmcm*-$U_3Ti$, *Immm*-$U_4Ti$, and *Cmmm*-$U_6Ti$ at 0 and 200 GPa, respectively.

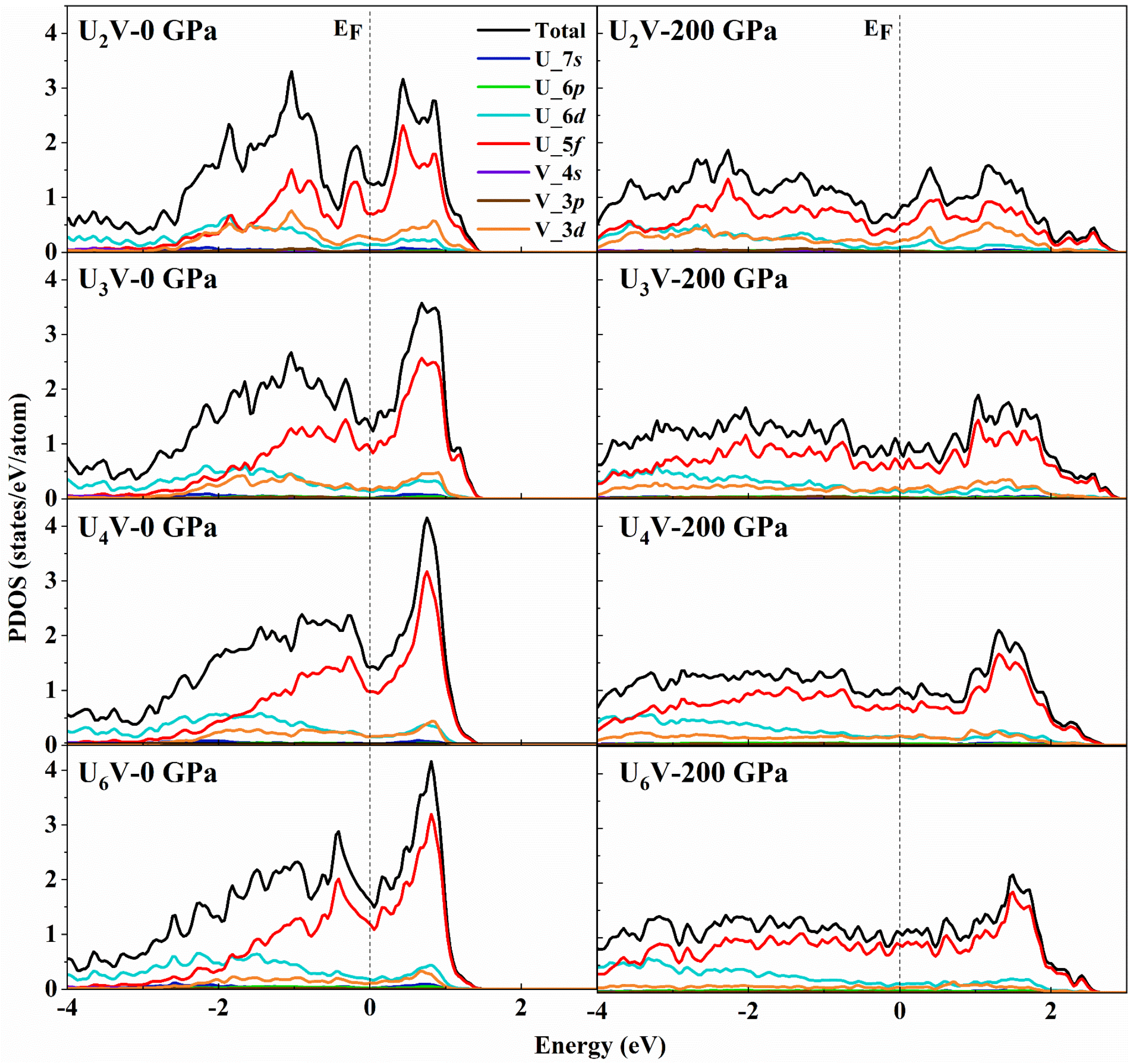

**Fig. S15** Calculated projected density of states (PDOS) for *P6/mmm*-$U_2V$, *Cmcm*-$U_3V$, *Immm*-$U_4V$, and *Cmmm*-$U_6V$ at 0 and 200 GPa, respectively.

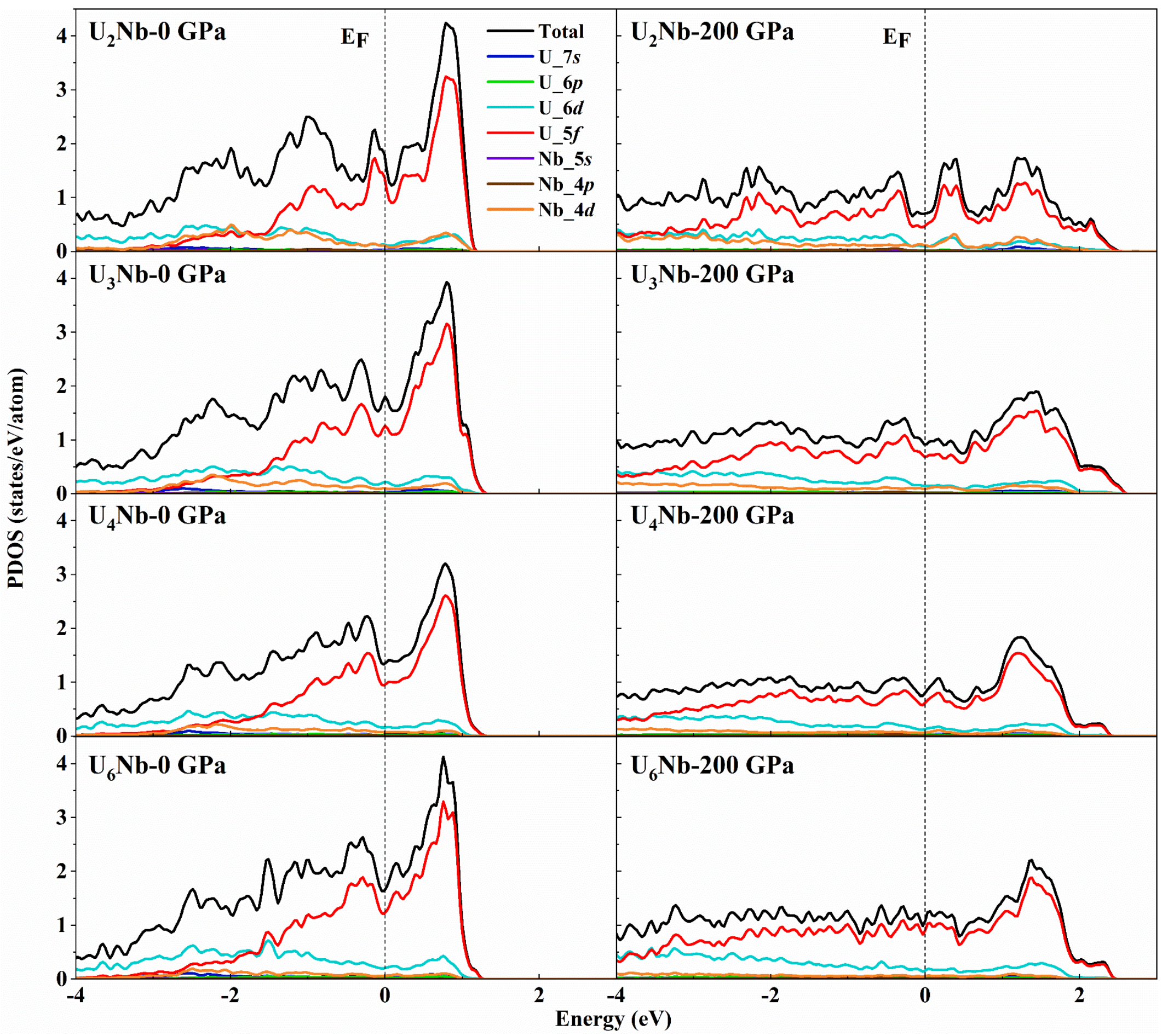

**Fig. S16** Calculated projected density of states (PDOS) for *P6₃/mmc*-$U_2Nb$ at 0 GPa and *P6/mmm*-$U_2Nb$ at 200 GPa, and *Cmcm*-$U_3Nb$, *Immm*-$U_4Nb$, and *Cmmm*-$U_6Nb$ at 0 and 200 GPa, respectively.

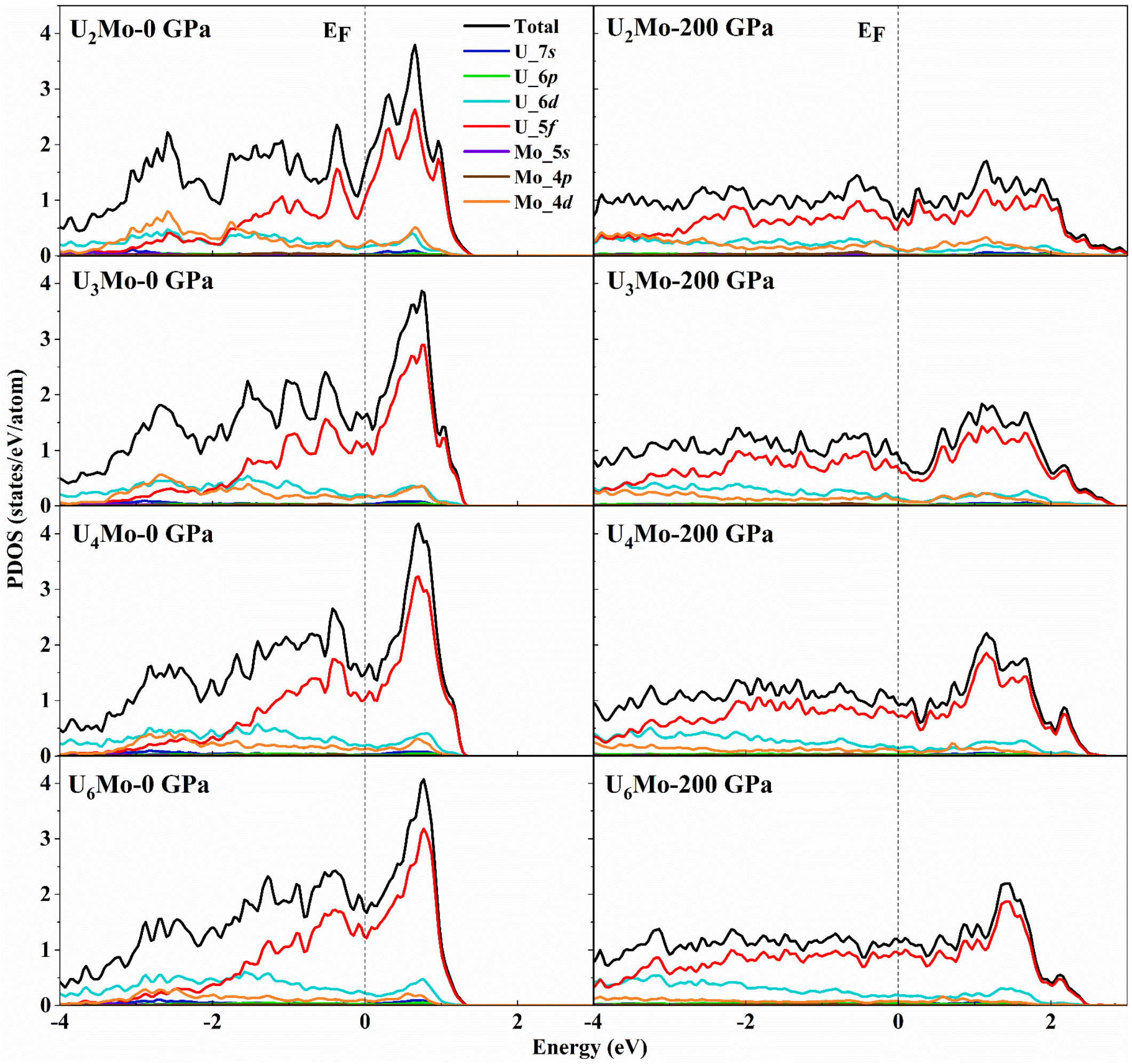

**Fig. S17** Calculated projected density of states (PDOS) for *P63/mmc*-$U_2Mo$ at 0 GPa and *P6/mmm*-$U_2Mo$ at 200 GPa, and *Cmcm*-$U_3Mo$, *Immm*-$U_4Mo$, and *Cmmm*-$U_6Mo$ at 0 and 200 GPa, respectively.

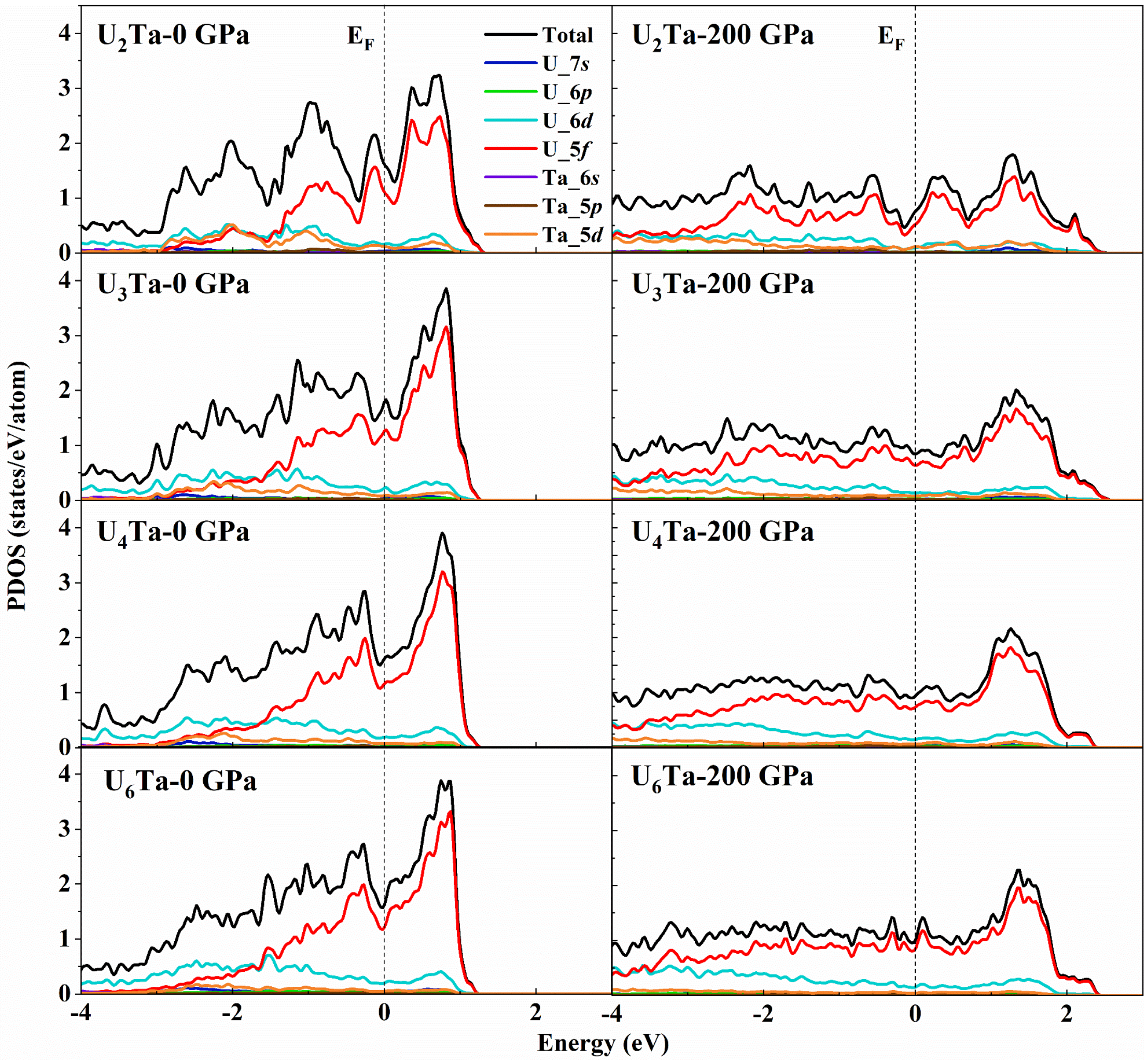

**Fig. S18** Calculated projected density of states (PDOS) for *P6₃/mmc*-$U_2Ta$ at 0 GPa and *P6/mmm*-$U_2Ta$ at 200 GPa, and *Cmcm*-$U_3Ta$, *Immm*-$U_4Ta$, and *Cmmm*-$U_6Ta$ at 0 and 200 GPa, respectively.

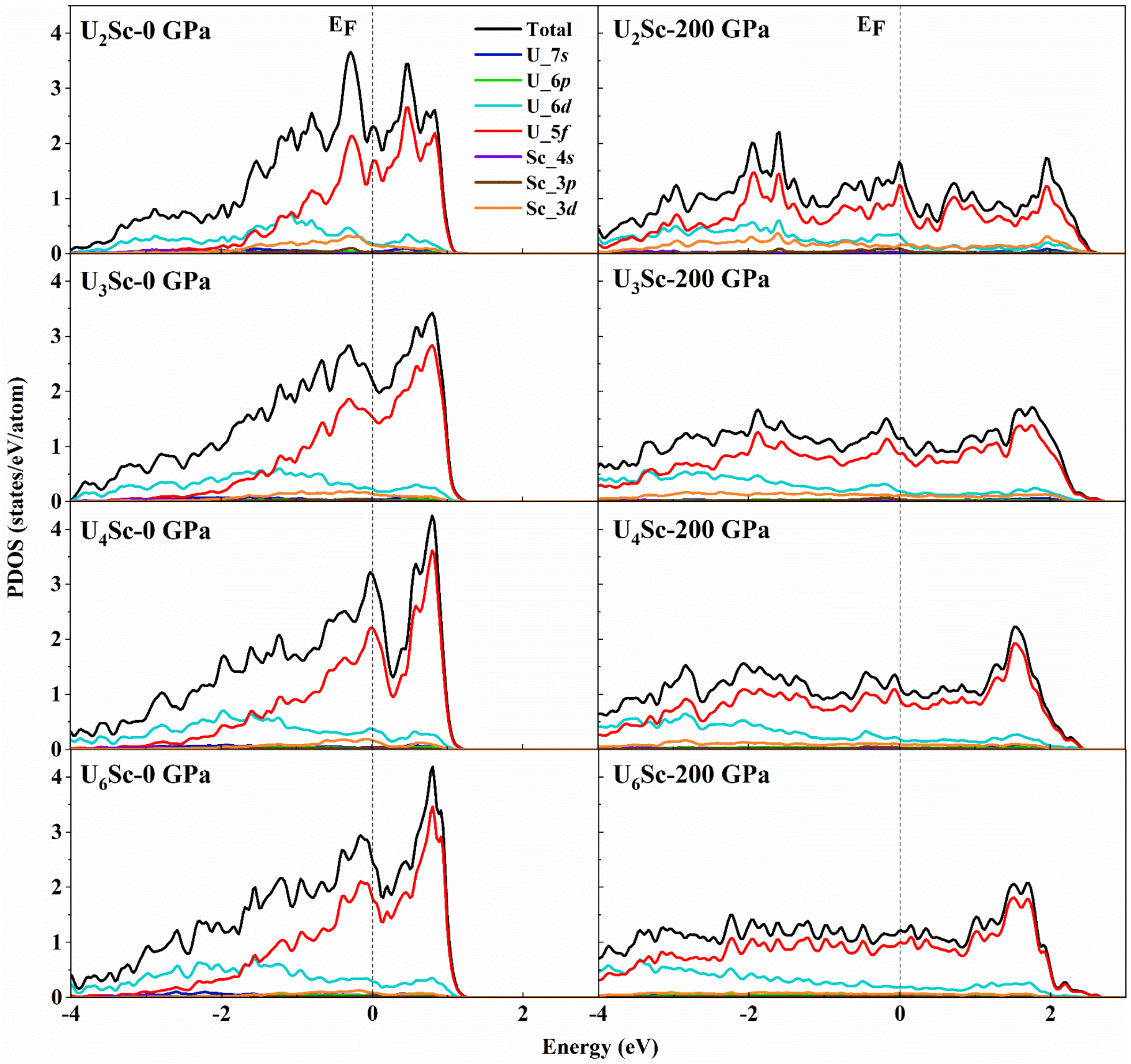

**Fig. S19** Calculated projected density of states (PDOS) for *P63/mmc*-$U_2Sc$ at 0 GPa and *P6/mmm*-$U_2Sc$ at 200 GPa, and *Cmcm*-$U_3Sc$, *Immm*-$U_4Sc$, and *Cmmm*-$U_6Sc$ at 0 and 200 GPa, respectively.

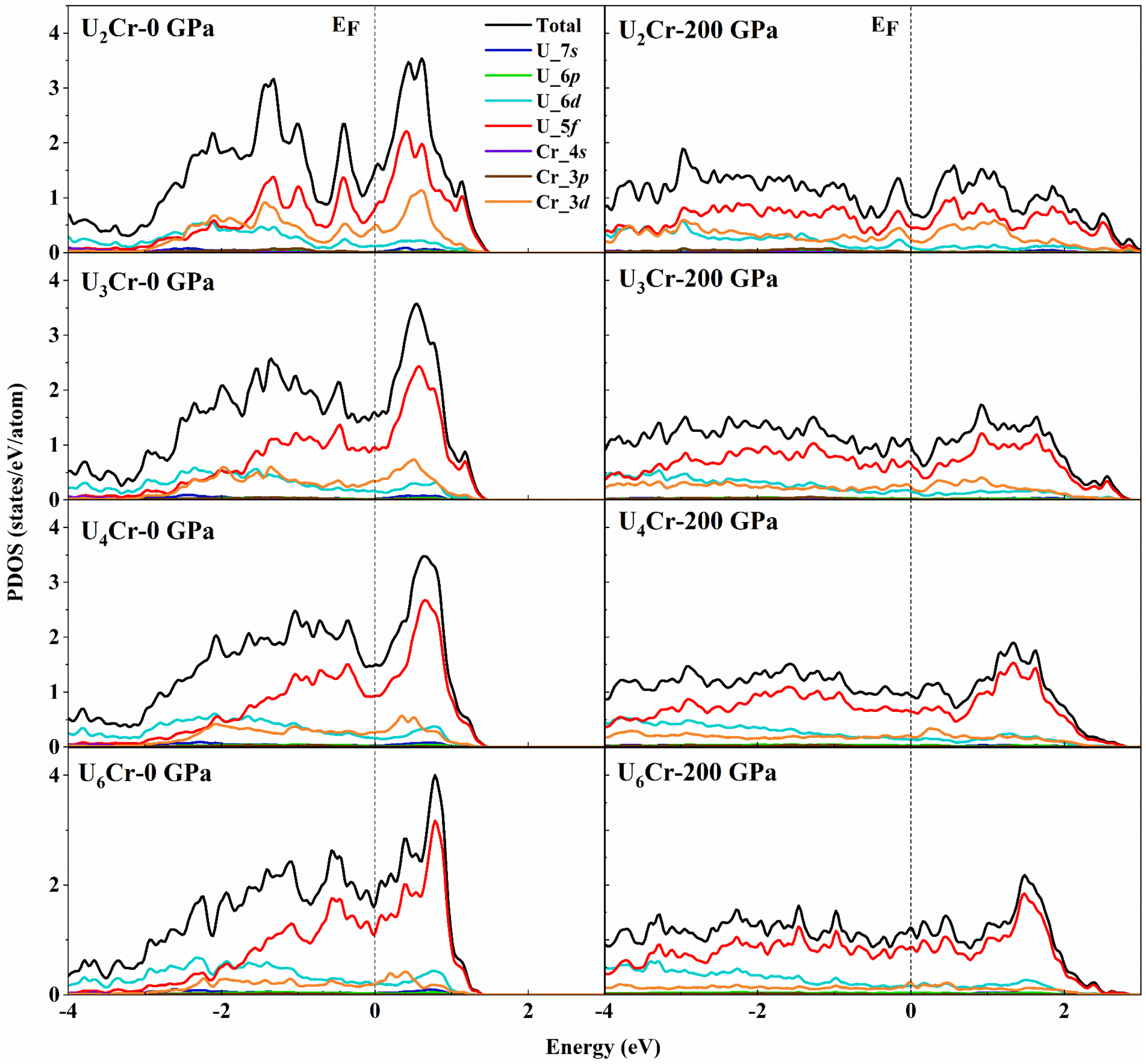

**Fig. S20** Calculated projected density of states (PDOS) for *P6/mmm*-$U_2Cr$, *Cmcm*-$U_3Cr$, *Immm*-$U_4Cr$, and *Cmmm*-$U_6Cr$ at 0 and 200 GPa, respectively.

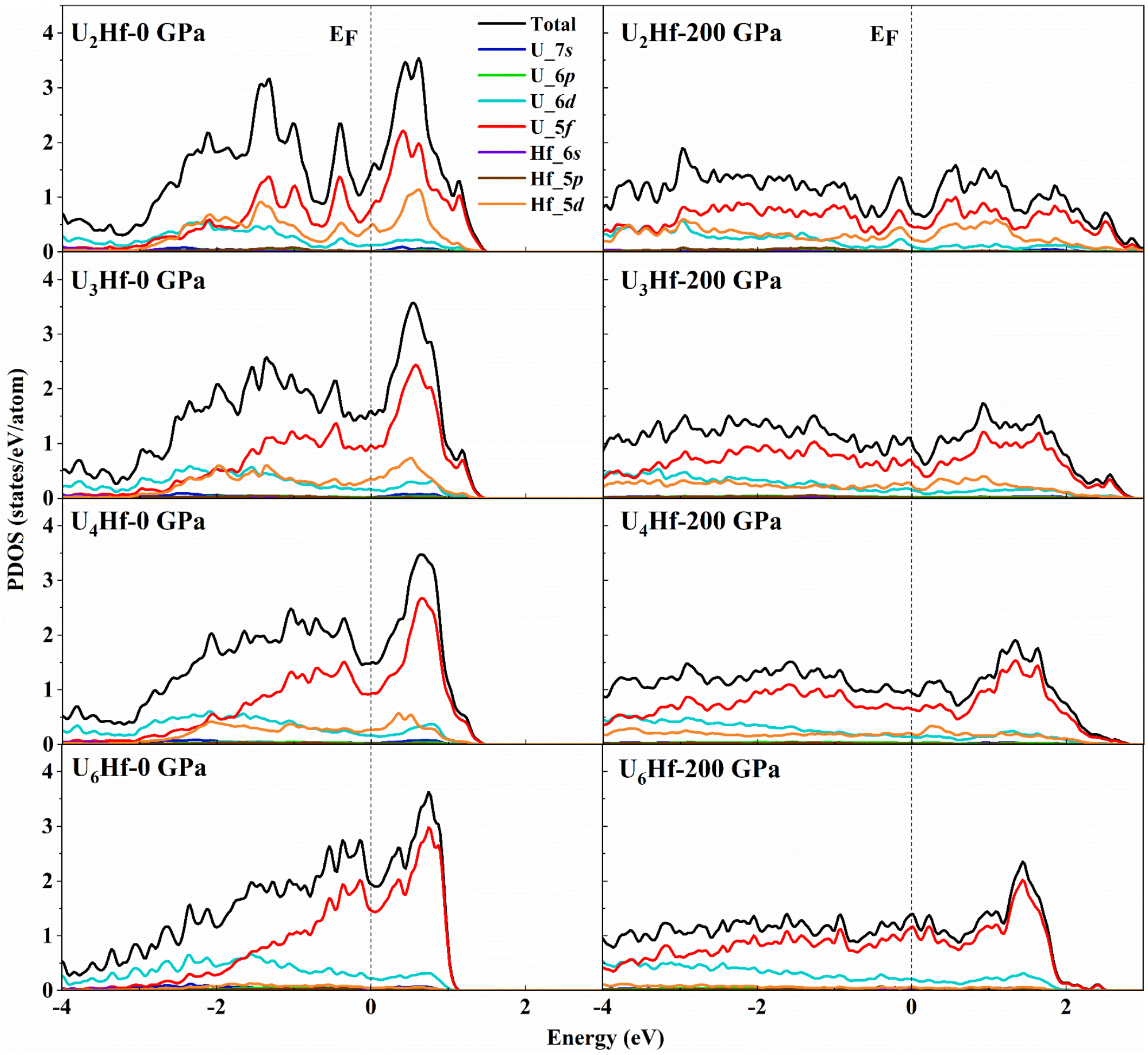

**Fig. S21** Calculated projected density of states (PDOS) for *P63/mmc*-$U_2Hf$ at 0 GPa and *P6/mmm*-$U_2Hf$ at 200 GPa, and *Cmcm*-$U_3Hf$, *Immm*-$U_4Hf$, and *Cmmm*-$U_6Hf$ at 0 and 200 GPa, respectively.

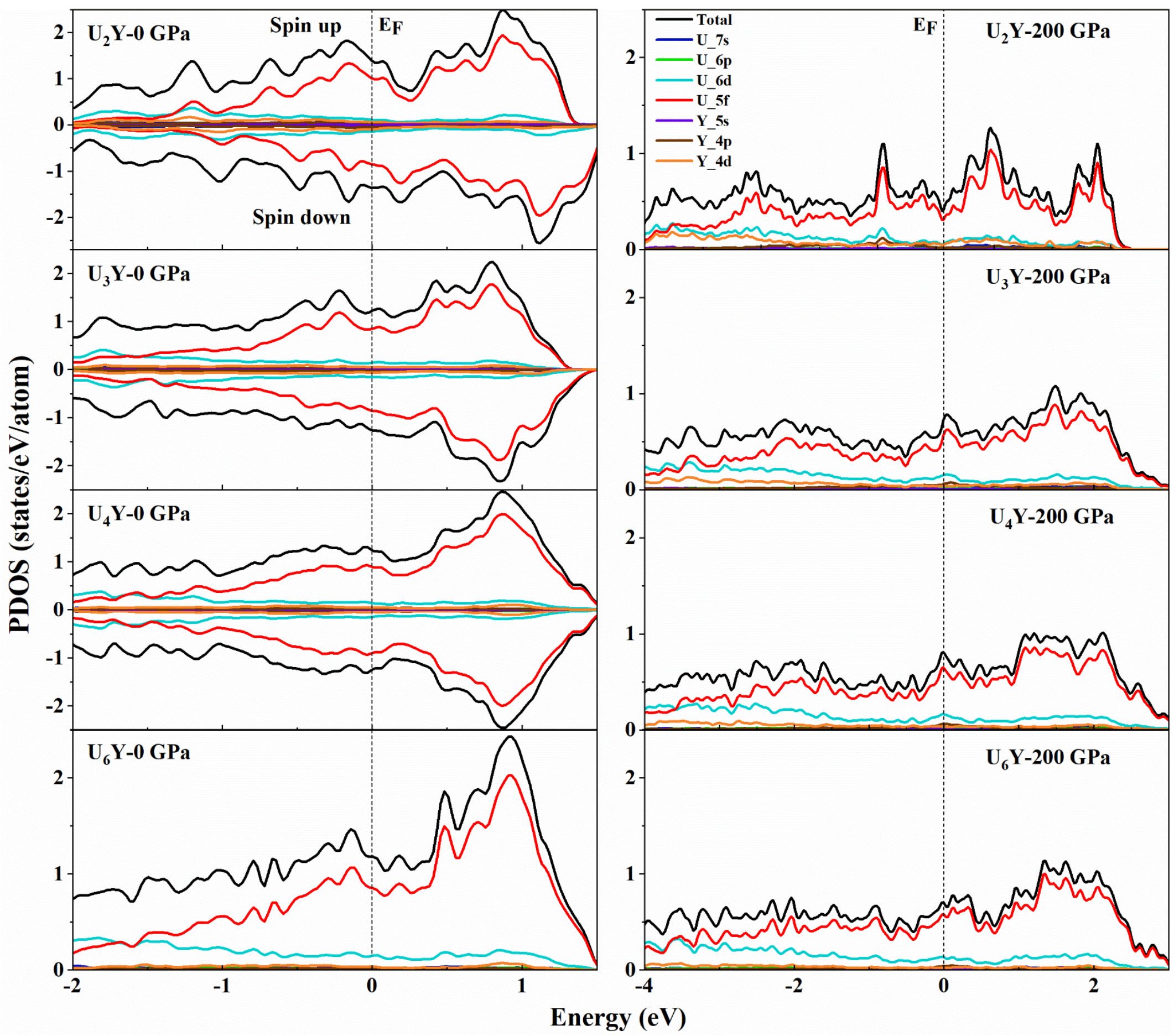

**Fig. S22** Calculated projected density of states (PDOS) for *P6₃/mmc*-U$_2$Y at 0 GPa and *P6/mmm*-U$_2$Y at 200 GPa, and *Cmcm*-U$_3$Y, *Immm*-U$_4$Y, and *Immm*-U$_6$Y at 0 and 200 GPa, respectively.

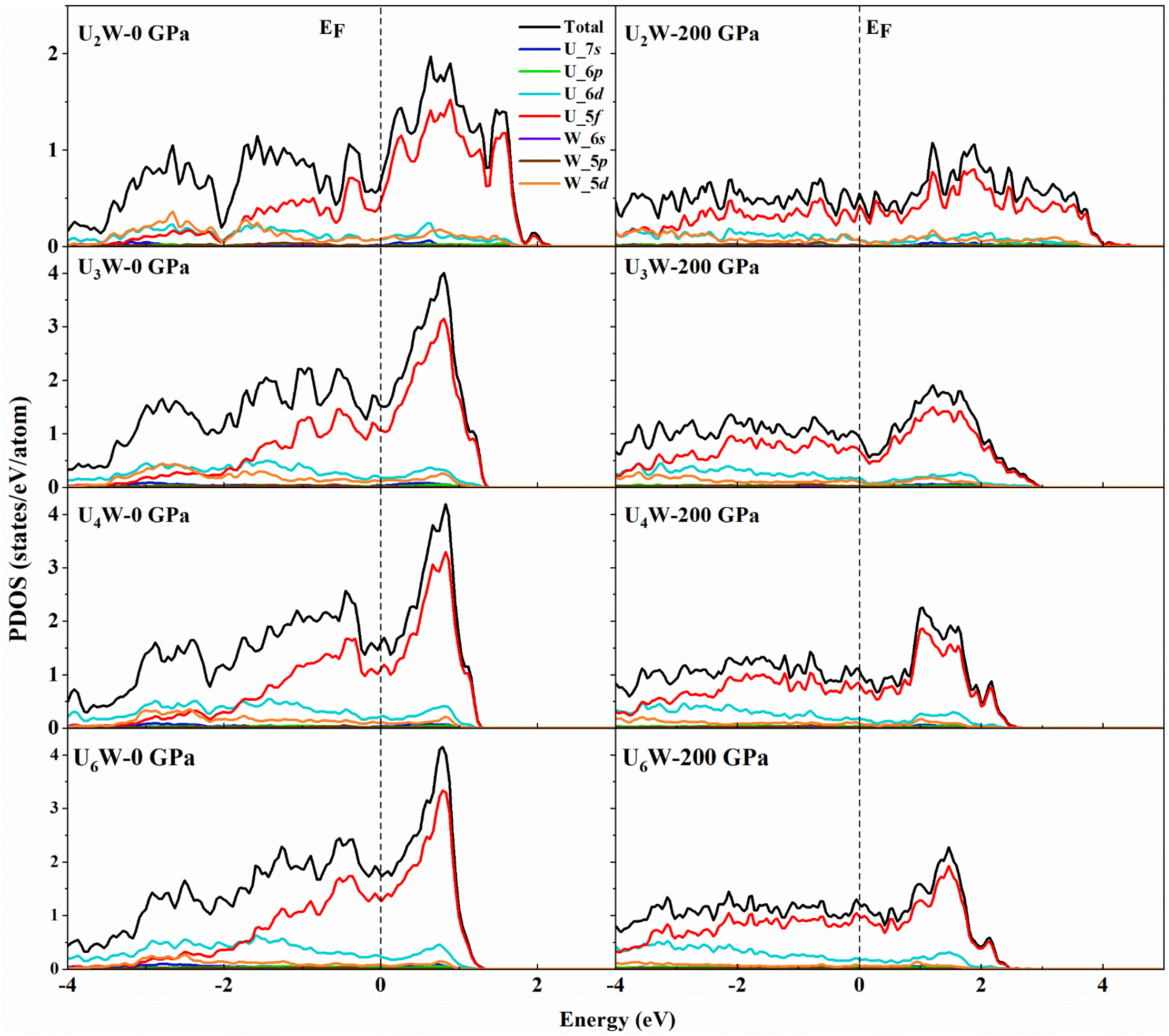

**Fig. S23** Calculated projected density of states (PDOS) for *P6/mmm*-$U_2W$ at 0 GPa and at 200 GPa, and *Cmcm*-$U_3W$, *Immm*-$U_4W$, and *Cmmm*-$U_6W$ at 0 and 200 GPa, respectively.

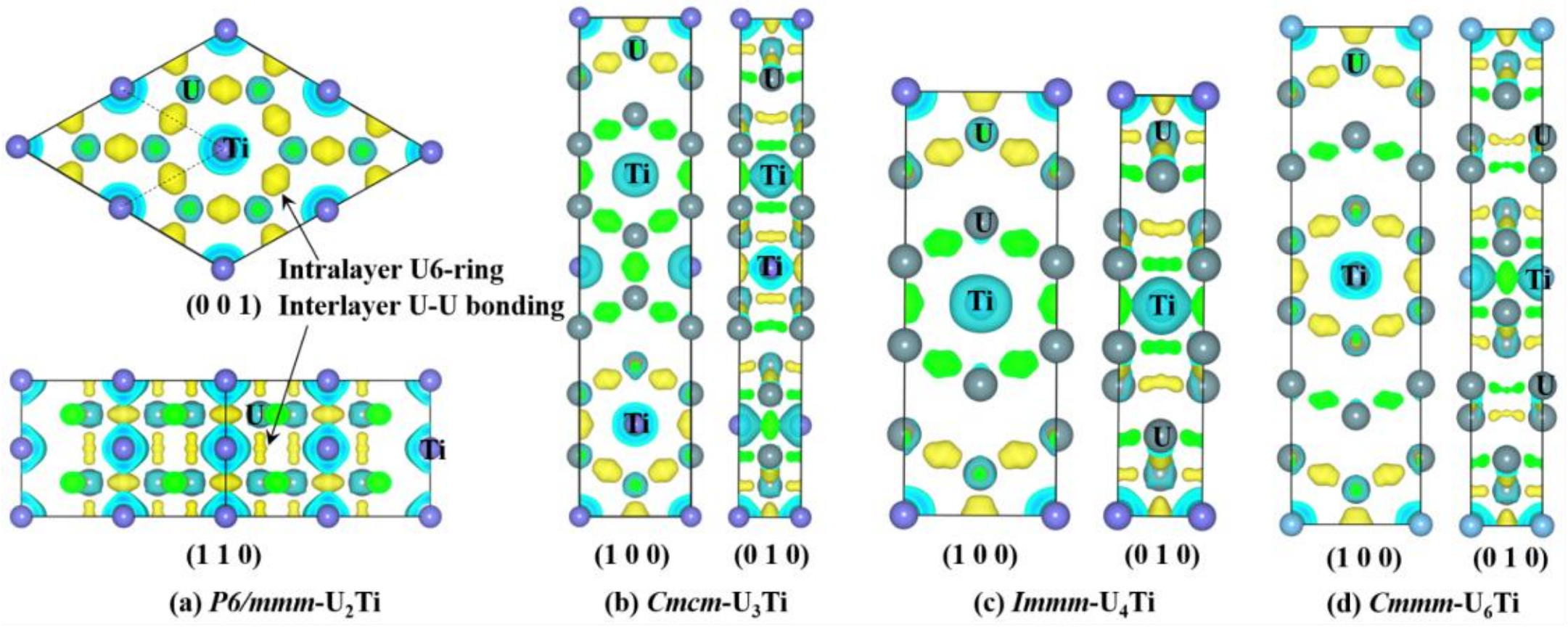

**Fig. S24** Differential charge density (isosurface = 0.02 e/Bohr$^3$) for (a) *P6/mmm*-$U_2Ti$, (b) *Cmcm*-$U_3Ti$, (c) *Immm*-$U_4Ti$, and (d) *Cmmm*-$U_6Ti$ at 200 GPa. The blue color

indicates loss of electrons and yellow color represents the gain of charge between U atoms.

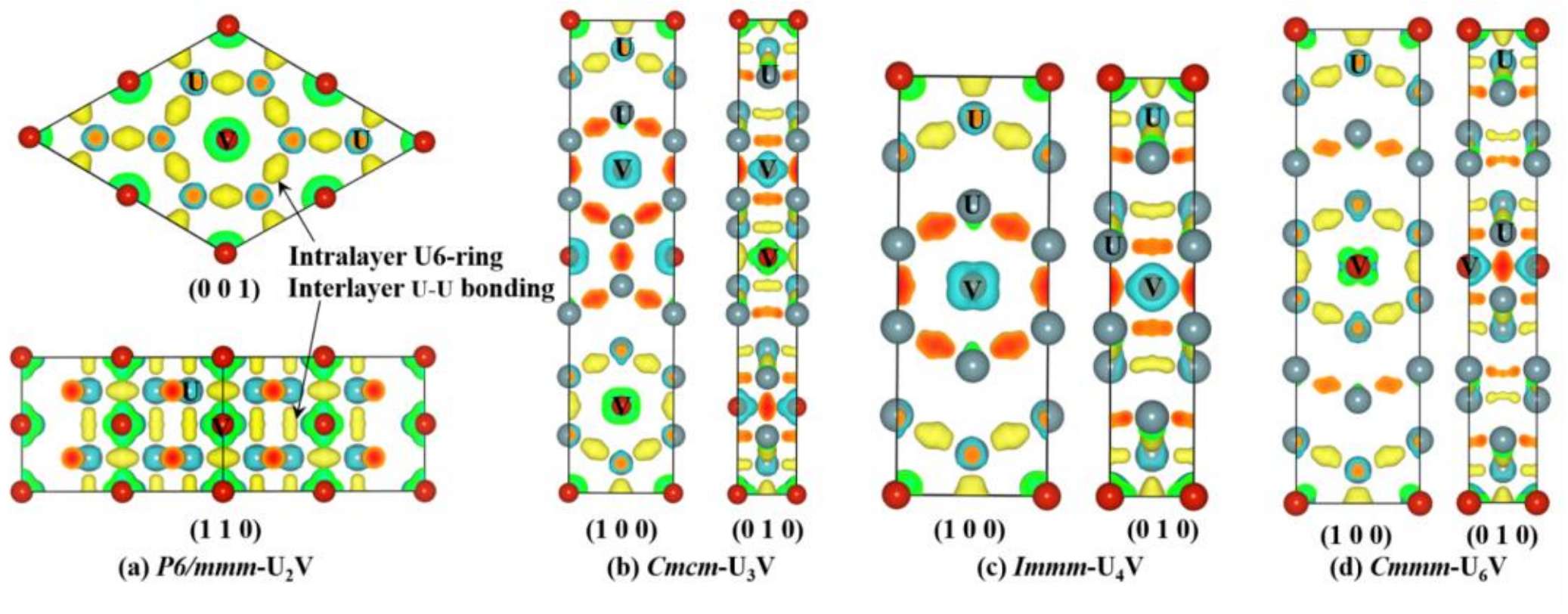

**Fig. S25** Differential charge density (isosurface = 0.02 e/Bohr$^3$) for (a) *P6/mmm*-$U_2V$, (b) *Cmcm*-$U_3V$, (c) *Immm*-$U_4V$, and (d) *Cmmm*-$U_6V$ at 200 GPa. The blue color indicates loss of electrons and yellow color represents the gain of charge between U atoms.

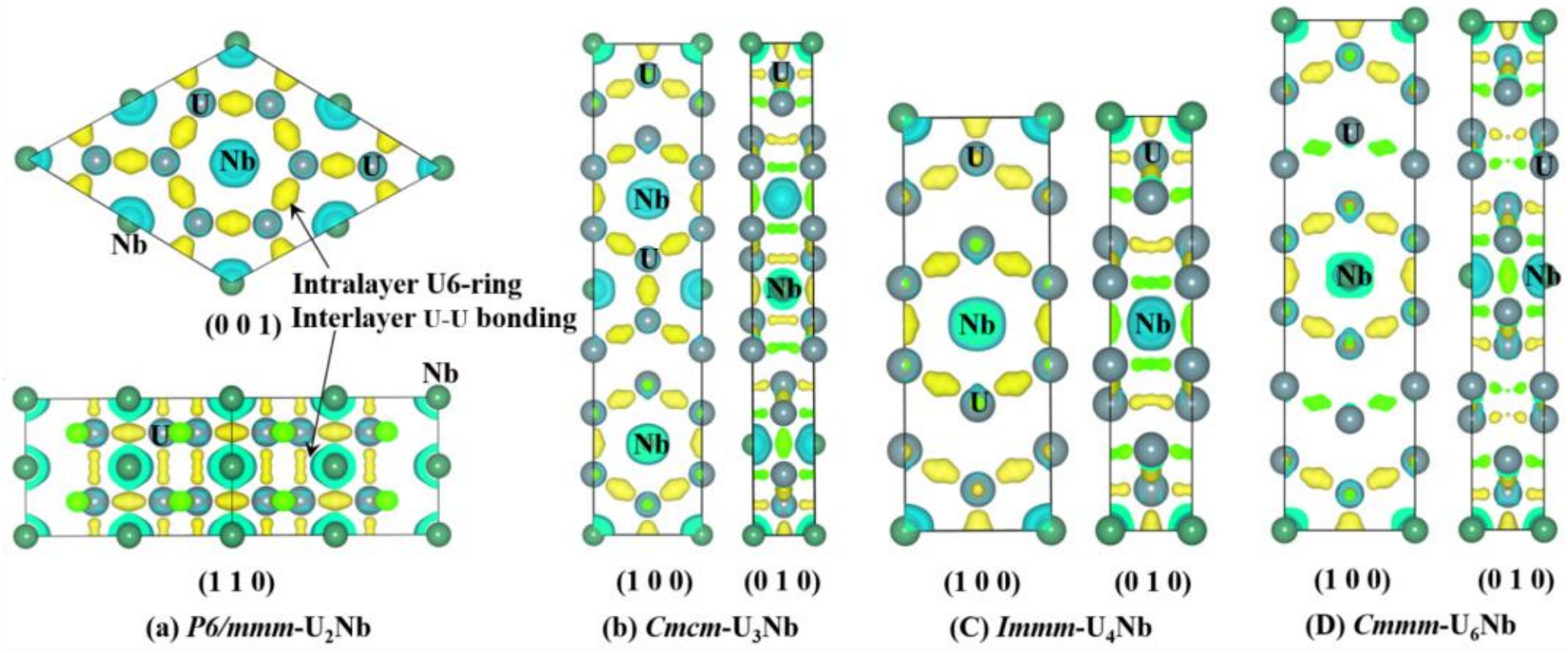

**Fig. S26** Differential charge density (isosurface = 0.02 e/Bohr$^3$) for (a) *P6/mmm*-$U_2Nb$, (b) *Cmcm*-$U_3Nb$, (c) *Immm*-$U_4Nb$, and (d) *Cmmm*-$U_6Nb$ at 200 GPa. The blue color indicates loss of electrons and yellow color represents the gain of charge between U atoms.

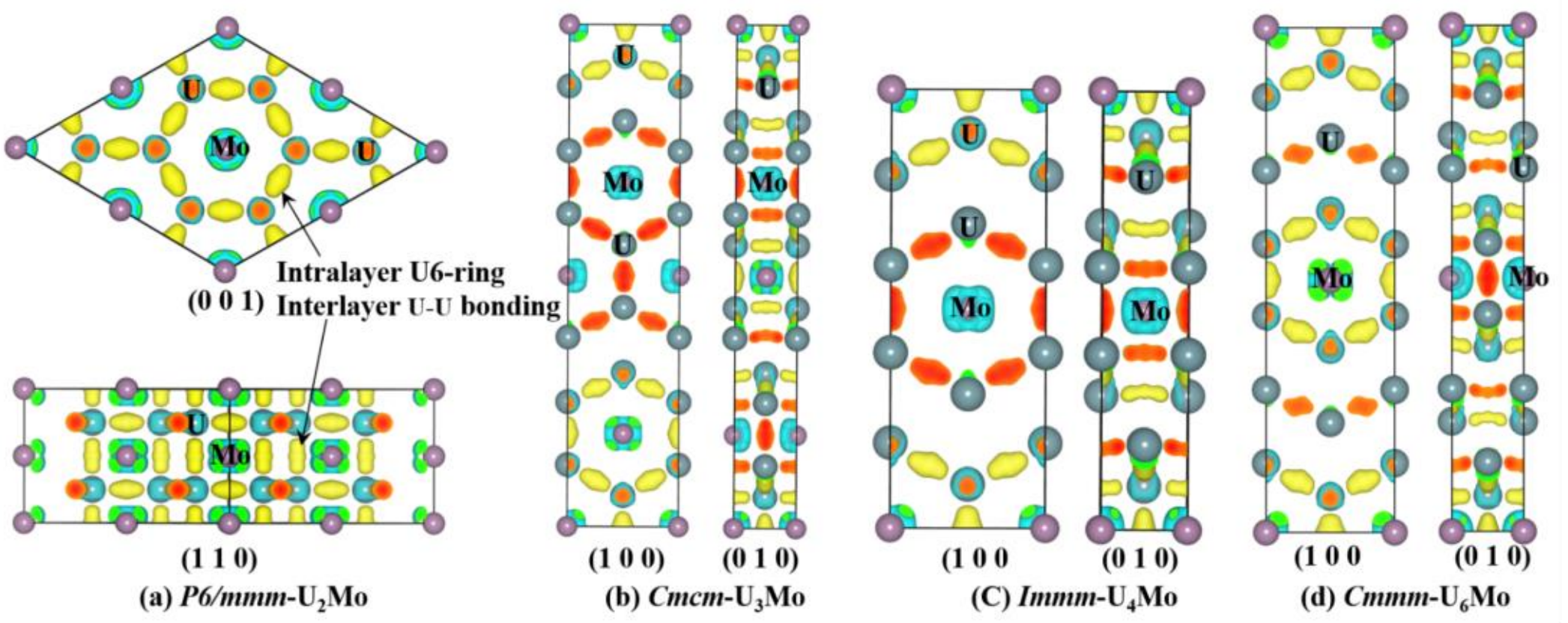

**Fig. S27** Differential charge density (isosurface = 0.02 e/Bohr$^3$) for (a) *P6/mmm*-$U_2Mo$, (b) *Cmcm*-$U_3Mo$, (c) *Immm*-$U_4Mo$, and (d) *Cmmm*-$U_6Mo$ at 200 GPa. The blue color indicates loss of electrons and yellow color represents the gain of charge between U atoms.

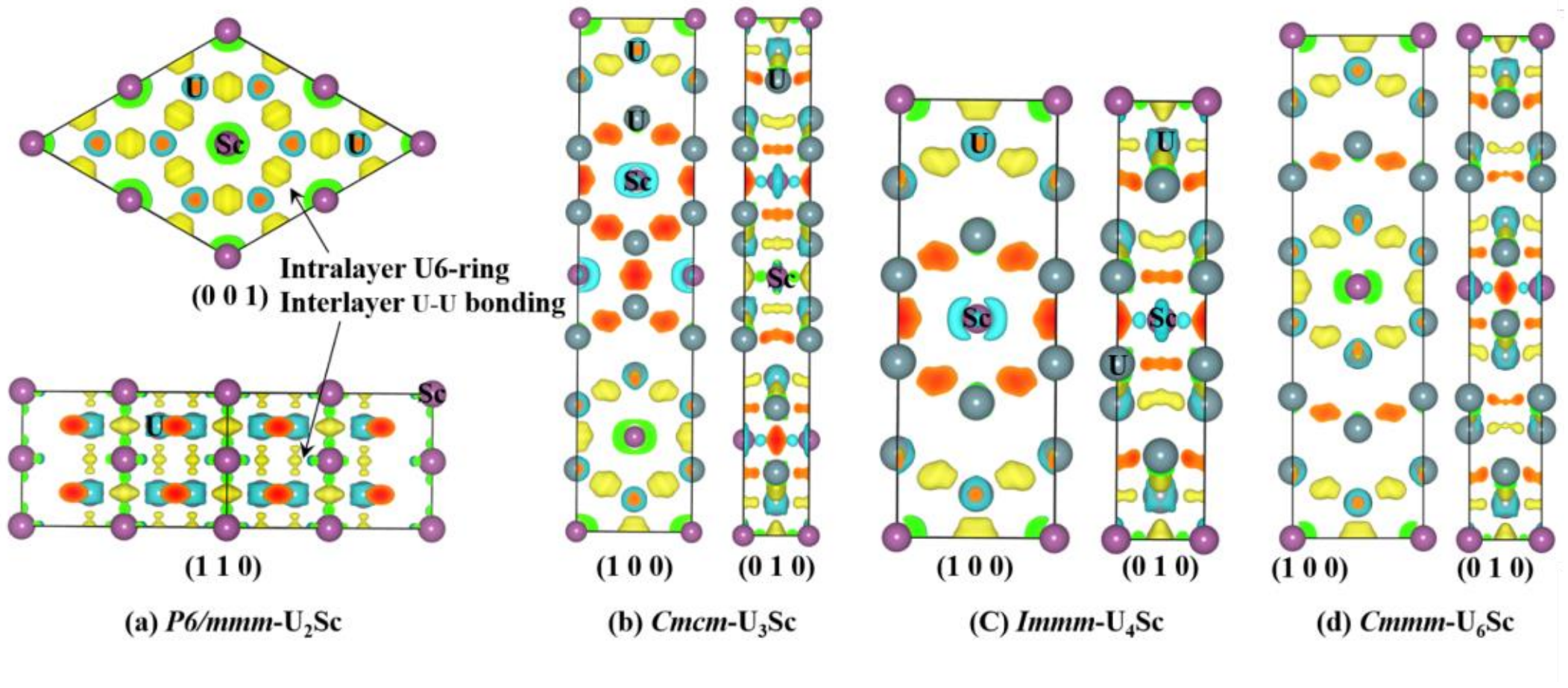

**Fig. S28** Differential charge density (isosurface = 0.02 e/Bohr$^3$) for (a) *P6/mmm*-$U_2Sc$, (b) *Cmcm*-$U_3Sc$, (c) *Immm*-$U_4Sc$, and (d) *Cmmm*-$U_6Sc$ at 200 GPa. The blue color indicates loss of electrons and yellow color represents the gain of charge between U atoms.

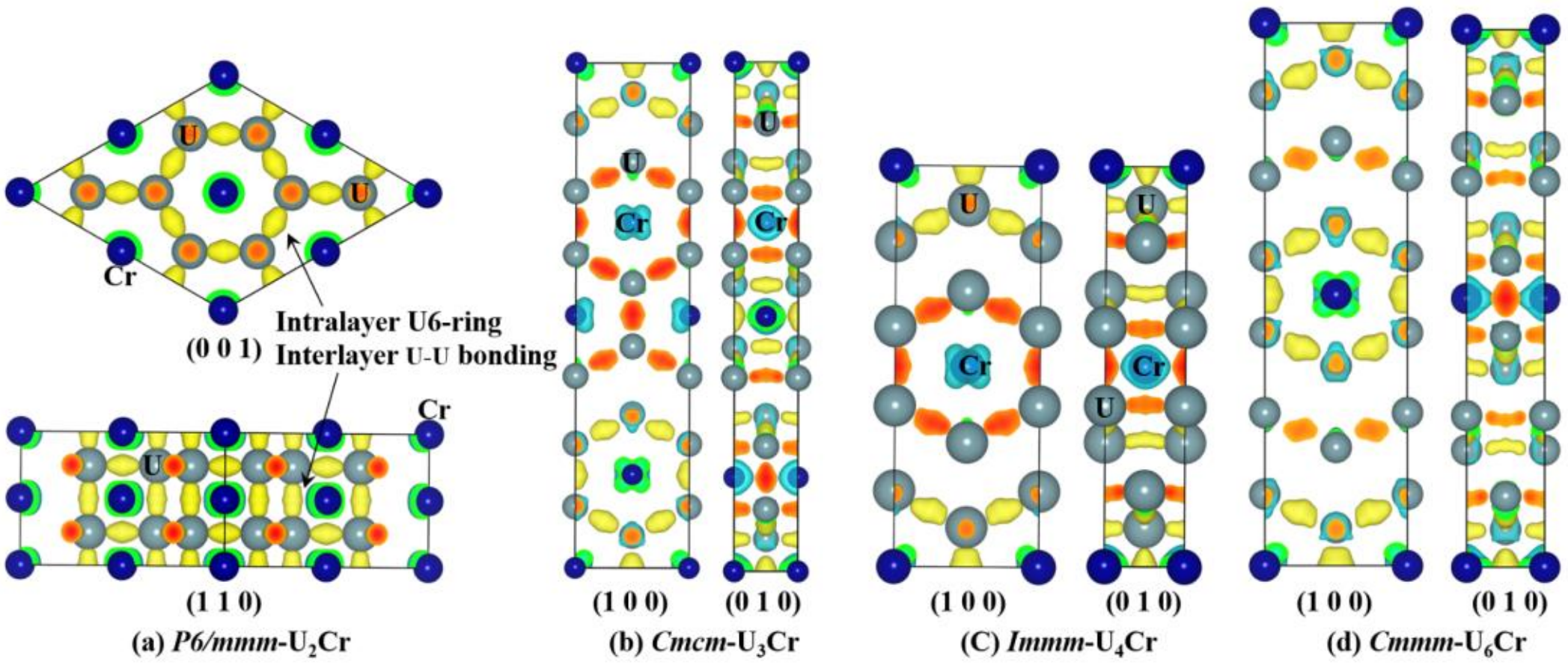

**Fig. S29** Differential charge density (isosurface = 0.02 e/Bohr$^3$) for (a) *P6/mmm*-$U_2Cr$, (b) *Cmcm*-$U_3Cr$, (c) *Immm*-$U_4Cr$, and (d) *Cmmm*-$U_6Cr$ at 200 GPa. The blue color indicates loss of electrons and yellow color represents the gain of charge between U atoms.

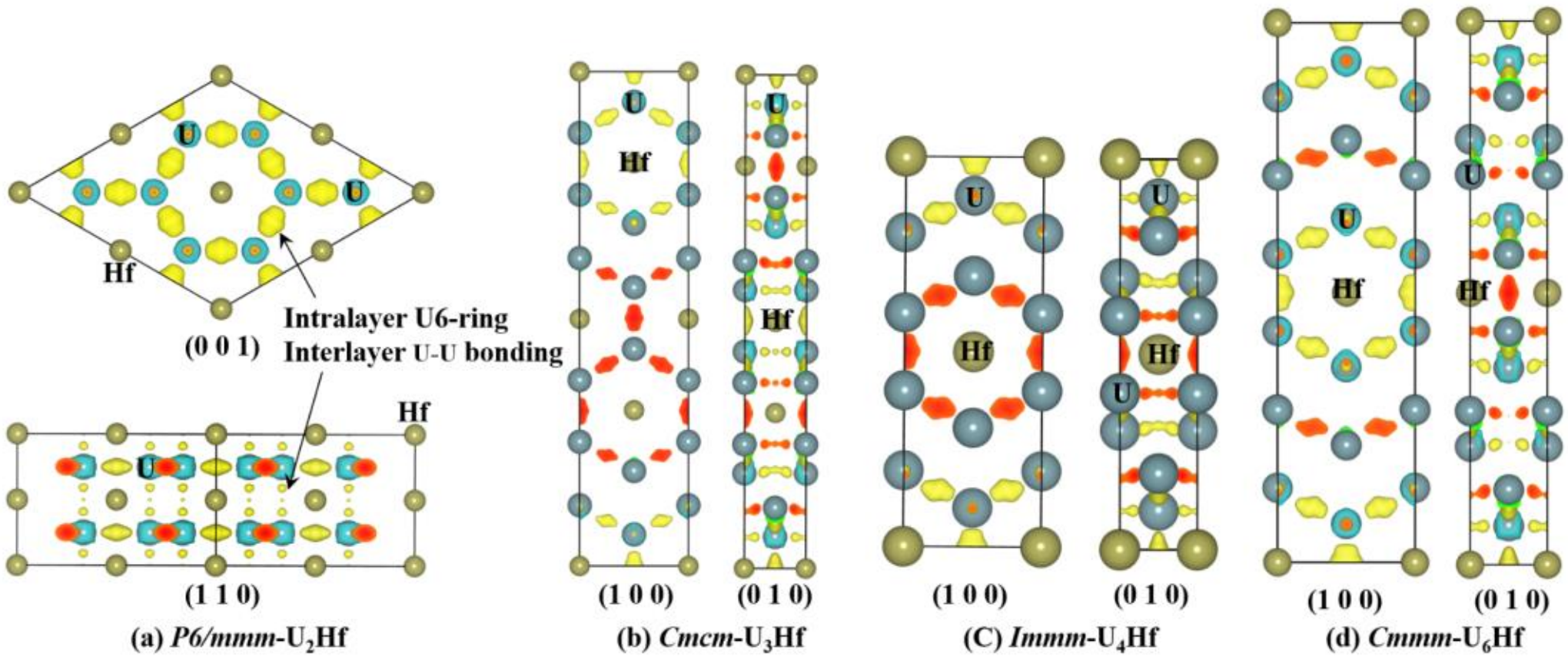

**Fig. S30** Differential charge density (isosurface = 0.02 e/Bohr$^3$) for (a) *P6/mmm*-$U_2Hf$, (b) *Cmcm*-$U_3Hf$, (c) *Immm*-$U_4Hf$, and (d) *Cmmm*-$U_6Hf$ at 200 GPa. The blue color indicates loss of electrons and yellow color represents the gain of charge between U atoms.

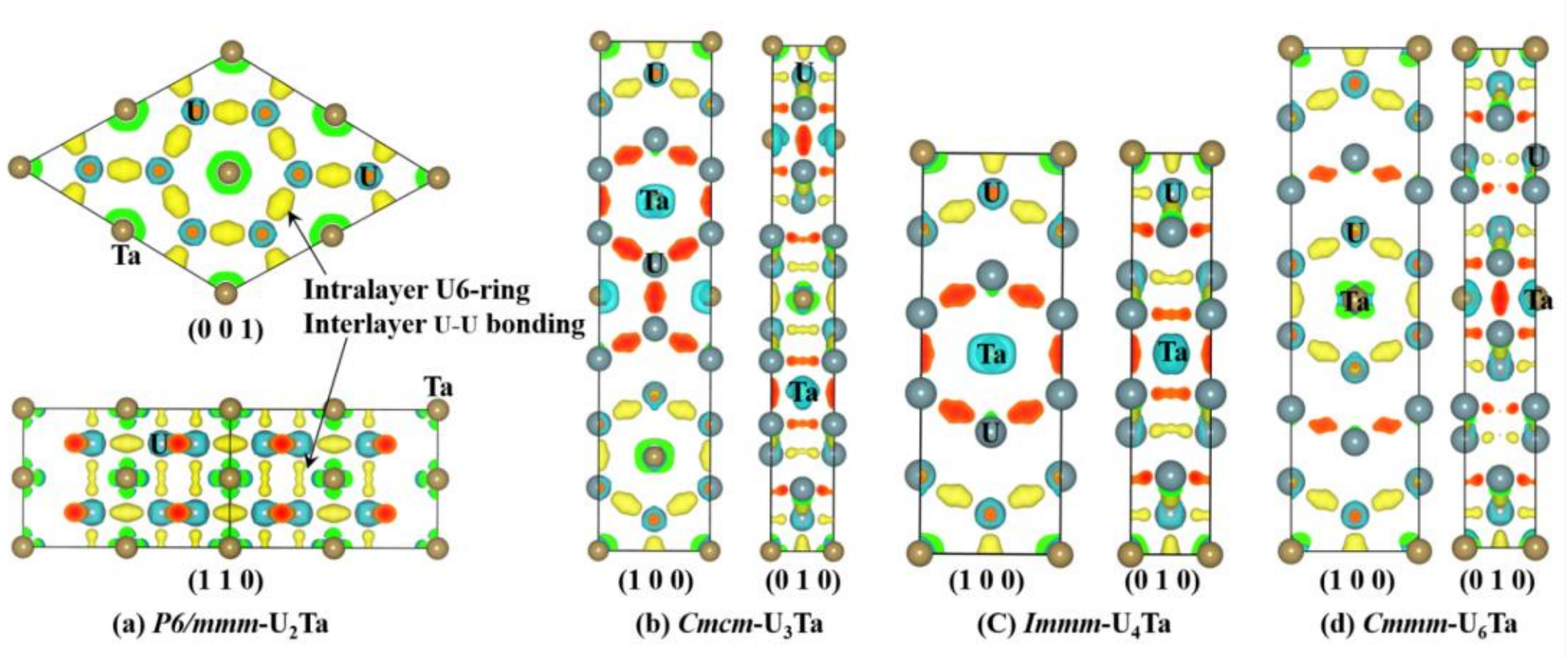

**Fig. S31** Differential charge density (isosurface = 0.02 e/Bohr$^3$) for (a) *P6/mmm*-$U_2Ta$, (b) *Cmcm*-$U_3Ta$, (c) *Immm*-$U_4Ta$, and (d) *Cmmm*-$U_6Ta$ at 200 GPa. The blue color indicates loss of electrons and yellow color represents the gain of charge between U atoms.

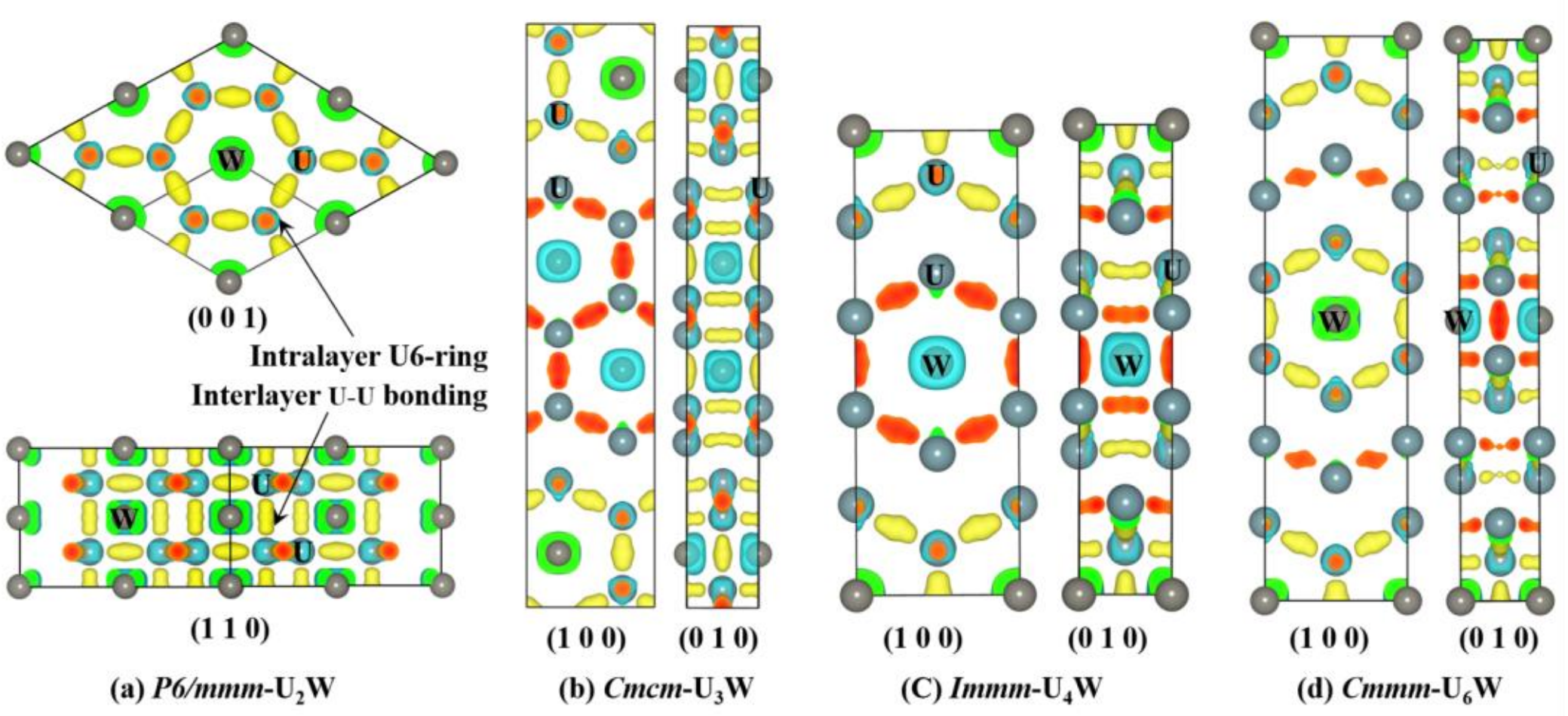

**Fig. S32** Differential charge density (isosurface = 0.02 e/Bohr$^3$) for (a) *P6/mmm*-$U_2W$, (b) *Cmcm*-$U_3W$, (c) *Immm*-$U_4W$, and (d) *Cmmm*-$U_6W$ at 200 GPa. The blue color indicates loss of electrons and yellow color represents the gain of charge between U atoms.

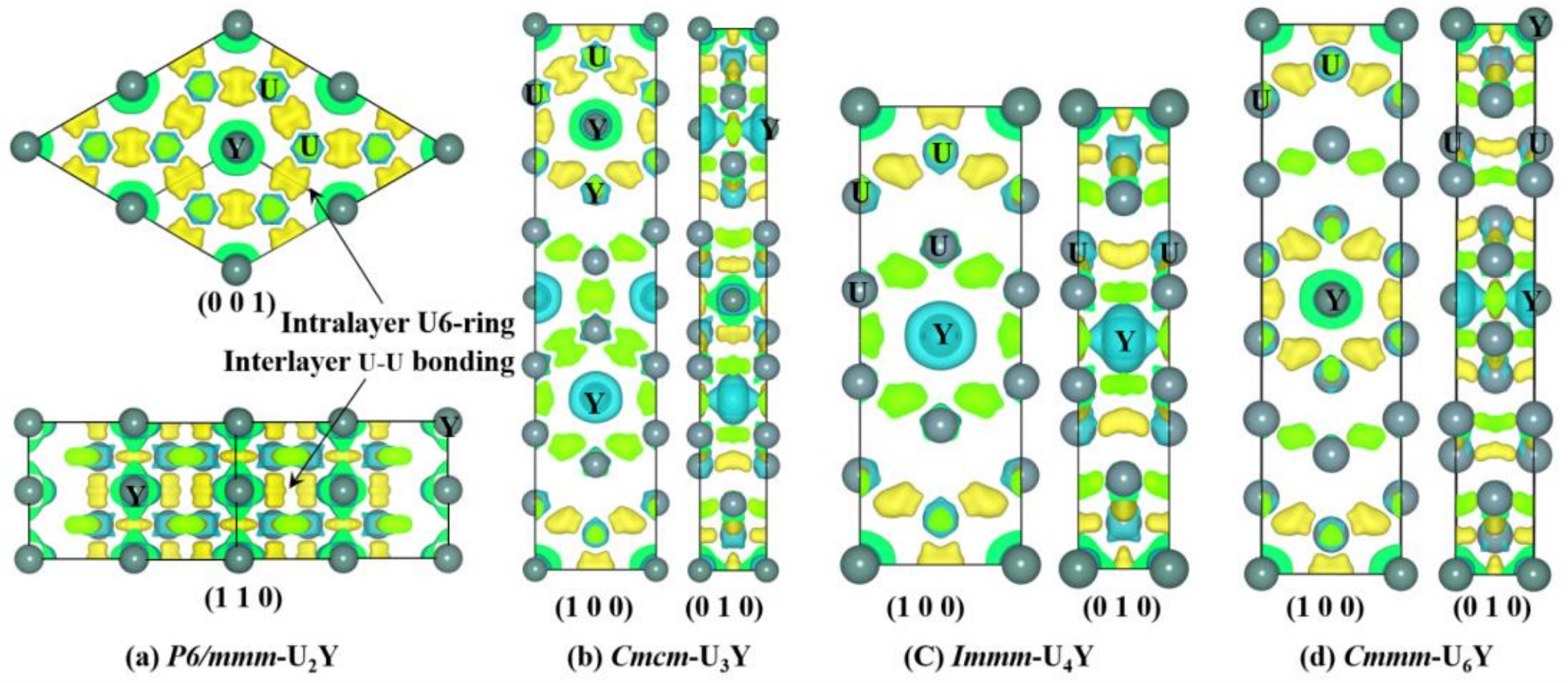

**Fig. S33** Differential charge density (isosurface = 0.02 e/Bohr$^3$) for (a) *P6/mmm*-$U_2Y$, (b) *Cmcm*-$U_3Y$, (c) *Immm*-$U_4Y$, and (d) *Cmmm*-$U_6Y$ at 200 GPa. The blue color indicates loss of electrons and yellow color represents the gain of charge between U atoms.

## SIV. Effect of core-core overlap on pseudopotential

The "Zr_sv" pseudopotential of Zr with a core radius of 2.5 bohr (1.27 Å) and "U" pseudopotential of U with a core radius of 2.8 bohr (1.482 Å) in VASP library were used in our calculations, which will produce 37.5% core-core overlap in *P6/mmm*-$U_2Zr$ at 200 GPa. Therefore, in order to quantify the influence of this core-core overlap on the calculated results under high pressure, we also employed other pseudopotentials with a smaller core radius for comparison. They include the PAW pseudopotential of Quantum Espresso (QE) software with a core radius of 1.5 bohr (0.794 Å) for Zr and 2.1 bohr (1.11 Å) for U. In addition, the ultra-soft pseudopotentials (USPPs) of CASTEP software with a core radius of 2.1 bohr (1.111 Å) for Zr and 2.1 bohr (1.111 Å) for U were also used for comparison.

The nearest-neighboring U-U, U-Zr and Zr-Zr distance in *P6/mmm*-$U_2Zr$ at 200 GPa in our calculation are 2.408, 2.721 and 2.408 Å, respectively. Thus, neither the PAW pseudopotential of QE nor USPPs of CASTEP has a core-core overlap. Other U2X (X= Sc, Ti, V, Cr, Y, Nb, Mo, Hf, Ta, W) systems are similar, and the corresponding pseudopotential data and U-U bond lengths are shown in Table S2. We used these pseudopotentials to calculate the equation of state (EOS) for *P6/mmm*-$U_2X$. The results are presented in Fig. S34. It can be seen that the EOS calculated by VASP is very consistent with those of QE and CASTEP, with the maximum difference less than 2%, which guarantees the reliability of the results calculated by VASP under high pressure. In general, for the PAW potential implemented in the VASP code, a slight core-core overlap has negligible impact on the calculated results.

In fact, when constructing the PAW potential using spherical Bessel functions, as implemented in the VASP code, a large overlap between atomic spheres is permitted. An extreme illustration of core-core overlap can be seen in hydrogen molecules at low pressures. The core radius of the standard PAW pseudopotential for hydrogen in VASP is 0.58 Å, while the H-H distance in an $H_2$ molecule is 0.74 Å. This results in an overlap of nearly 60% in the pseudopotential. However, this substantial overlap has minimal impact on its applicability in hydrogen-related research.

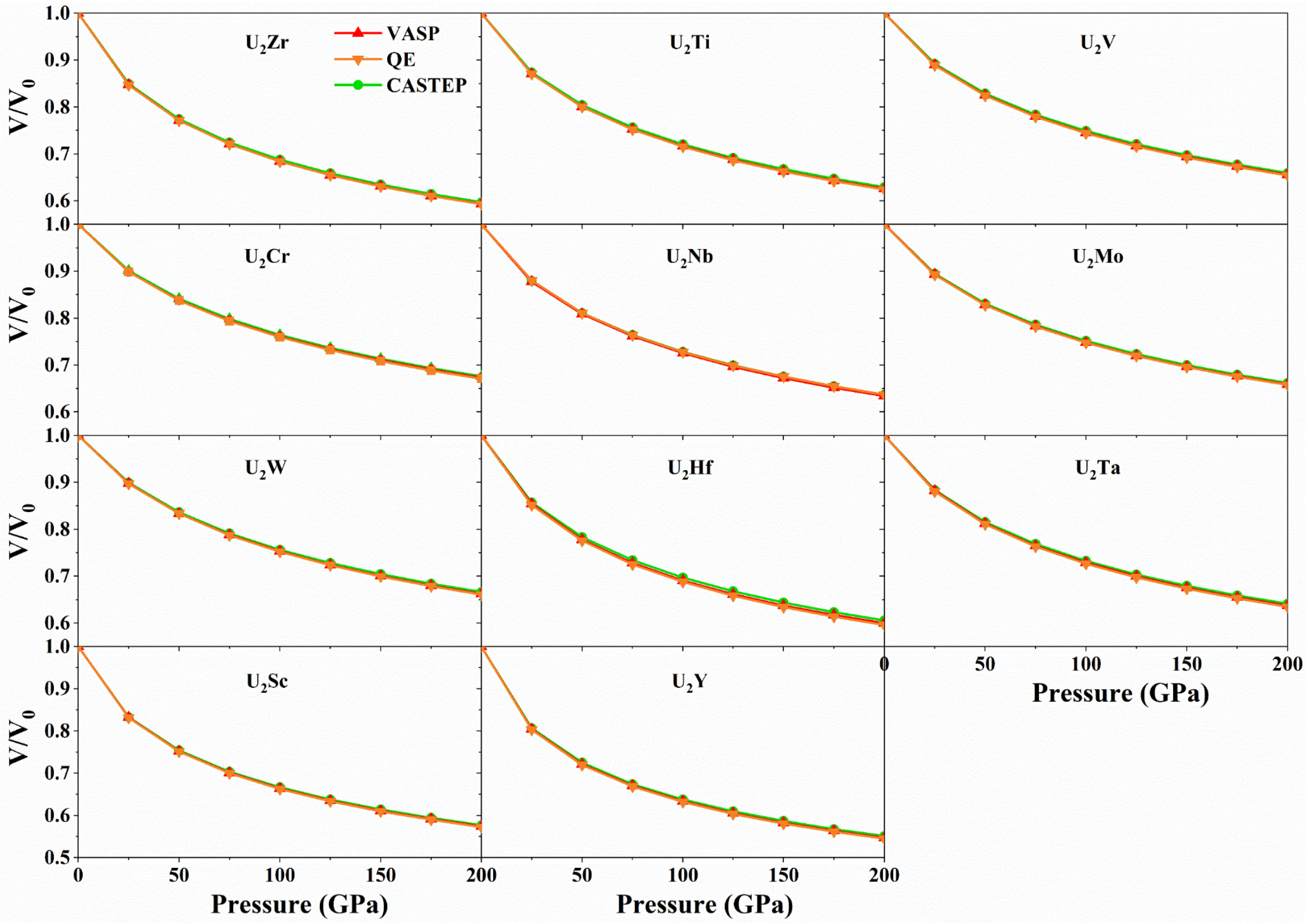

**Fig. S34** Calculated equation of state of *P6/mmm*-$U_2X$ (X= Zr, Ti, V, Cr, Nb, Mo, W, Hf, Ta, Sc, Y) at zero Kelvin with different pseudopotentials.

**Table S2.** Pseudopotential data (core radius in Å), the nearest-neighboring U-U bond (Å) and its core-core overlap in *P6/mmm*-$U_2X$ (X= Zr, Ti, V, Cr, Nb, Mo, W, Hf, Ta, Sc, Y) in VASP at 200 GPa.

| Atom | VASP | | QE | CASTEP |
|---|---|---|---|---|
| | GGA PBE | Radius | (PAW) Radius | (USPP) Radius |
| U | U | 1.482 | 1.111 | 1.111 |
| Zr | Zr_sv | 1.323 | 0.794 | 1.111 |
| Sc | Sc_sv | 1.323 | 0.635 | 0.953 |
| Ti | Ti_sv | 1.217 | 0.847 | 0.953 |
| V | V_sv | 1.217 | 0.794 | 0.953 |
| Cr | Cr_pv | 1.217 | 0.741 | 0.953 |
| Nb | Nb_sv | 1.270 | 0.794 | 0.847 |
| Mo | Mo_sv | 1.323 | 0.741 | 0.847 |
| W | W_sv | 1.323 | 1.111 | 1.270 |
| Hf | Hf_pv | 1.376 | 0.900 | 1.111 |
| Ta | Ta_pv | 1.323 | 0.794 | 1.270 |
| Y | Y_sv | 1.482 | 0.847 | 1.058 |

| Phase | U-U bond (Å) | Overlap in VASP (%) |
|---|---|---|
| - | - | - |
| $U_2Zr$ | 2.408 | 37.5 |
| $U_2Sc$ | 2.367 | 40.2 |
| U2Ti | 2.373 | 39.8 |
| $U_2V$ | 2.363 | 40.5 |
| $U_2Cr$ | 2.324 | 43.1 |
| $U_2Nb$ | 2.418 | 36.8 |
| $U_2Mo$ | 2.373 | 39.9 |
| $U_2W$ | 2.387 | 38.9 |
| $U_2Hf$ | 2.416 | 36.9 |
| $U_2Ta$ | 2.430 | 36.0 |
| $U_2Y$ | 2.386 | 39.0 |

# SV. Supplementary Figures

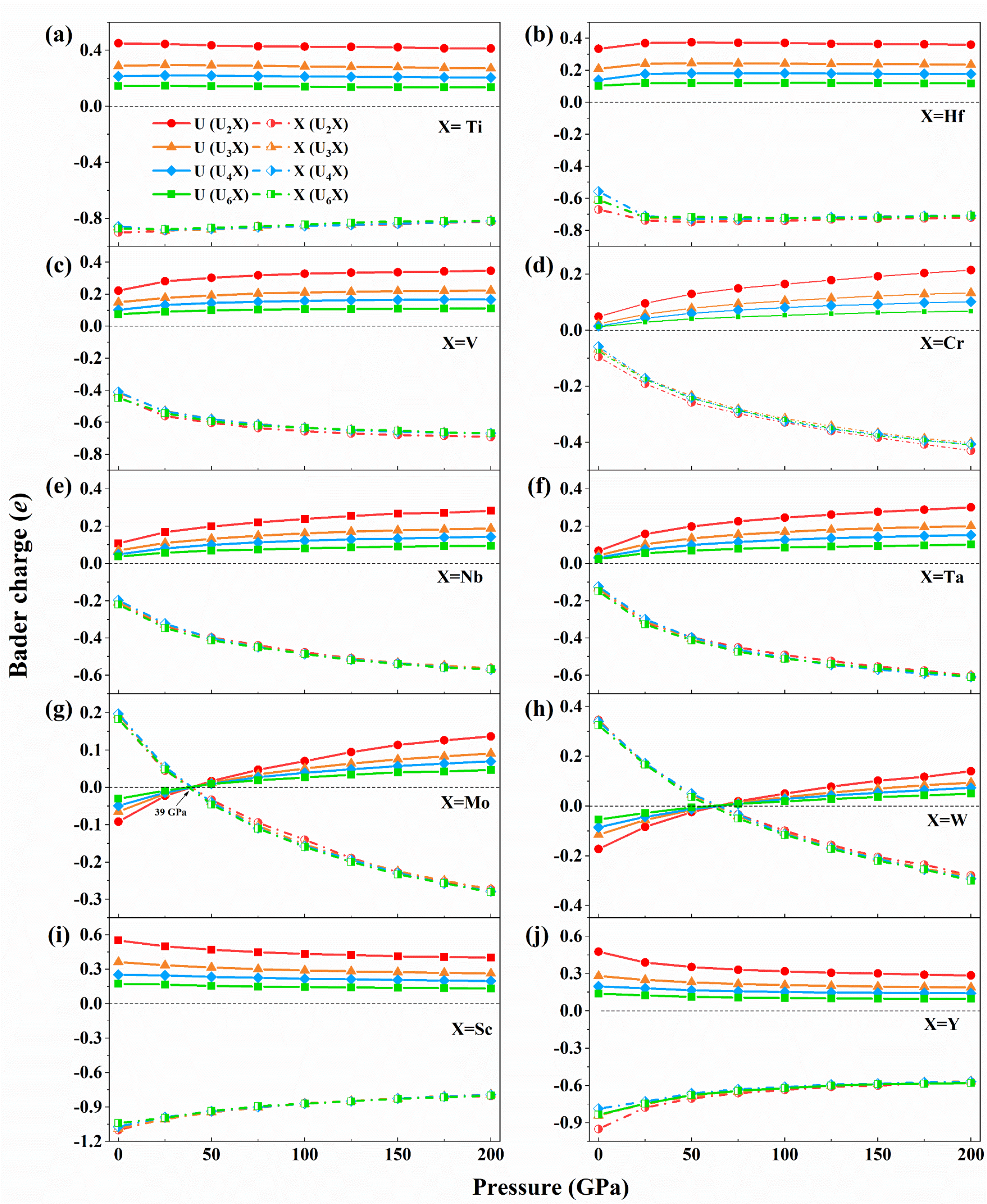

**Fig. S35** (color online) Pressure dependence of the averaged Bader charge of U and X atoms for *P6mmm*-$U_2$X, *Cmcm*-$U_3$X, *Immm*-$U_4$X, and *Cmmm*-$U_6$X phases (X=Ti, V, Hf, Nb, Cr, Ta, Mo, Sc): (a) X=Ti, (b) X=Hf, (c) X=V, (d) X=Cr, (e) X=Nb, (f) X=Ta, (g) X=Mo, (h) X=W, (i) X=Sc, and (j) X=Y.

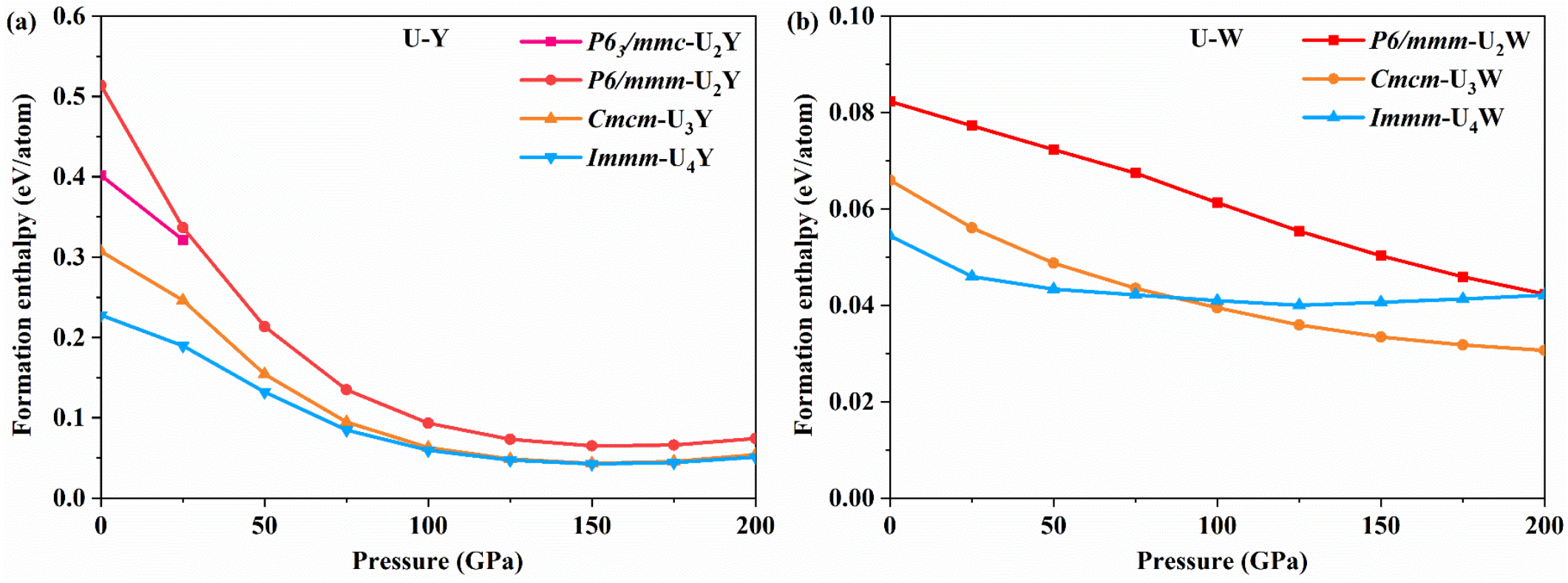

**Fig. S36** (color online) Calculated formation enthalpies of ordered phases in (a) U-Y and (b) U-W systems. The formation enthalpies of these ordered phases are all positive from 0 to 200 GPa. It is noteworthy that the *P6₃/mmc* structure in U-W system spontaneously relaxes into the *P6/mmm* structure.

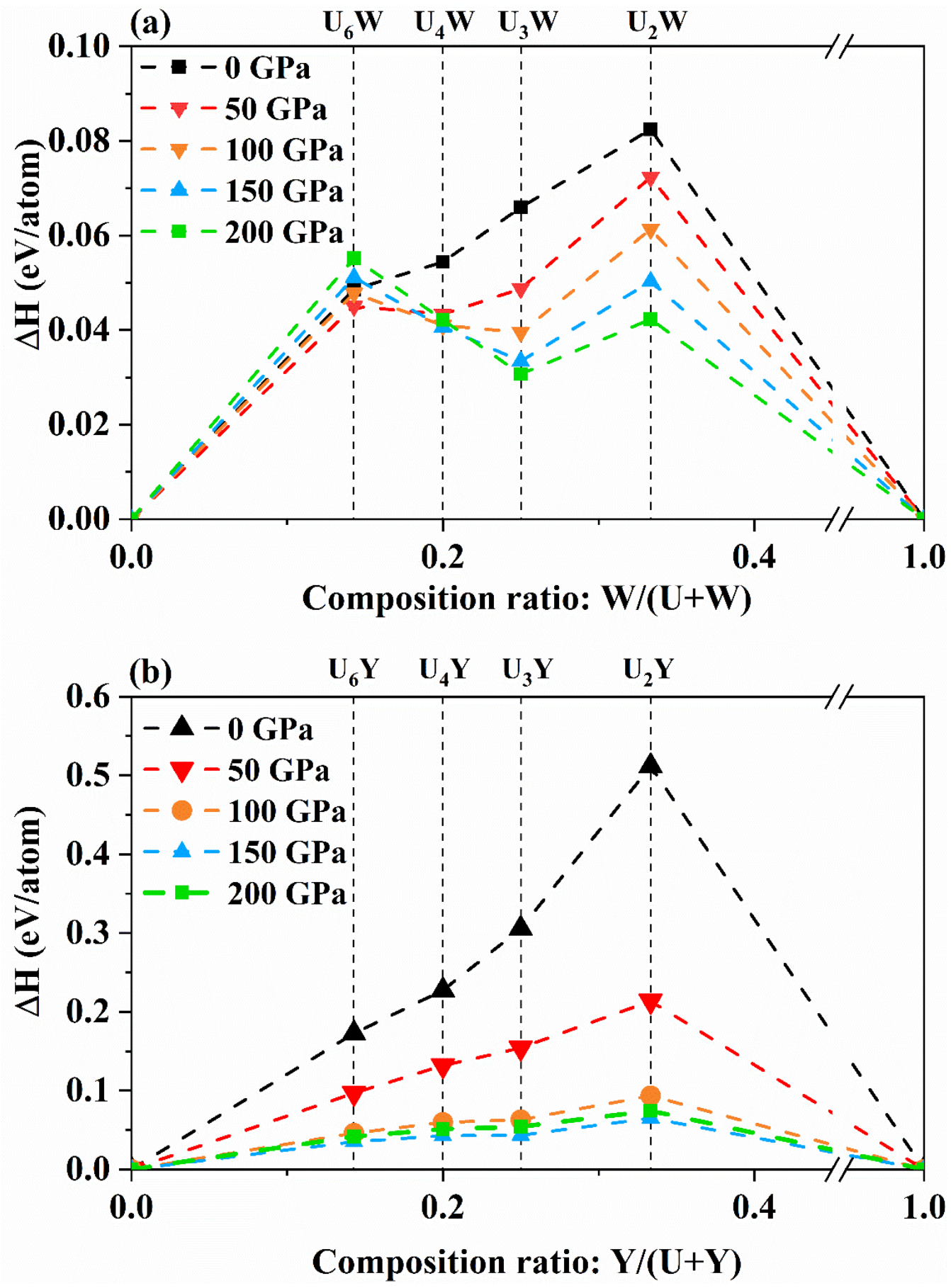

**Fig. S37** (color online) Thermodynamic convex hulls of (a) U-Y and (b) U-W alloys. Formation enthalpies of all stoichiometries we studied in both U-Y and U-W alloys are all positive in the whole pressure range we studied, indicating that the *P6₃/mmc*-$U_2X$, *P6/mmm*-$U_2X$, *Cmcm*-$U_3X$, *Immm*-$U_4X$, and *Cmmm*-$U_6X$ (X= Y, W) structures might be only metastable in either system.

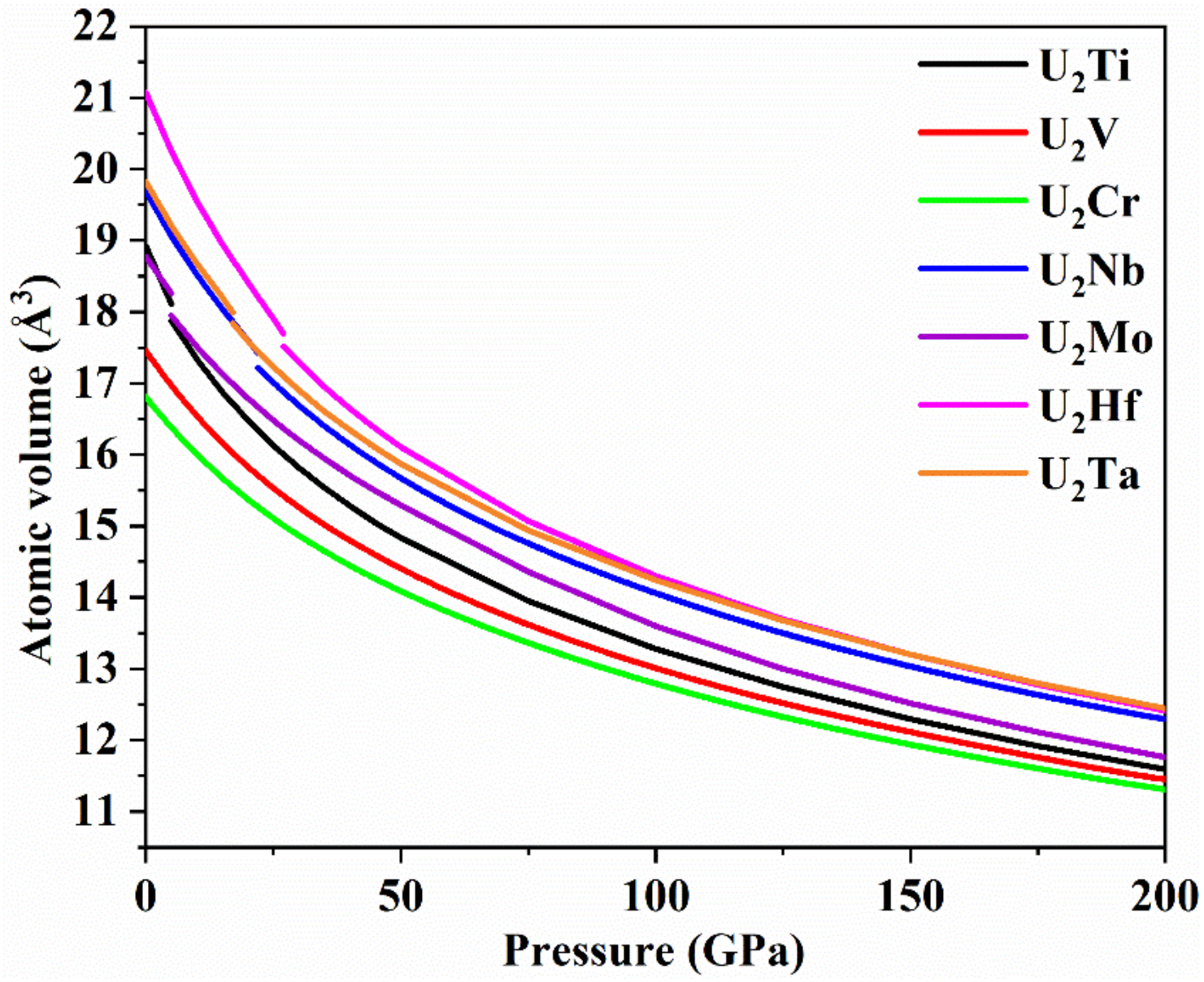

**Fig. S38** (color online) Calculated equation of state of $U_2X$ (X= Ti, V, Cr, Nb, Mo, Hf, Ta) at 0 K. The volume collapses at the phase transition pressure are 1.28%, 1.18%, 1.72%, 1.00% and 0.83% for $U_2X$ (X=Ti, Nb, Mo, Hf, Ta), respectively.

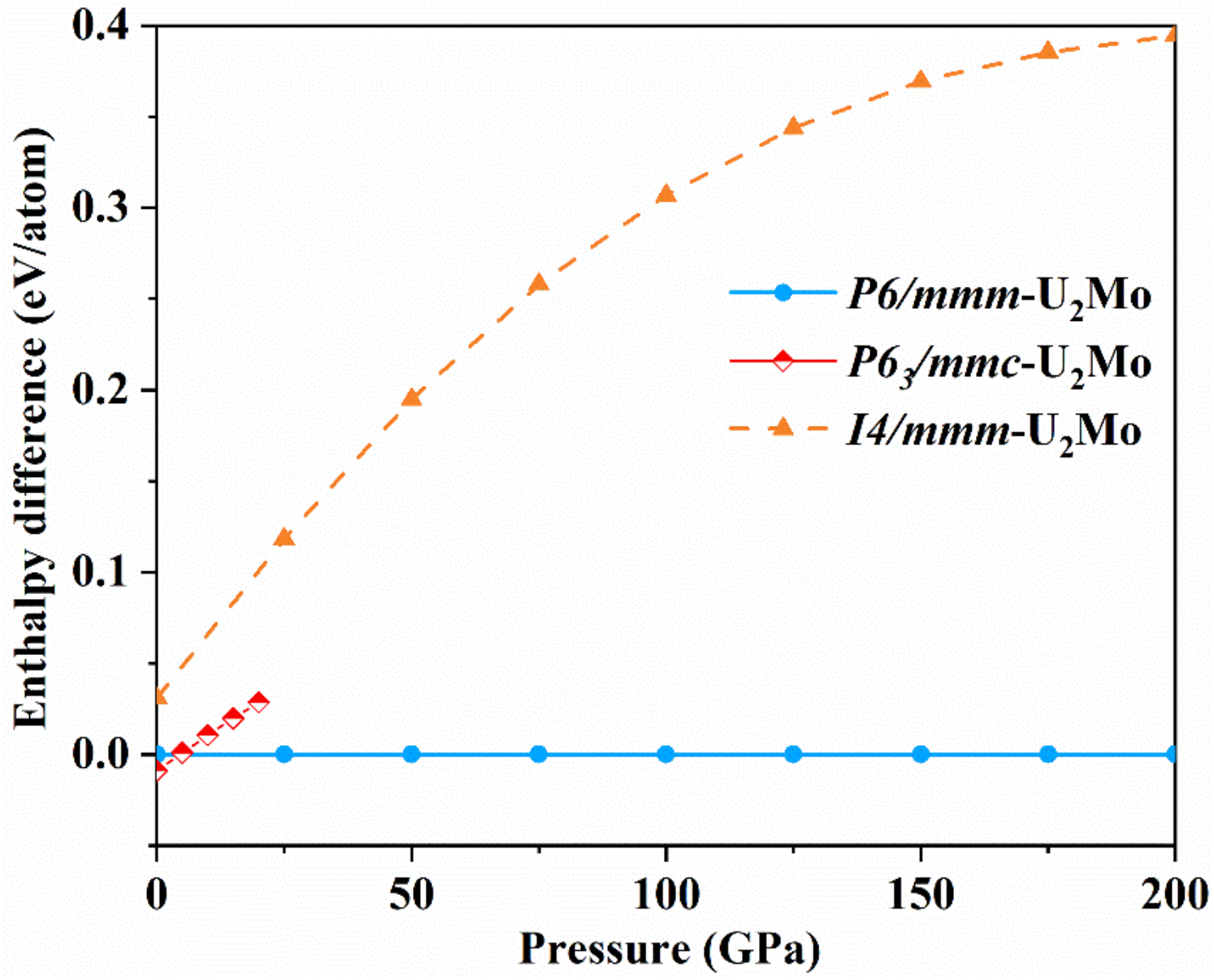

**Fig. S39** (color online) Enthalpy difference of *P6₃/mmc*-$U_2Mo$ and *I4/mmm*-$U_2Mo$ phases with respect to *P6/mmm*-$U_2Mo$ phase. The *P6/mmm*-$U_2Mo$ phase predicted by the DFT calculation is more stable than the *I4/mmm*-$U_2Mo$ phase proposed by the experimental study [3] between 0 and 200 GPa, while the new *P6₃/mmc*-$U_2Mo$ phase is identified as the most stable phase between 0 and 4.5 GPa.

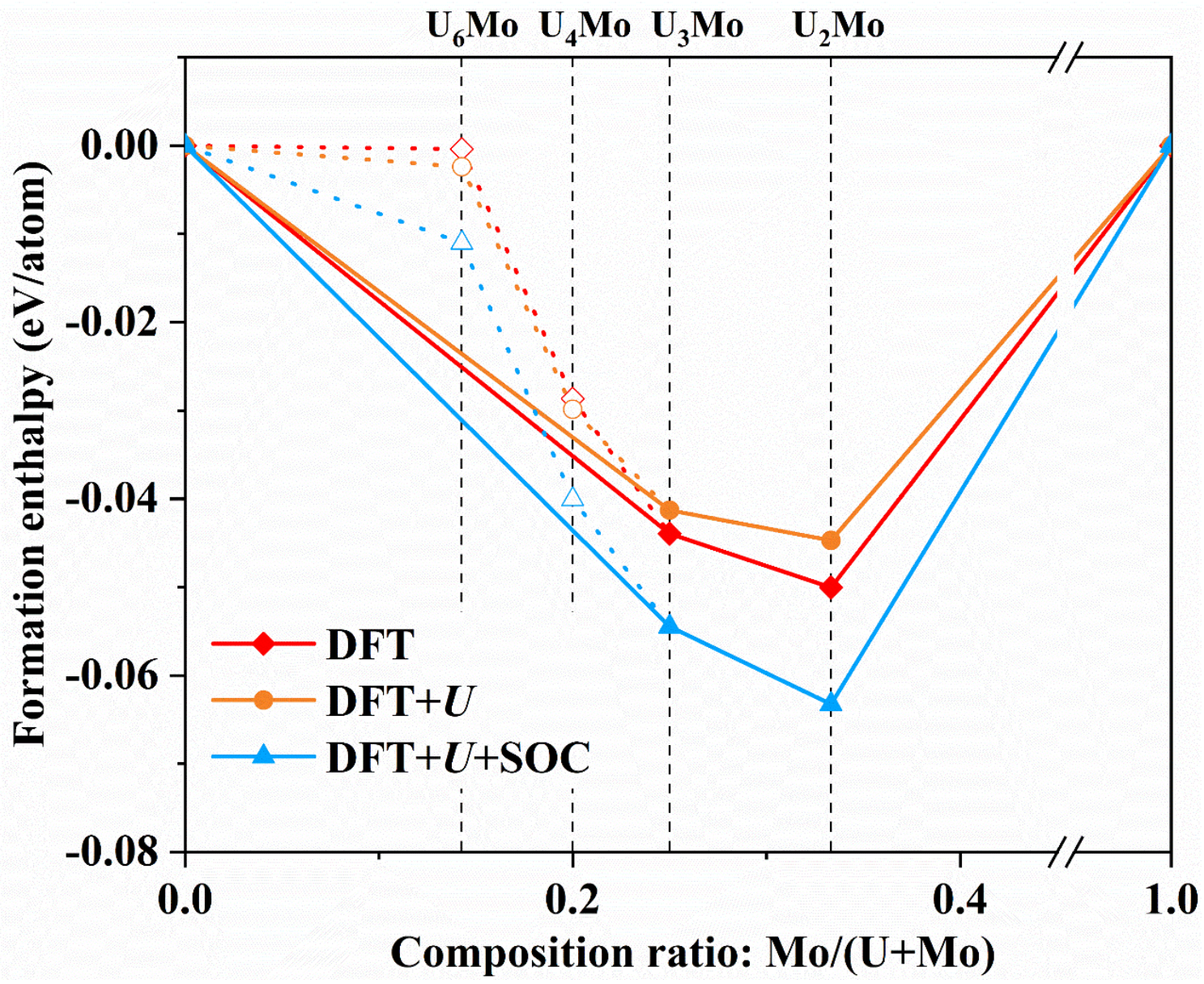

**Fig. S40** Convex hull diagram of U-Mo alloy at 100 GPa calculated by different methods: DFT, DFT+*U*, and DFT+*U*+SOC (*U*= 1.24 eV).

## SVI. Supplementary Tables

The elastic constants of ordered phases in U-Zr alloy at the given pressure are shown in Table S3. According to the criterion of mechanical stability, they all are mechanically stable. The lattice parameters and equilibrium atomic volumes of $U_2Ti$ and $U_2Mo$ at zero pressure calculated by this work are compared with the previous theoretical and experimental data in Table S4. The detailed structural information of the newly discovered ordered phases is presented in Tables S5-S15.

**Table S3.** The calculated elastic constants (in GPa) for $U_2Zr$, $U_3Zr$, $U_4Zr$, and $U_6Zr$ at the given pressures.

| Phase | $C_{11}$ | $C_{22}$ | $C_{33}$ | $C_{12}$ | $C_{13}$ | $C_{23}$ | $C_{44}$ | $C_{55}$ | $C_{66}$ |
|---|---|---|---|---|---|---|---|---|---|
| *P6₃/mmc*-$U_2Zr$ (0 GPa) | 196.92 | 196.92 | 260.90 | 95.47 | 34.41 | 34.41 | 68.26 | 68.26 | 50.72 |
| *P6/mmm*-$U_2Zr$ (100 GPa) | 207.87 | 207.87 | 220.22 | 91.96 | 38.89 | 38.89 | 87.85 | 87.85 | 57.96 |
| *Cmcm*-$U_3Zr$ (100 GPa) | 245.24 | 164.78 | 209.16 | 118.38 | 42.87 | 46.81 | 78.07 | 91.05 | 64.04 |
| *Immm*-$U_4Zr$ (100 GPa) | 291.49 | 168 | 225.55 | 96.86 | 59.73 | 54.28 | 60.73 | 101.41 | 66.62 |
| *Cmmm*-$U_6Zr$ (100 GPa) | 317.61 | 161.83 | 238.65 | 114.31 | 40.01 | 52.64 | 83.80 | 112.25 | 89.22 |
| Partially ordered $\delta$-$UZr_2$ (0 GPa) | 165.75 | 173.97 | 176.96 | 74.04 | 70.02 | 61.34 | 51.02 | 65.99 | 48.20 |
| Fully ordered $\delta$-$UZr_2$ (0 GPa) | 174.67 | 174.67 | 185.46 | 93.01 | 45.43 | 45.43 | 42.88 | 42.88 | 40.83 |

**Table S4.** Lattice parameters (*a* and *c* in Å) and equilibrium atomic volumes (in Å$^3$/atom) of $U_2Ti$ and $U_2Mo$ at zero pressure, along with previous theoretical and experimental data for comparison.

| System | Phase | Method | *a* | *c* | Volume |
|---|---|---|---|---|---|
| $U_2Ti$ | *P6/mmm* | This work | 4.780 | 2.809 | 18.528 |
| | | GGA [4] | 4.773 | 2.805 | 18.457 |
| | | Expt. [5] | 4.828 | 2.847 | 19.157 |
| | *P6₃/mmc* | This work | 4.766 | 5.772 | 18.925 |
| | | GGA [4] | 4.761 | 5.766 | 18.865 |
| $U_2Mo$ | *P6/mmm* | This work | 4.818 | 2.754 | 18.455 |
| | | WIEN2k [6] | 4.821 | 2.773 | 18.605 |
| | *I4/mmm* | This work | 3.432 | 9.691 | 19.028 |
| | | WIEN2k [6] | 3.444 | 9.684 | 19.144 |
| | | Expt. [3] | 3.427 | 9.834 | 19.249 |
| | *P6₃/mmc* | This work | 4.678 | 5.944 | 18.774 |

**Table S5.** Lattice parameters, atomic coordinates and Wyckoff site occupation of *P6₃/mmc*-$U_2Zr$, *P6/mmm*-$U_2Zr$, *Cmcm*-$U_3Zr$, *Immm*-$U_4Zr$, and *Cmmm*-$U_6Zr$ at the given pressure, respectively.

| **Phase** | **Lattice parameters (Å)** | **Atom** | **Site** | **Atomic coordinates** |
|---|---|---|---|---|
| *P6₃/mmc*-$U_2Zr$ (0 GPa) | a=b=4.9368<br>c=6.0563<br>α=β=90°<br>γ=120° | U | *4f* | (0.33333, 0.66667, 0.53750) |
| | | Zr | *2b* | (0.00000, 0.00000, 0.25000) |
| *P6/mmm*-$U_2Zr$ (100 GPa) | a=b=4.4338<br>c=2.5644<br>α=β=90°<br>γ=120° | U | *2d* | (0.33333, 0.66667, 0.50000) |
| | | Zr | *1a* | (0.00000, 0.00000, 0.00000) |
| *Cmcm*-$U_3Zr$ (100 GPa) | a=2.5492<br>b=20.4659<br>c=4.4936<br>α=β=γ=90° | U | *4c* | (0.50000, 0.71454, 0.75000)<br>(0.00000, 0.03208, 1.25000)<br>(0.00000, 0.84293, 0.75000) |
| | | Zr | *4c* | (0.50000, 0.90839, 1.25000) |
| *Immm*-$U_4Zr$ (100 GPa) | a=2.5395<br>b=4.5040<br>c=12.8577<br>α=β=γ=90° | U | *4j*<br>*4i* | (0.00000, 0.50000, 0.69404)<br>(0.00000, 0.00000, 0.60033) |
| | | Zr | *2a* | (0.00000, 0.00000, 0.00000) |
| *Cmmm*-$U_6Zr$ (100 GPa) | a=17.8657<br>b=4.5133<br>c=2.5160<br>α=β=γ=90° | U | *4g*<br>*4h*<br>*4h* | (0.71875, 0.00000, 0.00000)<br>(0.42813, 0.00000, 0.50000)<br>(0.86273, 0.00000, 0.50000) |
| | | Zr | *2a* | (0.00000, 0.00000, 0.00000) |

**Table S6.** Lattice parameters, atomic coordinates and Wyckoff site occupation of *P6₃/mmc*-$U_2Sc$, *P6/mmm*-$U_2Sc$, *Cmcm*-$U_3Sc$, *Immm*-$U_4Sc$, and *Cmmm*-$U_6Sc$ at the given pressure, respectively.

| Phase | Lattice parameters (Å) | Atom | Site | Atomic coordinates |
|---|---|---|---|---|
| *P6₃/mmc*-$U_2Sc$ (0 GPa) | a=b=4.7660<br>c= 5.7722<br>α=β=90°<br>γ=120° | U | *4f* | (0.33333, 0.66667, 0.53505) |
| | | Sc | *2b* | (0.00000, 0.00000, 0.25000) |
| *P6/mmm*-$U_2Sc$ (100 GPa) | a=b= 4.3022<br>c= 2.4846<br>α=β=90°<br>γ=120° | U | *2d* | (0.33333, 0.66667, 0.50000) |
| | | Sc | *1a* | (0.00000, 0.00000, 0.00000) |
| *Cmcm*-$U_3Sc$ (100 GPa) | a= 2.49290<br>b=20.0632<br>c= 4.4329<br>α=β=γ=90° | U | *4c* | (0.00000, 0.28704, 0.75000)<br>(0.00000, 0.65555, 0.75000)<br>(0.00000, 0.53145, 0.75000) |
| | | Sc | *4c* | (0.50000, 0.40820, 0.75000) |
| *Immm*-$U_4Sc$ (100 GPa) | a= 2.4994<br>b=4.4700<br>c=12.5324<br>α=β=γ=90° | U | *4j*<br>*4i* | (0.00000, 0.50000, 0.69094)<br>(0.00000, 0.00000, 0.59914) |
| | | Sc | *2a* | (0.00000, 0.00000, 0.00000) |
| *Cmmm*-$U_6Sc$ (100 GPa) | a=17.6224<br>b=4.5055<br>c=2.5019<br>α=β=γ=90° | U | *4g*<br>*4h*<br>*4h* | (0.71909, 0.00000, 0.00000)<br>(0.42972, 0.00000, 0.50000)<br>(0.86545, 0.00000, 0.50000) |
| | | Sc | *2a* | (0.00000, 0.00000, 0.00000) |

**Table S7.** Lattice parameters, atomic coordinates and Wyckoff site occupation of *P6₃/mmc*-$U_2Ti$, *P6/mmm*-$U_2Ti$, *Cmcm*-$U_3Ti$, *Immm*-$U_4Ti$, and *Cmmm*-$U_6Ti$ at the given pressure, respectively.

| Phase | Lattice parameters (Å) | Atom | Site | Atomic coordinates |
|---|---|---|---|---|
| *P6₃/mmc*-$U_2Ti$ (0 GPa) | a=b= 4.917<br>c= 6.0091<br>α=β=90°<br>γ=120° | U | *4f* | (0.33333, 0.66667, 0.52611) |
| | | Ti | *2b* | (0.00000, 0.00000, 0.25000) |
| *P6/mmm*-$U_2Ti$ (100 GPa) | a=b= 4.3526<br>c= 2.4855<br>α=β=90° | U | *2d* | (0.33333, 0.66667, 0.50000) |
| | | Ti | *1a* | (0.00000, 0.00000, 0.00000) |
| | γ=120° | | | |
| *Cmcm*-$U_3Ti$ (100 GPa) | a= 2.4883 b=19.9704 c= 4.3896 α=β=γ=90° | U | *4c* | (0.00000, 0.28734, 0.75000) (0.00000, 0.65446, 0.75000) (0.00000, 0.53062, 0.75000) |
| | | Ti | *4c* | (0.50000, 0.40847, 0.75000) |
| *Immm*-$U_4Ti$ (100 GPa) | a= 2.4904 b=4.4259 c=12.5445 α=β=γ=90° | U | *4j* *4i* | (0.00000, 0.50000, 0.69045) (0.00000, 0.00000, 0.59838) |
| | | Ti | *2a* | (0.00000, 0.00000, 0.00000) |
| *Cmmm*-$U_6Ti$ (100 GPa) | a=17.6040 b=4.4833 c=2.4947 α=β=γ=90° | U | *4g* *4h* *4h* | (0.71878, 0.00000, 0.00000) (0.43008, 0.00000, 0.50000) (0.86639, 0.00000, 0.50000) |
| | | Ti | *2a* | (0.00000, 0.00000, 0.00000) |

**Table S8.** Lattice parameters, atomic coordinates and Wyckoff site occupation of *P6/mmm*-$U_2V$, *Cmcm*-$U_3V$, *Immm*-$U_4V$, and *Cmmm*-$U_6V$ at the given pressure, respectively.

| **Phase** | **Lattice parameters (Å)** | **Atom** | **Site** | **Atomic coordinates** |
|---|---|---|---|---|
| *P6/mmm*-$U_2V$ (0 GPa) | a=b= 4.7010 c= 2.7366 α=β=90° γ=120° | U | *2d* | (0.33333, 0.66667, 0.50000) |
| | | V | *1a* | (0.00000, 0.00000, 0.00000) |
| *Cmcm*-$U_3V$ (100 GPa) | a= 2.4746 b=19.8969 c= 4.3657 α=β=γ=90° | U | *4c* | (0.00000, 0.28766, 0.75000) (0.00000, 0.65405, 0.75000) (0.00000, 0.53023, 0.75000) |
| | | V | *4c* | (0.50000, 0.40882, 0.75000) |
| *Immm*-$U_4V$ (100 GPa) | a= 2.4766 b=4.4181 c=12.4979 α=β=γ=90° | U | *4j* *4i* | (0.00000, 0.50000, 0.68987) (0.00000, 0.00000, 0.59839) |
| | | V | *2a* | (0.00000, 0.00000, 0.00000) |
| *Cmmm*-$U_6V$ (100 GPa) | a=17.5762 b=4.4735 c=2.4854 α=β=γ=90° | U | *4g* *4h* *4h* | (0.71845, 0.00000, 0.00000) (0.43020, 0.00000, 0.50000) (0.86730, 0.00000, 0.50000) |
| | | V | *2a* | (0.00000, 0.00000, 0.00000) |

**Table S9.** Lattice parameters, atomic coordinates and Wyckoff site occupation of *P6/mmm*-$U_2Cr$, *Cmcm*-$U_3Cr$, *Immm*-$U_4Cr$, and *Cmmm*-$U_6Cr$ at the given pressure, respectively.

| Phase | Lattice parameters (Å) | Atom | Site | Atomic coordinates |
|---|---|---|---|---|
| *P6/mmm*-$U_2Cr$ (100 GPa) | a=b= 4.2824<br>c= 2.4164<br>α=β=90°<br>γ=120° | U | *2d* | (0.33333, 0.66667, 0.50000) |
| | | Cr | *1a* | (0.00000, 0.00000, 0.00000) |
| *Cmcm*-$U_3Cr$ (100 GPa) | a= 2.4471<br>b=19.7574<br>c= 4.3904<br>α=β=γ=90° | U | *4c* | (0.00000, 0.28819, 0.75000)<br>(0.00000, 0.65382, 0.75000)<br>(0.00000, 0.52951, 0.75000) |
| | | Cr | *4c* | (0.50000, 0.40858, 0.75000) |
| *Immm*-$U_4Cr$ (100 GPa) | a= 2.4624<br>b=4.4237<br>c=12.4362<br>α=β=γ=90° | U | *4j*<br>*4i* | (0.00000, 0.50000, 0.68902)<br>(0.00000, 0.00000, 0.59843) |
| | | Cr | *2a* | (0.00000, 0.00000, 0.00000) |
| *Cmmm*-$U_6Cr$ (100 GPa) | a=17.4875<br>b=4.4787<br>c=2.4791<br>α=β=γ=90° | U | *4g*<br>*4h*<br>*4h* | (0.71837, 0.00000, 0.00000)<br>(0.43036, 0.00000, 0.50000)<br>(0.86829, 0.00000, 0.50000) |
| | | Cr | *2a* | (0.00000, 0.00000, 0.00000) |

**Table S10.** Lattice parameters, atomic coordinates and Wyckoff site occupation of *P6₃/mmc*-$U_2Nb$, *P6/mmm*-$U_2Nb$, *Cmcm*-$U_3Nb$, *Immm*-$U_4Nb$, and *Cmmm*-$U_6Nb$ at the given pressure, respectively.

| Phase | Lattice parameters (Å) | Atom | Site | Atomic coordinates |
|---|---|---|---|---|
| *$P6_3$/mmc*-$U_2Nb$ (0 GPa) | a=b= 4.8066<br>c= 5.9055<br>α=β=90°<br>γ=120° | U | *4f* | (0.33333, 0.66667, 0.53130) |
| | | Nb | *2b* | (0.00000, 0.00000, 0.25000) |
| *P6/mmm*-$U_2Nb$ (100 GPa) | a=b= 4.3770<br>c= 2.5420 | U | *2d* | (0.33333, 0.66667, 0.50000) |
| | α=β=90° γ=120° | Nb | *1a* | (0.00000, 0.00000, 0.00000) |
| *Cmcm*-$U_3Nb$ (100 GPa) | a= 2.5258 b=20.2965 c= 4.4328 | U | *4c* | (0.00000, 0.28564, 0.75000) (0.00000, 0.65644, 0.75000) (0.00000, 0.53146, 0.75000) |
| | α=β=γ=90° | Nb | *4c* | (0.50000, 0.40851, 0.75000) |
| *Immm*-$U_4Nb$ (100 GPa) | a=2.5138 b=4.4509 | U | *4j* *4i* | (0.00000, 0.50000, 0.69382) (0.00000, 0.00000, 0.59994) |
| | c=12.7669 α=β=γ=90° | Nb | *2a* | (0.00000, 0.00000, 0.00000) |
| *Cmmm*-$U_6Nb$ (100 GPa) | a=17.818 b=4.4983 c=2.5135 | U | *4g* *4h* *4h* | (0.71853, 0.00000, 0.00000) (0.42852, 0.00000, 0.50000) (0.86337, 0.00000, 0.50000) |
| | α=β=γ=90° | Nb | *2a* | (0.00000, 0.00000, 0.00000) |

**Table S11.** Lattice parameters, atomic coordinates and Wyckoff site occupation of *P6₃/mmc*-$U_2Mo$, *P6/mmm*-$U_2Mo$, *Cmcm*-$U_3Mo$, *Immm*-$U_4Mo$, and *Cmmm*-$U_6Mo$ at the given pressure, respectively.

| **Phase** | **Lattice parameters (Å)** | **Atom** | **Site** | **Atomic coordinates** |
|---|---|---|---|---|
| *$P6_3/mmc$*-$U_2Mo$ (0 GPa) | a=b= 4.6778 c= 5.9443 | U | *4f* | (0.33333, 0.66667, 0.53024) |
| | α=β=90° γ=120° | Mo | *2b* | (0.00000, 0.00000, 0.25000) |
| *P6/mmm*-$U_2Mo$ (100 GPa) | a=b= 4.3966 c= 2.4750 | U | *2d* | (0.33333, 0.66667, 0.50000) |
| | α=β=90° γ=120° | Mo | *1a* | (0.00000, 0.00000, 0.00000) |
| *Cmcm*-$U_3Mo$ (100 GPa) | a= 2.4906 b=20.2787 c= 4.4424 | U | *4c* | (0.00000, 0.28563, 0.75000) (0.00000, 0.65605, 0.75000) (0.00000, 0.53102, 0.75000) |
| | α=β=γ=90° | Mo | *4c* | (0.50000, 0.40798, 0.75000) |
| *Immm*-$U_4Mo$ (100 GPa) | a=2.4960 b=4.4468 | U | *4j* *4i* | (0.00000, 0.50000, 0.69360) (0.00000, 0.00000, 0.59989) |
| | c=12.7527 α=β=γ=90° | Mo | *2a* | (0.00000, 0.00000, 0.00000) |
| *Cmmm*-$U_6Mo$ (100 GPa) | a=17.8734 b=4.4795 c=2.5013 | U | *4g* *4h* *4h* | (0.71830, 0.00000, 0.00000) (0.42887, 0.00000, 0.50000) (0.86354, 0.00000, 0.50000) |
| | α=β=γ=90° | Mo | *2a* | (0.00000, 0.00000, 0.00000) |

**Table S12.** Lattice parameters, atomic coordinates and Wyckoff site occupation of *P6₃/mmc*-$U_2Hf$, *P6/mmm*-$U_2Hf$, *Cmcm*-$U_3Hf$, *Immm*-$U_4Hf$, and *Cmmm*-$U_6Hf$ at the given pressure, respectively.

| Phase | Lattice parameters (Å) | Atom | Site | Atomic coordinates |
|---|---|---|---|---|
| *$P6_3/mmc$*-$U_2Hf$ (0 GPa) | a=b= 4.9282<br>c= 6.0117<br>α=β=90°<br>γ=120° | U | *4f* | (0.33333, 0.66667, 0.53518) |
| | | Hf | *2b* | (0.00000, 0.00000, 0.25000) |
| *P6/mmm*-$U_2Hf$ (100 GPa) | a=b= 4.4119<br>c= 2.5450<br>α=β=90°<br>γ=120° | U | *2d* | (0.33333, 0.66667, 0.50000) |
| | | Hf | *1a* | (0.00000, 0.00000, 0.00000) |
| *Cmcm*-$U_3Hf$ (100 GPa) | a= 2.5325<br>b=20.3713<br>c= 4.4612<br>α=β=γ=90° | U | *4c* | (0.00000, 0.28542, 0.75000)<br>(0.00000, 0.65690, 0.75000)<br>(0.00000, 0.53218, 0.75000) |
| | | Hf | *4c* | (0.50000, 0.40828, 0.75000) |
| *Immm*-$U_4Hf$ (100 GPa) | a=2.5232<br>b=4.4717<br>c=12.7801<br>α=β=γ=90° | U | *4j*<br>*4i* | (0.00000, 0.50000, 0.69411)<br>(0.00000, 0.00000, 0.60004) |
| | | Hf | *2a* | (0.00000, 0.00000, 0.00000) |
| *Cmmm*-$U_6Hf$ (100 GPa) | a=17.8742<br>b=4.5054<br>c=2.5190<br>α=β=γ=90° | U | *4g*<br>*4h*<br>*4h* | (0.71882, 0.00000, 0.00000)<br>(0.42840, 0.00000, 0.50000)<br>(0.86259, 0.00000, 0.50000) |
| | | Hf | *2a* | (0.00000, 0.00000, 0.00000) |

**Table S13.** Lattice parameters, atomic coordinates and Wyckoff site occupation of *$P6_3/mmc$*-$U_2Ta$, *P6/mmm*-$U_2Ta$, *Cmcm*-$U_3Ta$, *Immm*-$U_4Ta$, and *Cmmm*-$U_6Ta$ at the given pressure, respectively.

| Phase | Lattice | Atom | Site | Atomic coordinates |
|---|---|---|---|---|

| | parameters (Å) | | | |
|---|---|---|---|---|
| *P6₃/mmc*-$U_2Ta$ (0 GPa) | a=b= 4.8342<br>c= 5.8759<br>α=β=90°<br>γ=120° | U | *4f* | (0.33333, 0.66667, 0.52966) |
| | | Ta | *2b* | (0.00000, 0.00000, 0.25000) |
| *P6/mmm*-$U_2Ta$ (100 GPa) | a=b= 4.3964<br>c= 2.5520<br>α=β=90°<br>γ=120° | U | *2d* | (0.33333, 0.66667, 0.50000) |
| | | Ta | *1a* | (0.00000, 0.00000, 0.00000) |
| *Cmcm*-$U_3Ta$ (100 GPa) | a= 2.5343<br>b=20.3932<br>c= 4.4404<br>α=β=γ=90° | U | *4c* | (0.00000, 0.28535, 0.75000)<br>(0.00000, 0.65673, 0.75000)<br>(0.00000, 0.53169, 0.75000) |
| | | Ta | *4c* | (0.50000, 0.40845, 0.75000) |
| *Immm*-$U_4Ta$ (100 GPa) | a=2.5205<br>b=4.4528<br>c=12.8301<br>α=β=γ=90° | U | *4j*<br>*4i* | (0.00000, 0.50000, 0.69443)<br>(0.00000, 0.00000, 0.60024) |
| | | Ta | *2a* | (0.00000, 0.00000, 0.00000) |
| *Cmmm*-$U_6Ta$ (100 GPa) | a=17.9098<br>b=4.4918<br>c=2.5188<br>α=β=γ=90° | U | *4g*<br>*4h*<br>*4h* | (0.71853, 0.00000, 0.00000)<br>(0.42820, 0.00000, 0.50000)<br>(0.86250, 0.00000, 0.50000) |
| | | Ta | *2a* | (0.00000, 0.00000, 0.00000) |

**Table S14.** Lattice parameters, atomic coordinates and Wyckoff site occupation of *P6₃/mmc*-$U_2Y$, *P6/mmm*-$U_2Y$, *Cmcm*-$U_3Y$, *Immm*-$U_4Y$, and *Cmmm*-$U_6Y$ at the given pressure, respectively.

| Phase | Lattice parameters (Å) | Atom | Site | Atomic coordinates |
|---|---|---|---|---|
| *P6₃/mmc*-$U_2Y$ (0 GPa) | a=b= 5.1205<br>c= 6.2178<br>α=β=90°<br>γ=120° | U | *4f* | (0.33333, 0.66667, 0.54632) |
| | | Y | *2b* | (0.00000, 0.00000, 0.25000) |
| *P6/mmm*-$U_2Y$ (100 GPa) | a=b= 4.4803<br>c= 2.4952<br>α=β=90°<br>γ=120° | U | *2d* | (0.33333, 0.66667, 0.50000) |
| | | Y | *1a* | (0.00000, 0.00000, 0.00000) |
| *Cmcm*-$U_3Y$ (100 GPa) | a= 2.5011<br>b=20.5129<br>c= 4.5209 | U | *4c* | (0.00000, 0.28590, 0.75000)<br>(0.00000, 0.65893, 0.75000)<br>(0.00000, 0.53351, 0.75000) |
| | α=β=γ=90° | Y | *4c* | (0.50000, 0.40912, 0.75000) |
| *Immm*-$U_4Y$ (100 GPa) | a=2.5033 b=4.4999 c=12.8906 α=β=γ=90° | U | *4j* *4i* | (0.00000, 0.50000, 0.69477) (0.00000, 0.00000, 0.60181) |
| | | Y | *2a* | (0.00000, 0.00000, 0.00000) |
| *Cmmm*-$U_6Y$ (100 GPa) | a=18.0004 b=4.5159 c=2.5071 α=β=γ=90° | U | *4g* *4h* *4h* | (0.71900, 0.00000, 0.00000) (0.42738, 0.00000, 0.50000) (0.86184, 0.00000, 0.50000) |
| | | Y | *2a* | (0.00000, 0.00000, 0.00000) |

**Table S15.** Lattice parameters, atomic coordinates and Wyckoff site occupation of *P6/mmm*-$U_2W$, *Cmcm*-$U_3W$, *Immm*-$U_4W$, and *Cmmm*-$U_6W$ at the given pressure, respectively.

| **Phase** | **Lattice parameters (Å)** | **Atom** | **Site** | **Atomic coordinates** |
|---|---|---|---|---|
| *P6/mmm*-$U_2W$ (0 GPa) | a=b= 4.8392 c= 2.7654 α=β=90° γ=120° | U | *2d* | (0.33333, 0.66667, 0.50000) |
| | | W | *1a* | (0.00000, 0.00000, 0.00000) |
| *Cmcm*-$U_3W$ (100 GPa) | a= 2.5003 b=20.4130 c= 4.4568 α=β=γ=90° | U | *4c* | (0.00000, 0.28520, 0.75000) (0.00000, 0.65655, 0.75000) (0.00000, 0.53129, 0.75000) |
| | | W | *4c* | (0.50000, 0.40787, 0.75000) |
| *Immm*-$U_4W$ (100 GPa) | a=2.5041 b=4.4495 c=12.8454 α=β=γ=90° | U | *4j* *4i* | (0.00000, 0.50000, 0.69456) (0.00000, 0.00000, 0.60042) |
| | | W | *2a* | (0.00000, 0.00000, 0.00000) |
| *Cmmm*-$U_6W$ (100 GPa) | a=17.9936 b=4.4769 c=2.5063 α=β=γ=90° | U | *4g* *4h* *4h* | (0.71839, 0.00000, 0.00000) (0.42830, 0.00000, 0.50000) (0.86245, 0.00000, 0.50000) |
| | | W | *2a* | (0.00000, 0.00000, 0.00000) |

## Supplementary references